\documentclass[oneside,11pt]{article}
\usepackage{jheppub} 
\usepackage{accents}

\usepackage{amsfonts,amsmath,amssymb,ascmac,bm,tensor}
\usepackage{fnpct} 
\usepackage{comment}
\usepackage{ifpdf}
\usepackage{slashed}
\usepackage{color}
\usepackage[mathscr]{eucal}
\usepackage[utf8]{inputenc}
\usepackage{physics}
\usepackage{cancel}
\usepackage{booktabs}
\usepackage{longtable}

\makeatletter
\g@addto@macro\bfseries{\boldmath}
\makeatother

\usepackage{blindtext} 

   \usepackage{graphicx}

\definecolor{red}{rgb}{1,0,0}
\definecolor{dblue}{RGB}{0, 0, 128}

\newcommand{\NL}{\text{NL} }
\newcommand{\ee}{\mathrm{e}}

\newcommand{\fo}{{(1)}}
\newcommand{\so}{{(2)}}
\newcommand{\rr}{{\text{r}}}
\newcommand{\mm}{{\text{m}}}
\newcommand{\New}{{\text{N}}}
\newcommand{\Syn}{{\text{S}}}

\newcommand{\bz}{{\boldsymbol{0}}}

\newcommand{\conv}{{T_\mm (p_1)T_\mm (p_2)\zeta_I^{(1)}(\p_1)\zeta_I^{(1)}(\p_2)}}

\newcommand{\convr}{{\zeta_{I}^{(1)}(\p_1)\zeta_{I}^{(1)}(\p_2)}}
\newcommand{\convimpl}[1]{\underline{ #1 }}

\newcommand{\tot}{{\text{tot}}}

\newcommand{\Si}{\text{Si}}
\newcommand{\Ci}{\text{Ci}}

\newcommand \cB {{\mathcal B}}
\newcommand \cC {{\mathcal C}}
\newcommand \cD {{\mathcal D}}
\newcommand \cL {{\mathcal L}}
\newcommand \cH {{\mathcal H}}
\newcommand \cI {{\mathcal I}}
\newcommand \cJ {{\mathcal J}}
\newcommand \cK {{\mathcal K}}

\def \O {{\mathcal O}}
\newcommand \hE {\hat{E}}
\newcommand \hN {\hat{N}}

\newcommand{\yy}{{\tilde y}}

\newcommand{\Dm}{D_{\rm m}}
\newcommand{\Dr}{D_{\rm r}}
\newcommand{\Dmr}{D_{{\rm m}/{\rm r}}}

\newcommand{\Tm}{T_{\rm m}}
\renewcommand{\Tr}{T_{\rm r}}
\newcommand{\Tmr}{T_{{\rm m}/{\rm r}}}
\newcommand{\SU}{{\rm SU}}

\renewcommand{\b}{{\mathbf b}}
\newcommand{\x}{{\mathbf x}}
\renewcommand{\k}{{\mathbf k}}
\newcommand{\q}{{\mathbf q}}
\newcommand{\p}{{\mathbf p}}

\def\nn{\nonumber\\}

\newcommand{\onetwo}{{ 1\! \leftrightarrow\!  2}}

\newcommand{\dblue}[1]{{\color{dblue}{#1}}}

\makeatletter
\newcommand\footnoteref[1]{\protected@xdef\@thefnmark{\ref{#1}}\@footnotemark}
\makeatother

\allowdisplaybreaks[1]

\begin{document}

 
\title{
Synchronizing the Consistency Relation
}

\preprint{}
\makeatletter
\def\@fpheader{\ }
\makeatother

\author[1]{Keisuke Inomata,}
\author[1]{Hayden Lee,}
\author[1,2]{Wayne Hu}

\affiliation[1]{Kavli Institute for Cosmological Physics and Enrico Fermi Institute, The University of Chicago, Chicago, IL 60637, USA}
\affiliation[2]{Department of Astronomy \& Astrophysics, The University of Chicago, Chicago, IL 60637, USA}

\emailAdd{inomata@uchicago.edu, haydenl@uchicago.edu, \\
whu@background.uchicago.edu} 

\abstract{
 We study the $N$-point function of the density contrast to quadratic order in the squeezed limit during the matter-dominated (MD) and radiation-dominated (RD) eras in synchronous gauge.
	Since synchronous gauge follows the free-fall frame of  observers,  the equivalence principle dictates that in the gradient approximation for the long-wavelength mode there is only a  single, manifestly time-independent consistency relation for the $N$-point function. This simple form is dictated by the {\it initial} mapping between synchronous and local coordinates, unlike Newtonian gauge and its correspondingly separate dilation and Newtonian consistency relations. 
    Dynamical effects only appear at quadratic order in the squeezed limit and are again characterized by a change in the local background, also known as the separate universe approach.  We show that for the 3-point function the compatibility between these squeezed-limit relations and second-order perturbation theory requires both the initial and dynamical contributions to match, as they do in single-field inflation.  This clarifies the role of evolution or late-time projection effects in establishing the consistency relation for observable bispectra, which is especially important for radiation acoustic oscillations and for establishing consistency below the matter-radiation equality scale in the MD era.  Defining an appropriate angle and time average of these oscillations is also important for making separate universe predictions of spatially varying local observables during the RD era, which can be useful for a wider range of cosmological predictions beyond $N$-point functions. 
}
\maketitle

\clearpage
\section{Introduction}

The inflation era in the early universe sets the initial condition of the cosmological perturbations, which seed the cosmic microwave background (CMB) anisotropies and the large scale structure (LSS)~\cite{Mukhanov:1981xt,Mukhanov:1982nu,Starobinsky:1982ee, Guth:1982ec,Panagiotakopoulos:1982rn,Bardeen:1983qw}.
The cosmological perturbations are often characterized by the initial comoving curvature perturbations\footnote{We put the subscript $I$ on $\zeta_I$ to indicate that it is the initial value of the curvature perturbation at the end of inflation.} $\zeta_I$. 
Their power spectrum $P_{\zeta_I}$ has now been precisely measured on the CMB scales~\cite{Planck:2018vyg}, revealing a nearly scale-invariant shape. However, this information about the initial condition is essentially kinematic, following from the approximate de Sitter symmetry of inflation, and the underlying dynamics of inflation remains largely untested.

To gain a deeper understanding of the microphysics of inflation, it is essential to measure higher-order correlations or non-Gaussianity of the curvature perturbations. Such measurements provide a more direct window into the dynamical content of inflation, and can help distinguish between different inflationary models~\cite{Achucarro:2022qrl}.
Primordial non-Gaussianity can be observed in the CMB bispectrum~\cite{Spergel:1999xn,Goldberg:1999xm,Komatsu:2001rj,Bartolo:2004if} and the statistics of large-scale structure including scale-dependent halo bias~\cite{Dalal:2007cu,Slosar:2008hx}, and upper limits on its presence have already ruled out some multi-field and low sound speed inflationary parameter space.

In single-field inflation models, the three-point correlation function of the curvature perturbations in the squeezed limit satisfies the following consistency relation~\cite{Maldacena:2002vr} (see also \cite{Creminelli:2004yq,Cheung:2007sv, Creminelli:2012ed,Hinterbichler:2012nm,Hinterbichler:2013dpa}):
\begin{align}
	{\expval{\zeta_I(\mathbf q) \zeta_I({\mathbf k_1}) \zeta_I({\mathbf k_2})}} &= -   (n_s(k_S)-1)  P_{\zeta_I}(q)  P_{\zeta_I}(k_S) \,(2\pi)^3 \delta_D({\mathbf q}+\mathbf k_1 + \mathbf k_2)  + \mathcal O(q^2)\,,
	\label{eq:original_cons_rel}
\end{align}	
where $\delta_D$ is the Dirac delta function and
\begin{equation}
n_s(k_S)-1 = \frac{\dd \ln k_S^3 P_{\zeta_I}(k_S)}{\dd \ln k_S}\,,
\end{equation}
characterizes the tilt, with ${\mathbf k}_S = ({\mathbf k}_1 - {\mathbf k}_2)/2$ and $q \ll k_1, k_2$. 
This relation stems from the fact that the long-wavelength mode, $\zeta_I(\mathbf q)$, acts just as the rescaling of the spatial coordinates up to $\mathcal O(q)$.
This consistency relation can be extended to observable correlation functions after inflation, e.g., the squeezed bispectrum $\expval{\zeta_I \delta \delta}$, which describes the correlation of the late-time density perturbations $\delta$ in the presence of the long-wavelength mode $\zeta_I$.
If future measurements of the squeezed three-point function in the CMB anisotropies and the LSS discover a deviation from this evolved (late-time) consistency relation, it would rule out single-field inflation models.

Previous studies of the late-time consistency relation have so far focused on Newtonian gauge~\cite{Creminelli:2013mca, Horn:2014rta} and often in the context of a separate ``Newtonian consistency relation'' at ${\cal O}(q)$~\cite{Kehagias:2013yd,Peloso:2013zw}. In this paper, we present a new analysis of the consistency relation in synchronous gauge, which we show has the same form as the rescaling of coordinates in the inflationary consistency relation.
Since the synchronous gauge is defined as the free-fall frame of initially comoving observers, the equivalence principle requires that the consistency relations are trivial/kinematic spatial coordinate transformations up to $\mathcal O(q)$. 
In addition, synchronous time slicing clarifies the separate roles of initial non-Gaussianity versus the evolution of short-wavelength fluctuations in establishing the observable late-time consistency relation.\footnote{In synchronous gauge, constant-coordinate-time hypersurfaces coincide with constant-proper-time hypersurfaces. An advantageous feature of this property is that the linear halo/galaxy bias relation can be extended to scales near the horizon size~\cite{Baldauf:2011bh,Jeong:2011as,dePutter:2015vga}.}

Moreover, the synchronous calculations at $\mathcal O(q^2)$ reveal the dynamical effect of the evolution of short-wavelength density perturbations in the presence of a long-wavelength mode, often approximated in the so-called separate universe approach where it is considered as a local background cosmology~\cite{Sirko:2005uz,Gnedin:2011kj,Baldauf:2011bh,Li:2014sga,Hu:2016ssz}.
This is because the separate universe construction is conceptually related to a free-fall frame and contains the dynamics of a long-wavelength adiabatic mode.
Consequently, synchronous gauge has the computational advantages in the derivation of the consistency relation. 
Unlike in Newtonian gauge, the consistency relation in synchronous gauge can be derived with time-independent coordinate transformations, which are manifestly non-dynamical and highlight the equivalence principle by maintaining the same free-fall frame throughout the superhorizon evolution.\footnote{
An alternative is to construct Fermi normal coordinates~\cite{Baldauf:2011bh} and its generalization to the conformal case for superhorizon perturbations~\cite{Pajer:2013ana,Dai:2015rda} is closely related to the synchronous construction.
}
As a result, the adiabatic or gauge conditions for the long mode associated with the coordinate transformation become simple in synchronous gauge, and do not require any conditions on the matter content of the universe, such as the neglect of the anisotropic stress of the neutrinos, as is common in other techniques.

To verify that the synchronous gauge consistency relation is satisfied for single-field inflation and show how it can be violated under other types of initial non-Gaussianity, we compare it to second-order density perturbation calculations assuming a perfect fluid during matter-dominated (MD) and radiation-dominated (RD) eras. 
Second-order perturbations have been extensively studied in the context of both the CMB~\cite{Hu:1993tc,Dodelson:1993xz,Pyne:1995bs,Mollerach:1997up,Matarrese:1997ay,Creminelli:2004pv,Bartolo:2004ty,Tomita:2005et,Bartolo:2005fp,Bartolo:2005kv,Bartolo:2006cu,Bartolo:2006fj,Pitrou:2008ak,Khatri:2008kb} and the LSS~\cite{Tomita1967a,1981MNRAS.197..931J,1983MNRAS.203..345V,1984MNRAS.209..139J,Suto:1990wf,Makino:1991rp,Bartolo:2005xa,Koyama:2018ttg}.
In general, the second-order density perturbations can be divided into the homogeneous part, whose evolution is the same as the first-order density perturbations, and the inhomogeneous part, whose evolution is sourced by the convolution integrals of the first-order perturbations.
Importantly, the homogeneous part includes the primordial non-Gaussianity, whereas the inhomogeneous part leads to a purely non-primordial, late-time non-Gaussianity due to the non-linear evolution of the perturbations. 
We show both the homogeneous and inhomogeneous parts are needed for the consistency relation to hold and that the inhomogeneous evolution in RD is responsible for establishing the consistency relation in MD below the horizon scale at matter-radiation equality.  Finally we show that the separate universe approach gives the same result as the second-order calculation at $\mathcal O(q^2)$, if we consider an angle and time average  of the latter.

This paper is organized as follows. 
In \S\ref{sec:cons_rel}, we derive the late-time consistency relation in synchronous gauge and show how its extension to $\mathcal O(q^2)$ can be interpreted using the separate universe approach.
Then, we discuss the compatibility of  these relations with the evolution of second-order perturbations, first from arbitrary initial non-Gaussianity conditions, and then from single-field initial non-Gaussianity in \S\ref{sec:second_order}. 
We devote \S\ref{sec:conclusion} to the conclusion. A number of appendices contain additional technical details and derivations. 
In Appendix~\ref{app:remove_delta}, we review the removal of the delta function from the consistency relation.
In Appendix~\ref{app:secondorder}, we solve for second-order perturbations in synchronous gauge. In Appendix~\ref{app:cons_rel_newtonian}, we present and review the analogous results in Newtonian gauge, and in Appendix~\ref{app:new_to_syn} we perform gauge transformations to relate the results in Newtonian and synchronous gauges as well as the initial conditions from comoving and uniform density gauge. Appendix~\ref{app:notation} collects the main variables used in this paper.

\section{Consistency Relations and Separate Universe in Synchronous Gauge}
\label{sec:cons_rel}

The physical essence of the cosmological consistency relations is the equivalence principle. These relations express the fact that a local observer's experience is insensitive to particular gravitational perturbations---those that can be generated by coordinate transformations. 
As such, it is natural to expect that the kinematic constraints that the consistency relations capture will be most simply stated in coordinates adapted to locally freely falling observers, which are precisely those of synchronous gauge. 

In this section, we derive the kinematic consequences of these spatial coordinate transformations in synchronous gauge that mimic the effect of a long-wavelength curvature perturbation of wavenumber $q$ to $\O(q)$, and translate them into soft theorems satisfied by correlation functions of local observables. These same  techniques of absorbing the long-wavelength mode into the background, now thought of as defining a separate universe or local Friedmann-Lema\^{i}tre-Robertson-Walker (FLRW) background, also mimic the truly dynamical effects of the long-wavelength curvature perturbation on local observables at $\O(q^2)$.

\subsection{Synchronous Gauge and Free-Fall Frame}
\label{sec:synchfreefall}
We are interested in the behavior of fluctuations around a spatially flat FLRW background metric 
$\delta g_{\mu\nu} = g_{\mu\nu}- \bar g_{\mu\nu}$, where $\bar g_{\mu \nu}$ is the conformally flat background metric with the scale factor $a$.
Synchronous gauge is defined by the requirement that the perturbations of the metric satisfy $\delta g_{0\mu} = 0$. The line element can therefore be parameterized as
\begin{align}
 \dd s^2 
& = a^2(\eta) \Big[ {-} \dd \eta^2 + \big( (1-2 \Psi) \delta_{ij} + 2 E_{ij} \big) \dd x^i \dd x^j \Big]\,,
\label{eq:def_metric_pertb}
\end{align}
where the metric perturbations are solely in the spatial components.
Note that the metric perturbations here denote (long-wavelength) first-order perturbations, while later in \S\ref{sec:second_order} we will discuss the higher-order impact on short-wavelength observables in the presence of long-wavelength mode.
In general, $E_{ij}$ contains components that transform as a scalar, vector, and tensor under the spatial rotation symmetries of the background. 
While the long-wavelength tensor mode in $E_{ij}$ induces locally observable tidal effects as soon as it enters the horizon~\cite{Dai:2013kra,Dai:2015rda},
we focus on the dynamics of scalar perturbations (in particular the second-order scalar perturbations induced by the first-order scalar perturbations) in this paper.
Then, we restrict to the scalar component of $E_{ij}$, which we write as
\begin{align}
	E_{ij} \equiv \left( \frac{\partial_i \partial_j}{\nabla^2} - \frac{1}{3} \delta_{ij} \right) \hat{E}\,,
	\label{eq:e_ij_e_hat}
\end{align}
where $\hat E$ is a 3-scalar function.\footnote{Note that our definition differs from that of Malik \& Wands in~\cite{Malik:2008im}, which uses $E_{ij}|_{\rm MW}=\partial_i\partial_j E|_{\rm MW}$. Instead, we have split the spatial metric perturbation into the trace and traceless pieces, and absorbed all of the trace into $\Psi$.}
Throughout this work, indices of $3$-vectors and 3-tensors, such as spatial vectors, momentum vectors, and their spatial derivatives, are raised or lowered by the comoving background spatial metric $\delta_{ij}$, which is the same convention as in \cite{Malik:2008im}.  
Repeated spatial indices of $3$-vectors and 3-tensors are likewise summed over even when they both are lowered or raised.

Synchronous gauge is so named because it is the coordinate system that is established by a set of free-fall observers who initially synchronize their clocks from their initial spatial coordinate positions. As such, synchronous gauge retains the freedom to specify the time slice of the synchronization and those initial spatial coordinates, leaving the residual coordinate freedom $x^\mu \rightarrow x^\mu + \xi^\mu$ with
\begin{align}
\xi^0 = \frac{\gamma(\mathbf x)}{a},\qquad \xi^i = {\partial^i \gamma(\mathbf x) \int \frac{\dd\eta}{a} + h^i(\mathbf x)}\,.
\end{align}
Notice that the residual gauge freedom represented by the spatial functions $\gamma({\mathbf x})$ and $h^i(\mathbf x)$ is non-dynamical, reflecting its origin in the choice of initial observers.
It is advantageous to choose for the time slicing synchronous observers that are initially at rest with respect to the background expansion as $a\rightarrow 0$, which fixes the residual temporal freedom $\gamma({\mathbf x})=0$.  Since these test observers are in free fall, as long as gravity is the only relevant force this means that their frame coincides with the one that moves with the matter, namely comoving gauge where the matter stress-energy tensor is set to $T^0_{\hphantom{0}i} =0$ (see Eq.~\eqref{eq:CotoS}).   In general, 
this approximation holds above the Jeans scale of the matter $r_s$ (see \cite{Hu:2016wfa} for the generalization beyond general relativity).  For example, in the RD era where this scale is comparable to the horizon, $r_s \sim \eta$, corrections for a Fourier wavenumber $q$ due to the change in slicing are suppressed by $(q\eta)^2$.  \
In the MD era and linear theory, the Jeans scale of the pressureless matter vanishes, and so the correction is entirely negligible.
Specifically, the 3-curvatures on comoving and synchronous slicings physically coincide. 
In linear theory, since the background is homogeneous, they coincide at each coordinate point as well, regardless of the choice of synchronous spatial coordinates $h^i(\mathbf x)$.  In Fourier space, this implies that the comoving curvature perturbation $\zeta$ satisfies\footnote{The comoving curvature perturbation is denoted by $\cal R$ in some literature (e.g. Ref.~\cite{Baumann:2009ds}).}
\begin{align}
	\zeta = - \Psi - \frac{1}{3} \hat E + \mathcal O(q^2)\,,
 \label{eq:zetasynch}
\end{align}
where the notation $\O(q^n)$ throughout denotes corrections of order $(q r_s)^n$ when ${q r_s \ll 1}$.  Notice that in the formal limit $\eta\rightarrow 0$, we have $r_s\rightarrow 0$ and the quadratic correction vanishes.  In the inflationary context, the minimum value of $r_s$ corresponds to the horizon at the end of inflation.  Since this is negligible compared to wavelengths considered here, we will denote this initial value as $\zeta=\zeta_I$ and drop any $\O(q^2)$ corrections initially.

In Fourier space, the anisotropic metric perturbation is expressed as
\begin{align}
	E_{ij}(\mathbf q) = \left( \frac{q_i q_j}{ q^2} - \frac{1}{3} \delta_{ij} \right) \hat{E}(\mathbf q)\,.
	\label{eq:e_ij_hat_e_ij}
\end{align}
While a common convention is to choose $h^i$ such that $\lim_{\mathbf q\rightarrow \bz} \hat E =0$ at finite $\zeta$, for pedagogy we retain the residual gauge freedom.  Without loss of generality, we can parameterize it by the fraction of the invariant curvature perturbation that is carried individually by $\Psi$ or $\hat E$ as
\begin{align}
 \lim_{q\rightarrow 0} \frac{\Psi}{\zeta} =	- (1-f) ,\qquad  \lim_{q\rightarrow 0}  \frac{\frac{1}{3}\hat E}{\zeta} =	- f\,.
 \label{eq:f_def}
\end{align}
See Eq.~(\ref{eq:f_gauge_trans}) for the specific relationship between the spatial coordinate threading for different $f$.

\subsection{Consistency Relation up to 
\texorpdfstring{${\cal O}(q)$}{O(q)}}

 To introduce the consistency relation, we adopt the approach used in \cite{Creminelli:2012ed,Creminelli:2013mca}. 
For the adiabatic mode, the long-wavelength perturbation with the wavenumber $q$ above the Jeans scale should act on short-wavelength perturbations as the coordinate transformation  with deviations only at  $\mathcal O(q^2)$.
If the perturbations originate from single-field inflation, then this ansatz applies to the initial conditions for the short-wavelength perturbations as well.

In this case, the whole correlation function of the short-mode density perturbations in the presence of the long mode, represented by the initial comoving curvature perturbation $\zeta_{I}$, is the same as the one without the long-wavelength curvature perturbation after an appropriate coordinate transformation ${\bf x} \rightarrow \tilde {\bf x}$:
\begin{align}
	\expval{\delta(\mathbf x_1, \eta_1) \cdots \delta(\mathbf x_N, \eta_N)|\zeta_{I}} = \expval{\delta(\tilde{\mathbf x}_1, {\eta}_1) \cdots \delta(\tilde{\mathbf x}_N, {\eta}_N)} + {\cal O}(q^2)\, ,
 \label{eq:npt_delta}
\end{align}
where the notation
 ${\cal O}(q^n)$ in correlators implicitly  means that the Fourier modes that are included in this long-wavelength $\zeta_I$ satisfy  $|{\bf q}\cdot ({\bf x}_i - {\bf x}_j )|\ll 1$ in addition to $q r_s\ll 1$, i.e., that we can Taylor expand $\zeta_I({\mathbf x}_i-{\mathbf x}_j)$ and treat its effects as a background modulation for the short-wavelength modes order by order.

Supposing that the coordinate transformation induced by $\zeta_{I}$ is perturbatively small, we obtain to ${\cal O}(q)$
\begin{align}
	\expval{\zeta_{I}(\mathbf x) \delta(\mathbf x_1, \eta_1) \cdots \delta(\mathbf x_N, \eta_N)} 
	&= \Big\langle\zeta_{I}(\mathbf x) \sum_{a=1}^N  \delta(\mathbf x_1, \eta_1) \cdots \Delta \delta(\mathbf x_a, \eta_a) \cdots \delta(\mathbf x_N, \eta_N)\Big\rangle\,,
	\label{eq:cons_rel_real}
\end{align}
where we have used $\langle \zeta_{I} \rangle=0$.
Here 
$\Delta \delta(\mathbf x_a, \eta_a) \equiv \delta(\tilde {\mathbf x}_a, \eta_a) - \delta(\mathbf x_a, \eta_a)$, which correlates with $\zeta_{I}$ due to the coordinate change it represents.   
In  Fourier space, the consistency relation can be expressed as
\begin{equation}
	\hskip -10pt\lim_{\mathbf q \rightarrow \bz} \expval{\zeta_{I}(\mathbf q) \delta(\mathbf k_1,\eta_1) \cdots \delta(\mathbf k_N, \eta_N)} = \Big\langle \zeta_{I}(\mathbf q) \sum_{a=1}^N  \delta(\mathbf k_1,\eta_1) \cdots \Delta \delta(\mathbf k_a,\eta_a) \cdots \delta(\mathbf k_N, \eta_N)\Big\rangle\,,
	\label{eq:cons_rel_fourier}
\end{equation}
where $\Delta \delta(\mathbf k_a,\eta_a) \equiv \int \dd^3 x_a\, \ee^{-i \mathbf k_a \cdot \mathbf x_a} \Delta \delta(\mathbf x_a,\eta_a)$. 
In synchronous gauge, the task therefore is to characterize the coordinate change that models the ``background wave''  influence of $\zeta_{I}$.
In the passive coordinate transform approach, the coordinate shift
\begin{equation}
\tilde x^i = x^i + \xi^i\, 
\end{equation} 
corresponds to a change in the small-scale density perturbation as
\begin{align}
	\Delta \delta(\mathbf x, \eta) =  \xi^i{\partial_i}\delta\,.
\end{align}
Note that under this transformation and keeping only first-order terms in $\xi^i$, the spatial metric transforms as 
\begin{align}
	\widetilde {\delta g_{ij}} =&\  \delta g_{ij} - (\partial_\lambda \bar g_{ij}) {\xi^\lambda} - \bar g_{i\lambda} (\partial_j \xi^\lambda) -  \bar g_{\lambda j} (\partial_i \xi^\lambda)\,.
	\label{eq:spatialmetrictransform}
\end{align}
In general, we seek to model this change at the level of a gradient approximation for $\zeta_{I}$ as
\begin{equation}
\zeta_{I}({\bf x}) \approx \zeta_{I}(\bz) + x^i\partial_i \zeta_{I}{(\bz)}\, ,
\end{equation}
so that the Fourier-space consistency relation is maintained to $\mathcal O(q)$.

Unlike Newtonian gauge (see Appendix \ref{app:cons_rel_newtonian}) and comoving gauge (where the translation of spatial coordinates is time and anisotropic stress dependent \cite{Pajer:2017hmb}), this coordinate change is purely spatial and time independent, but also carries the residual gauge freedom parameterized by $f$ in Eq.~(\ref{eq:f_def}).
We account for the most general case by separating this coordinate change as
\begin{equation}
\xi^i = (1-f) \xi^i_{T} + f \xi^i_{\bar T}\, ,
\end{equation}
where $T$ denotes $\Psi$ (the trace component) and $\bar T$ denotes $\hat E$ (the trace-free component).  We can consider each piece independently, and then compose the most general consistency relation from this linear combination, as we shall show explicitly for the three-point function in \S\ref{sec:3point}.  We loosely call the spatial threading with  $f=0$ and $f=1$ as the (initially) ``isotropic'' and ``anisotropic'' cases, respectively, even though evolution makes $\hat E\ne 0$ and $\Psi\ne 0$ for any $f$ as the perturbations cross the horizon scale.

\subsubsection{Isotropic-Synchronous Gauge ($f = 0$)}
\label{subsub:zero_tracefree_f1}

In this case, the initial comoving curvature perturbation is represented by the synchronous gauge metric $\Psi$ as $\zeta_I = - \Psi + \mathcal O(q^2)$.  Given the isotropy of the spatial metric, we can only allow a local rescaling of the coordinates, consisting of spatial dilation and special conformal transformations (SCT), with
\begin{align}
\xi^i|_{f=0} =	\xi_{T}^i &\equiv  \lambda x^i + 2 \mathbf b \cdot \mathbf x\, x^i -  x^2 b^i\, ,
\end{align}
where $\mathbf b \cdot \mathbf x = b^i x^i$. 
From Eq.~(\ref{eq:spatialmetrictransform}), it follows that $\zeta_{I}$ transforms as 
\begin{align}
	\label{eq:zeta_L_trans}
		\tilde \zeta_{I} = \zeta_{I} - \lambda - 2 \mathbf b \cdot \mathbf x\,,
\end{align}
from which we can associate the dilation $\lambda$ and SCT ${\mathbf b}$ parameters that would be required to absorb $\zeta_I$ into the coordinate change: $\lambda = \zeta_{I}(\bz)$ and $2b^i = \partial_i \zeta_{I}(\bz)$.   
Since this is purely a time-independent spatial coordinate redefinition, this transformation behaves exactly as $\zeta$ would for the growing mode of adiabatic perturbations\footnote{A coordinate transformation cannot  change the local number density ratios of different particle species, which would be required to mimic an isocurvature mode.}
to $\mathcal O(q)$ without further restrictions due to the gauge conditions or the matter content, including any anisotropic stress it carries (see Appendix~\ref{app:cons_rel_newtonian} for the Newtonian case, where further restrictions must be taken into account).

The above coordinate change then transforms the small-scale density perturbation as 
\begin{align}
	\Delta \delta(\x, \eta) =  (\lambda x^i + 2 \b \cdot \x\, x^i - x^2 b^i )\partial_i\delta(\x,\eta)\, .
\end{align}
In Fourier space, this transformation can be expressed as
\begin{align}
	\Delta \delta(\k,\eta) = -\Big[ \lambda(3 + k^i \partial_{k^i}) + i( 6 \b \cdot \partial_{\k} + 2b^i k^j \partial_{k^i} \partial_{k^j} - \mathbf b \cdot \k \nabla_k^2)  \Big] \delta(\k,\eta)\,.
\end{align}
Substituting this into Eq.~(\ref{eq:cons_rel_fourier}), we obtain the consistency relation 
\begin{align}
	&\lim_{\mathbf q \rightarrow \bz} \expval{\zeta_{I}(\mathbf q) \delta({\mathbf k_1},\eta_1) \cdots \delta({\mathbf k_N},\eta_N)}	\label{eq:cons_rel_f1} \\
	& = -P_{\zeta_I}(q) \sum_{a=1}^{N}\left(3 + k^i_a \partial_{k^i_a} + 3 \mathbf q \cdot \partial_{\mathbf k_a} + q^i k^j_a \partial_{k^i_a} \partial_{k^j_a} - \frac{1}{2}\mathbf q \cdot \mathbf k_a \nabla_{k_a}^2\right) 
\expval{
 \delta({\mathbf k_1},\eta_1) \cdots \delta({\mathbf k_N},\eta_N)},\nonumber
\end{align}
where we have neglected the $\mathcal O(q^2)$ contributions and used 
\begin{align}
\expval{\zeta_I(\q) \zeta_{I}(\x =\bz)} &= \lim_{\x \rightarrow \bz}\expval{\zeta_I(\q)  \int \frac{\dd^3 k}{(2\pi)^3} \ee^{i \k \cdot \x} \zeta_I(\k)} = P_{\zeta_I}(q)\,, \nonumber \\
\expval{\zeta_I(\mathbf q) \partial_i \zeta_{I}(\x =\bz)} &= \lim_{\x \rightarrow \bz}\expval{\zeta_I(\q)  \int \frac{\dd^3 k}{(2\pi)^3} (i k^i) \ee^{i \k \cdot \x} \zeta_I(\k)} = -i q^i P_{\zeta_I}(q)\,.
\end{align}
Note that the correlators on both sides of
Eq.~(\ref{eq:cons_rel_f1}) contain momentum-conserving delta functions. Using standard operations to remove the momentum-conserving delta functions which we review in Appendix~\ref{app:remove_delta},
we obtain
\begin{align}
	\lim_{\mathbf q \rightarrow \bz} \expval{\zeta_I(\mathbf q) \delta({\mathbf k_1},\eta_1) \cdots \delta({\mathbf k_N},\eta_N)}' &= -P_{\zeta_I}(q)({\cal D}+q^i{\cal K}^i )\expval{\delta({\mathbf k_1},\eta_1) \cdots \delta({\mathbf k_N},\eta_N)}'\, ,
	\label{eq:f_1_gauge_cons_rel}
\end{align}
where $\expval{\cdots}= (2\pi)^3 \delta_D\left({\bf k}_{\rm tot}\right)\expval{\cdots}'$
and
the differential operators
\begin{align}
	\cD &= 3(N-1)+ \sum_{a=1}^{N} k^i_a \partial_{k^i_a}\, , \label{Dop}\\
 {\cal K}^i &= \sum_{a=1}^N \left(3 \partial_{k^i_a} + k^j_a \partial_{k^j_a} \partial_{k^i_a} - \frac{1}{2} k^i_a \nabla^2_{k_a} \right)
	 \label{SCTop}
\end{align}
correspond to the dilation and SCT generators (for the scaling weight of zero). As we shall discuss in \S \ref{sec:3point}, this removal of the momentum-conserving delta function does not mean that we remove momentum conservation. The consistency relation still enforces conservation, but there are various ways to send ${\bf q}\rightarrow \bz$ that can be used to simplify its construction and interpretation, especially for the ${\mathcal{O}}(q)$ term.

\subsubsection{Anisotropic-Synchronous Gauge ($f = 1$)}
We now discuss the case of $f = 1$, for which $\zeta_I = - \hat E/3 + \mathcal O(q^2)$.  Since $\hat E$ is an  anisotropic metric perturbation, this case requires an anisotropic rescaling of coordinates to model. This leads us to consider 
\begin{align}
	\xi^i|_{f=1} = \xi^i_{\bar T} &\equiv 3\left[ \left( \frac{b^i b_j}{b^2} - \frac{1}{3} \delta^{i}_{\,j} \right) ( -\lambda - 2\mathbf x \cdot \mathbf b ) x^j + \frac{b^i}{b^2} (\mathbf x \cdot \mathbf b)^2 - \frac{1}{3} b^i \mathbf x^2 \right],
	\label{eq:xi_zero_trace}
\end{align}
where $\lambda$ and $b^i$ are a constant scalar and vector, respectively.
The derivative of $\xi^i_{\bar T}$ is given by
\begin{align}
	\partial_j\xi^i_{\bar T} = 3 \left( \frac{b^i b_j}{b^2} - \frac{1}{3} \delta^{i}_{\,j} \right) \left( -\lambda - 2 \mathbf x \cdot \mathbf b \right) -2  \left( b^i x_j - b_j x^i\right).
\end{align}
This implies that $\partial_i\xi^i_{\bar T} = 0$, which keeps $\tilde \Psi^\fo = \Psi^\fo =0$ up to $\mathcal O(q)$. 
With this transformation, the trace-free part of the spatial metric transforms as 
\begin{align}
	\tilde E_{ij} = E_{ij} + 3 \left( \frac{b_i b_j}{b^2} - \frac{1}{3} \delta_{ij} \right) \left( \lambda + 2 \mathbf x \cdot \mathbf b \right).
	\label{eq:e_ij_trans}
\end{align}
To see the transformation of $\hat E$, let us focus on the perturbation at finite momentum $q$ in the gradient approximation around $x^i=0$,
\begin{align}
E_{ij}(\mathbf x; \mathbf q) &= \int \frac{\dd^3 k}{(2\pi)^3} \left( \frac{k_i k_j}{k^2} -\frac{1}{3} \delta_{ij} \right) \hat E({\mathbf k})\ee^{i {\mathbf k}\cdot {\mathbf x}}\nn
&= \left( \frac{q_i q_j}{q^2} -\frac{1}{3} \delta_{ij} \right) A\ee^{i {\mathbf q}\cdot {\mathbf x}}\approx \left( \frac{q_i q_j}{q^2} -\frac{1}{3} \delta_{ij} \right) A( 1 + i \mathbf q \cdot \mathbf x)\, ,
\end{align}
where we used $ \hat E({\mathbf k}) = A (2\pi)^3 \delta_D({\mathbf k}-{\mathbf q})$. 
We note that, for the single Fourier long mode, we can rewrite the consistency relation, Eq.~(\ref{eq:cons_rel_real}), as
\begin{align}
	\expval{E_{ij}(\mathbf x;\mathbf q) \delta(\mathbf x_1,\eta_1) \cdots \delta(\mathbf x_N,\eta_N)} = \expval{E_{ij}(\mathbf x;\mathbf q) \expval{\delta(\mathbf x_1,\eta_1) \cdots \delta(\mathbf x_N,\eta_N)|E_{ij}(\mathbf x;-\mathbf q)}}.
\end{align}
The reason for the minus sign of $-\mathbf q$ in the right-hand-side (RHS) is that it gives a nonzero value after the ensemble average with $E_{ij}(\mathbf x; \mathbf q)$; for a spectrum of modes this is equivalent to conjugation and the condition that $E_{ij}(\mathbf x)$ is real.
To erase $E_{ij}(\mathbf x; -\mathbf q)$ with the coordinate transformation, we then equate $\lambda = -A/3$ and $2 b^i = i q^i A/3$.
 Since $\hat E$ corresponds to the comoving curvature perturbation $\zeta$ in the case of $f = 1$ using Eq.~(\ref{eq:zetasynch}), we find
\begin{equation}
 \zeta_{I}({\mathbf x}) =-\frac{1}{3}  \hat E({\mathbf x}) = \lambda + 2 \mathbf x \cdot \mathbf b\, ,
\end{equation}
so that this transformation models $\zeta_{I}$ as desired.

On the other hand, under the coordinate shift of Eq.~(\ref{eq:xi_zero_trace}), the density perturbation transforms as 
\begin{align}
	\Delta \delta(\mathbf x,\eta) &=  \dblue{3}\left[ \left( \frac{b_i b_j}{b^2} - \frac{1}{3} \delta_{ij} \right) (-\lambda - 2 \mathbf x \cdot \mathbf b ) x^j + \frac{b^i}{b^2} (\mathbf x \cdot \mathbf b)^2 - \frac{1}{3} b^i x^2 \right]\partial_i \delta(\x,\eta),\\ 
	\Delta \delta(\mathbf k,\eta) 
	 &= \dblue{3} \left( \frac{b_i b_j}{b^2} - \frac{1}{3} \delta_{ij} \right) \Big[\lambda k^i \partial_{k^j} + i(  k^i \partial_{k^j}(2 \mathbf b \cdot \partial_{\mathbf k}) - (\mathbf b \cdot \mathbf k) \partial_{k^i} \partial_{k^j}) \Big]\delta(\mathbf k)\, .
\end{align}
Substituting $\lambda = \zeta(\bz)$, $2 b^i = -i q^i \zeta(\bz)$ and using Eq.~(\ref{eq:cons_rel_fourier}), we obtain
\begin{align}
	&\lim_{\mathbf q \rightarrow \bz} \expval{\zeta_I(\mathbf q) \delta({\mathbf k_1},\eta_1) \cdots \delta({\mathbf k_N},\eta_N)} = 3 P_{\zeta_I}(q)  \label{eq:cons_rel_zero_trace} \\
	&\quad \times  \sum_{a=1}^{N} \left( \frac{q_i q_j}{q^2} - \frac{1}{3} \delta_{ij} \right) \left(k^i_a  + k^i_a (\partial_{\mathbf k_a} \cdot \mathbf q) - \frac{1}{2}(\mathbf k_a \cdot \mathbf q) \partial_{k^i_a} \right) \partial_{k^j_a}\expval{\delta({\mathbf k_1},\eta_1) \cdots \delta({\mathbf k_N},\eta_N)}.\nonumber
\end{align}
We then remove the delta function in both sides using a generalization of the standard procedure, which we review in Appendix~\ref{app:remove_delta}.
We finally obtain\footnote{Note that the consistency relation in the anisotropic-synchronous gauge \eqref{eq:f_0_gauge_cons_rel} takes the same form as the tensor consistency relation in comoving gauge~\cite{Creminelli:2012ed, Hinterbichler:2013dpa}, except that the traceless projector $\frac{q_iq_j}{q^2}-\frac{1}{3}\delta_{ij}$ is replaced with a traceless-transverse polarization tensor. In contrast, the anisotropic scalar perturbations in synchronous gauge are traceless but not transverse.}
\begin{align}
	&\lim_{\mathbf q \rightarrow \bz} \expval{\zeta_I(\mathbf q) \delta({\mathbf k_1},\eta_1) \cdots \delta({\mathbf k_N},\eta_N)}' = 3 P_{\zeta_I}(q)  \label{eq:f_0_gauge_cons_rel} \\
	&\quad \times \sum_{a=1}^{N} \left( \frac{q_i q_j}{q^2} - \frac{1}{3} \delta_{ij} \right) \left(k^i_a  + k^i_a (\partial_{\mathbf k_a} \cdot \mathbf q) - \frac{1}{2}(\mathbf k_a \cdot \mathbf q) \partial_{k^i_a}  \right)\partial_{k^j_a}\expval{\delta({\mathbf k_1},\eta_1) \cdots \delta({\mathbf k_N},\eta_N)}'\,.\nonumber	
\end{align}
In the case of a general mixture of trace and trace-free pieces, the consistency relation is simply a linear combination of Eqs.~(\ref{eq:f_1_gauge_cons_rel}) and (\ref{eq:f_0_gauge_cons_rel}), which we will illustrate next with the consistency relation for the three-point function.

\begin{figure}[t]
\begin{center}
\includegraphics[width=5in]{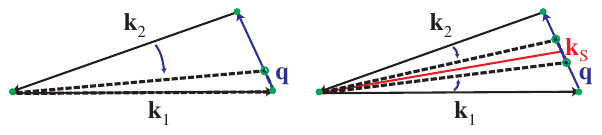}
\end{center}
\caption{Approach to the squeezed limit with ${\mathbf k}_1$ or ${\mathbf k}_S = ({\mathbf k}_1-{\mathbf k_2})/2$ fixed. }
\label{fig:triangle}
\end{figure}

\subsubsection{Three-Point Consistency Relation}
\label{sec:3point}

Let us illustrate the advantages of synchronous gauge by explicitly considering the simplest consistency relation, namely that of the 3-point function.  The 3-point function in the squeezed limit involves derivative operations on the 2-point function in momentum space, which require us to specify  how momentum is conserved in taking the squeezed limit. 

The suppression of the momentum-conserving delta functions does not mean that we break momentum conservation as ${\mathbf q}\rightarrow \bz$ but rather that we have a choice as to what momenta to leave fixed as we take this limit.    For example, consider the momentum-conserving triangles ${\bf k}_1+{\bf k}_2+{\bf q}= \bz$ in Fig.~\ref{fig:triangle}.   We can either take $\mathbf q\rightarrow \bz$ at fixed ${\bf k}_1$ or take the midpoint approach and fix ${\bf k}_S =  ( {\mathbf k}_1-{\mathbf k}_2)/2$.
While Eqs.~(\ref{eq:f_1_gauge_cons_rel}) and (\ref{eq:f_0_gauge_cons_rel}) hold for any such choice and the resulting expressions are the same once converted to the same associations for ${\bf k}$, the separate contributions of the $\mathcal O(q^0)$ and $\mathcal O(q)$ terms differ. 
We can use this fact to further simplify the consistency relation and highlight its physical content in synchronous gauge.  

As pointed out by Refs.~\cite{Creminelli:2011rh, Creminelli:2012ed}, originally in the context correlators in comoving gauge evaluated on the same time surface, the principal advantage of holding ${\bf k}_S$ fixed is that the $\mathcal O(q)$ terms vanish, leaving the relations for absorbing $\zeta_{I}$ in the gradient approximation exactly the same as for the homogeneous approximation.  
In the equal-time correlator case where $\eta=\eta_1=\eta_2$, the two-point function on the RHS becomes $P_\delta(k_S,\eta)$ and has symmetric $k_1$ and 
$k_2$ derivatives (since  $k_S = (k_1+k_2)/2 + \mathcal O(q^2)$), so that momentum conservation guarantees the vanishing of terms with $\sum_a q^i k^i_{a}=0 + \mathcal O(q^2)$.  
For unequal-time correlators the derivation of the consistency relation in comoving gauge from coordinate transformations is more involved due to the time-dependent translation required to maintain the gauge condition of an isotropic spatial metric~\cite{Pajer:2017hmb}, which we will refer to as ``isotropic threading'',   especially in the presence of neutrino anisotropic stress.\footnote{The neutrino anisotropic stress provides a {\it leading-order} correction to the time-dependent translation between synchronous and comoving or uniform density gauges, as can be seen from the linear gauge transformation that removes $\hat E$ (see Eq.~\eqref{eq:CotoS}).
}

In synchronous gauge, this notational advantage and conceptual simplicity applies to unequal-time ($\eta_1\ne\eta_2$) correlators as well, since the underlying coordinate
transformation has  no  time dependence.   
The explicit, most general, three-point consistency relation in synchronous gauge to $\O(q)$ takes the form\footnote{See~\cite{Mitsou:2021mgv} for the analogous result derived for the inflationary bispectra for comoving slicing and anisotropic threading.} 
\begin{equation}
	\boxed{\lim_{\mathbf q \rightarrow \bz} \frac{\expval{\zeta_I(\mathbf q) \delta({\mathbf k_1},\eta_1) \delta({\mathbf k_2},\eta_2)}'}{P_{\zeta_I}(q)P_\delta(k_S,\eta_1,\eta_2)}
	= - n_\delta(k_S,\eta_1,\eta_2) + 3f\big(\mu^2 n_\delta(k_S,\eta_1,\eta_2) + (1-3 \mu^2)\big)} \,,
\label{eq:general_gauge_cons_rel}
\end{equation}
where $\mu \equiv \hat{k}_S \cdot \hat q$ is the angle between the short and long modes, and the unequal-time power spectrum is defined as
\begin{align}
 \expval{\delta({\mathbf k_1},\eta_1)  \delta({\mathbf k_2},\eta_2)} &= (2\pi)^3 \delta_D({\mathbf k_1}+{\mathbf k_2}) P_\delta(k_1,\eta_1,\eta_2)\,, 
\end{align}
with its (time-dependent) spectral tilt
\begin{align}
	{n_\delta(k,\eta_1,\eta_2)} &\equiv \frac{\dd  \ln k^3 P_{\delta}(k_S,\eta_1,\eta_2) }{\dd \ln k}\,.
	\label{eq:n_delta_k_def_1}
\end{align}
Notice that for $f=0$, which is the usual choice for a fully gauge-fixed synchronous gauge, the entire consistency relation to ${\mathcal O}(q)$ is a dilation. Hence in this synchronous gauge, the so-called Newtonian consistency relation is also entirely subsumed into a dilation (cf.~Eq.~(\ref{eq:dilation_cons_rel_new_final})). 
Specifically, the consistency relation for $f=0$ can be rewritten compactly as 
\begin{align}
	\lim_{\mathbf q \rightarrow \bz} \expval{\zeta_I(\mathbf q) \delta({\mathbf k_1},\eta_1) \delta({\mathbf k_2},\eta_2)}' &= -
  n_\delta(k,\eta_1,\eta_2) P_{\zeta_I}(q)  P_\delta(k_S,\eta_1,\eta_2) \qquad (f=0)\,.
\label{eq:general_gauge_cons_rel_ndelta}
\end{align}
The synchronous-gauge interpretation is that local observers cannot tell if they are moving in  the gradient of the long-wavelength curvature  perturbation because they are freely falling, even when comparing at different times.   The local coordinates that they would establish would remove the dilation that Eq.~\eqref{eq:general_gauge_cons_rel_ndelta} represents, leaving no means of locally measuring any impact of the long-wavelength mode to ${\cal O}(q)$.

In fact, if we choose $f=1$ for which there is no isotropic rescaling of local spatial scales from the initial metric, the consistency relation becomes\footnote{The same anisotropic angular dependence is found in the squeezed three-point function of $\zeta$ in solid inflation~\cite{Endlich:2012pz, Endlich:2013jia}.}
\begin{align}
	\lim_{\mathbf q \rightarrow \bz} \frac{\expval{\zeta_I(\mathbf q) \delta({\mathbf k_1},\eta_1) \delta({\mathbf k_2},\eta_2)}'}{P_{\zeta_I}(q)P_\delta(k_S,\eta_1,\eta_2)}
	&= - (1-3 \mu^2)( n_\delta(k_S,\eta_1,\eta_2)-3) \qquad (f=1)\,  .
\label{eq:f1_gauge_cons_rel}
\end{align}
Since the angle average of $\mu^2$ is $1/3$, we have
\begin{equation}
	\lim_{\mathbf q \rightarrow \bz} \overline{\expval{\zeta_I(\mathbf q) \delta({\mathbf k_1},\eta) \delta({\mathbf k_2},\eta)}'} = 0 \qquad (f=1) \,,
	\label{eq:three_point_f1_angle_ave}
\end{equation}
where the overline indicates the angle average.
The angle-averaged consistency relation then becomes trivial, and this $f=1$ choice makes it clear that a local observer cannot detect the presence of a long-wavelength mode in the gradient approximation.

\subsection{Separate Universe at 
\texorpdfstring{$\mathcal O(q^2)$}{O(q\texttwosuperior)}}
\label{sec:SU}

The advantage of employing synchronous gauge for adiabatic perturbations becomes even more apparent at $\mathcal O(q^2)$.  Here the impact of the long-wavelength $\zeta_I$ cannot be merely absorbed into free-fall coordinates, since it changes  the local density and 3-curvature that determines short-wavelength observables (see Appendix~\ref{sec:3ricci}).  On the other hand, this background wave still appears as a local change in the FLRW background under which short-wavelength fluctuations evolve.   The isotropic impact of this background can be described by a change in cosmological parameters that is sometimes called the separate universe approach.

Here the advantage of synchronous gauge is that its coordinates are closely related to the description of separate-universe observers comoving with the local expansion, as discussed in \cite{Hu:2016ssz,Hu:2016wfa}.
In this case synchronous gauge remains the local frame established by free-fall observers, whereas comoving gauge can differ even in the choice of time-slicing if the matter fields experience non-gravitational forces. 

In the separate universe construction, synchronous perturbations are reabsorbed into the local background by 
a change in the background density $\rho_\SU = \rho(1+\delta_L)$ from the long-wavelength $\delta_L$, the local scale factor ${a_\SU}=a(1-\Psi)$, the local FLRW curvature 
\begin{equation}
	K_\SU =  \frac{2}{3}\nabla^2\left(\Psi + \frac{\hat E}{3}\right),
\label{eq:KL}
\end{equation}
where we have assumed that $K=0$ in the global background,  and the local Hubble parameter 
\begin{align}
	H_\SU = \left(1 -\frac{\dd \Psi}{\dd\ln a}\right) H\,.
\label{eq:HL}	
\end{align}
Note that the subscript ``SU'' indicates a local quantity in the separate universe.
  The small-scale density fluctuations relative to the local background density then evolve in this separate universe in a manner given by this change in these cosmological parameters.  This construction applies even if the small-scale observables are in the nonlinear regime, and has been successfully implemented in cosmological simulations \cite{Sirko:2005uz,Baldauf:2015vio,Li:2014sga,Wagner:2014aka}.

The only difference between the separate universe  and synchronous descriptions is the absorption of the synchronous metric fluctuation into the local scale factor.   In the separate universe construction this would introduce an additional {\it evolving} dilation of short-wavelength scales since $\Psi$ evolves with time at $\O(q^2)$ (see \cite{Li:2014sga}, Eq.~(44)), which is absent in synchronous coordinates and moreover cannot be absorbed into the initial definition of synchronous spatial coordinates with the gauge parameter $f$.  Effectively in the fluid mechanics analogy, synchronous spatial coordinates act as Lagrangian coordinates in contrast with the separate universe Eulerian coordinates.   

In addition, the absorption of the metric fluctuation into the scale factor only applies to its spatial trace $\Psi$ and so the impact of the anisotropic piece $\hat E$ is eliminated by angle averaging, in the same way as we treated the $f=1$ case in Eq.~(\ref{eq:three_point_f1_angle_ave}).  
  In fact, the separate universe description can be generalized to encapsulate the full angular dependence  by using an anisotropic background cosmology~\cite{Ip:2016jji}.  This has also been successfully implemented in simulations \cite{Schmidt:2018hbj,Akitsu:2020fpg,Masaki:2020drx,Stucker:2020fhk}, but for simplicity we employ the FLRW separate universe  and angle-averaged correlation functions here.

\section{Second-Order Perturbations in Synchronous Gauge}
\label{sec:second_order}

In this section, we assess the compatibility between the consistency relation and second-order perturbation theory results in synchronous gauge and its dependence on initial non-Gaussianity. 
Before going to the concrete calculations, we summarize the methodology of the second-order perturbations.
See Appendix~\ref{app:secondorder} for a detailed derivation.

 For simplicity we assume here a perfect fluid where the energy-momentum tensor is expressed as
\begin{align}
	T^\mu_{\ \nu} &=  (\rho + P) u^\mu u_\nu + P \delta^\mu_{\ \nu}\,,
\end{align}
where $\rho$ is the energy density, $P$ is the pressure, and $u_\mu$ is the four-velocity.
The general expression of the energy-momentum tensor with nonzero anisotropic stress is given in Appendix~\ref{app:secondorder}. However, an advantage of the synchronous gauge consistency relation is that anisotropic stress does not introduce any conceptual differences in its construction unlike Newtonian (see Appendix~\ref{app:cons_rel_newtonian}) or comoving gauges \cite{Pajer:2017hmb} (see also Eq.~(\ref{eq:CotoS})).
  Anisotropic stress from neutrinos or radiative viscosity does of course inhibit its analytic verification using second-order perturbation theory and allows only numerical checks in general, which is why we omit it here.
We expand the density perturbation up to second order as
\begin{align}
	\label{eq:rho_pertb_def}
	\rho &\approx \bar \rho + \rho^\fo + \frac{1}{2} \rho^\so\,, 
\end{align}
where the overbar indicates the background value and the superscripts denote the order in perturbations.
We then define the $n$-th order density perturbation $\delta^{(n)}$ as 
\begin{align}
    \label{eq:delta_def}
	\delta^{(n)} \equiv \frac{\rho^{(n)}}{\bar \rho}\,.
\end{align}
Similarly, we expand the metric perturbations up to second order as,
\begin{align}
\Psi &\approx \Psi^{(1)} + \frac{1}{2} \Psi^{(2)} \,, \quad E_{ij} \approx   E^\fo_{ij} + \frac{1}{2} E^\so_{ij}. \label{eq:e_pertb_def}
\end{align}
We also define the transfer functions of perturbations relative to the initial comoving curvature $\zeta_I$ into the initial $k$-dependence and  growth pieces, $\Tmr$ and $\Dmr$, respectively: 
\begin{align}	
	\label{eq:delta_zeta_rel}
	\delta^\fo (\k, \eta) &=    \Dmr(y) {\Tmr(k)}\zeta^\fo_I(\k)\,, 
\end{align}
where $y\equiv k \eta$ with $k \equiv |\mathbf k|$, and the subscripts, ``m'' and ``r'', represent a quantity during the MD and RD eras, respectively.  
In the RD era, $\Tr(k)=1$.  In the MD era, $\Tm(k)$ represents the change in the spectrum from the end of inflation to the epoch of MD ($\eta > \eta_\mm$) due to the sub-Jeans-scale evolution during the prior RD epoch.  In the literature, this is usually referred to as the transfer function for the large-scale structure and has the limiting forms
\begin{align}
\begin{array}{lc}
\Tm(k) = 1 \,,  & k\eta_{\rm eq} \ll 1\, , \nonumber\\[3pt]
\Tm(k) \propto k^{-2} \ln k \,,\quad &  k\eta_{\rm eq}\gg 1 \,.
\end{array}
\end{align}
The detailed form of $\Tm$ is not relevant for this work, but it can be computed using the standard Einstein-Boltzmann codes or be approximated with a fitting formula~\cite{Eisenstein:1997jh}.  
The density growth function $\Dmr$ then transfers these ``initial'' perturbations to the final perturbations at $\eta$, and its explicit form is given below. 
Unlike \cite{Creminelli:2013mca,Horn:2014rta}, we explicitly account for subhorizon evolution in the RD regime and its impact on separating the pieces of the consistency relation that come from initial non-Gaussianity and evolution.

Consequently, the power spectrum of the density perturbation $P_\delta$ is related to $P_{\zeta_I}$ by
\begin{align}
 P_{\delta}(k,\eta_1,\eta_2) = \Dmr(k\eta_1)\Dmr(k\eta_2) {\Tmr^2(k)} P_{\zeta_I}(k)\,.
 \label{eq:Pdelta}
\end{align}
At second order, the density perturbation can be expressed as 
\begin{align}
	\delta^\so (\mathbf k,\eta) &= 
\int \frac{\dd^3 p_1\dd^3 p_2}{(2\pi)^3}\delta_D(\k-\p_1-\p_2) \cI_{\mm/\rr}^{\rm inhom.}(u,v,y) T_{\mm/\rr}(p_1)T_{\mm/\rr}(p_2)\zeta^{(1)}_I(\p_1)\zeta^{(1)}_I(\p_2)
 \nn
 &\quad
  + \frac{\Dmr(k\eta)}{\Dmr(k\eta_{\mm/\rr})}
 \delta^\so (\mathbf k,\eta_{\mm/\rr})\,,\label{eq:delta_so_final_exp}
\end{align}
where $u \equiv p_1/k$, $v \equiv p_2/k$,
and $\mathcal I_{\mm/\rr}^{\rm inhom.}$ is the part of the second-order kernel that evolves the second-order density field from some initial surface $\eta_{\mm/\rr}$ to $\eta$ under the first-order sources. 
Note that we have used $\mathbf p_1 \cdot \mathbf p_2 = k^2(1-u^2-v^2)/2$ with $\mathbf p_2 = \mathbf k - \mathbf p_1$ to express the momentum dependence of the kernel $\cI$ in terms of $u$ and $v$.
The second-order contribution at the initial surface evolves under the homogeneous equations, which are the same as in linear theory.  In general, a fraction of the initial value at $\eta_{\mm/\rr}$ goes into each of the homogeneous modes as determined by matching at the boundary.  Here we have assumed all of the initial contributions are in the growing mode as appropriate for taking $\eta_{\mm/\rr} \rightarrow 0$.  More generally, when there are first-order-squared sources before $\eta_{\mm/\rr}$,  this expression holds for $\eta\gg \eta_{\mm/\rr}$ when only the growing mode component survives, with $\delta^{(2)}({\bf k},\eta_{\mm/\rr})$ as the initial value of that component  at $\eta_{\mm/\rr}$ (see Appendix \ref{app:secondorder} for details). 

In general, we can characterize the initial or homogeneous piece by specifying an initial non-Gaussianity.  For the three-point study, this involves the initial three-point function.
For the RD era, we can generically characterize the homogeneous term through the initial three-point function at the end of inflation or the start of radiation domination, $\eta=\eta_\rr$, by 
\begin{equation}
\lim_{\mathbf q \rightarrow \bz}{\expval{ \zeta_I(\mathbf q) \delta(\mathbf k_1,\eta_\rr) \delta(\mathbf k_2,\eta_\rr)}'} = {\cal H}_\rr(q,k_S,\mu) P_{\zeta_I}(q) P_{\delta}(k_S, \eta_\rr,\eta_\rr) \qquad ({\rm RD})\,,
\label{eq:RDinitial}
\end{equation}
assuming that all $k$-modes are outside the horizon at $\eta_\rr$.
For an initial non-Gaussianity obeying the single-field
inflationary consistency relation \eqref{eq:original_cons_rel}, the initial three-point function also satisfies the general consistency relation~\eqref{eq:general_gauge_cons_rel},
\begin{equation}
{\cal H}_\rr(q,k_S,\mu) = -n_\delta +   3f\big( \mu^2 n_\delta + (1-3 \mu^2)\big) + {\cal O}(q^2)\,,
\label{eq:h_rr}
\end{equation}
where $n_\delta = n_s+3$.  
Here ${\cal O}(q^2)$ generally contains $(q\eta_\rr)^2$ and for 
single-field inflation $(q/k_S)^2$ terms.  
It may also contain terms associated with the horizon scale at which slow-roll conditions are violated during inflation~\cite{Zegeye:2021yml}. 
For a contrasting case, where the consistency relation can be violated initially, we consider local non-Gaussianity from multi-field inflation in Appendix \ref{app:NewtonianlocalNG}.

Likewise, we characterize the effective initial condition for the calculation in the MD era as
\begin{equation}
\lim_{\mathbf q \rightarrow \bz}{\expval{ \zeta_I(\mathbf q) \delta(\mathbf k_1,\eta_\mm) \delta(\mathbf k_2,\eta_\mm)}'} =
{\cal H}_\mm(q,k_S,\mu) P_{{\zeta_I}}(q) P_{\delta}(k_S, \eta_\mm,\eta_\mm)\qquad (\text{MD})\,.
\label{eq:MDinitial}
\end{equation}
Here $\eta_\mm \gtrsim \eta_{\rm eq}$ is the beginning of the MD era whereafter we omit the radiation component.  We shall see that dynamical evolution in the RD universe through the inhomogeneous term preserves the consistency relation, so that single-field inflation models satisfy 
\begin{equation}
{\cal H}_\mm(q,k_S,\mu) = -n_\delta(k_S) +  3f\big( \mu^2 n_\delta(k_S) + (1-3 \mu^2)\big) + {\cal O}(q^2)\,,
\label{eq:h_mm}
\end{equation}
where likewise we shall see that the ${\cal O}(q^2)$ term contains $(q\eta_\mm)^2$ terms from the RD evolution as well as $(q/k_S)^2$ terms from 
single-field inflation. 
 The preceding RD epoch does change $n_\delta$ from $n_s+3$ 
since $P_\delta(k,\eta_1,\eta_2) \propto k^4 \eta_1^2 \eta_2^2 \Tm^2(k) P_{\zeta_I}(k)$ well after the equality but leaves $n_\delta$ independent of  $\eta_1$ and $\eta_2$ during that era ($\eta_1 {>} \eta_\mm$ and $\eta_2 {>} \eta_\mm$). 
Note that even for local non-Gaussianity where the inflationary consistency relation is violated, the dilation piece that arises from $n_\delta(k_S) -(n_s+3)$ or $\Tm$ will remain (see Appendix \ref{sec:NewtonianMD}, Eq.~(\ref{eq:3pt_new_q01})).

\subsection{Consistency Relation and Initial Non-Gaussianity}
\label{subsec:3pt_cr}

In the following, we will compare the second-order perturbations and the consistency relation by computing the three-point function up to $\mathcal O(q)$. 
Since we focus on the case where the long mode is always outside the Jeans scale, i.e., $q \ll \eta^{-1}$ in RD and $q \ll \eta_{\rm eq}^{-1}$ in MD, we set $\Tmr(q) = 1$ for the long mode.
We can then calculate the left-hand-side (LHS) of the consistency relation by expressing the second-order perturbations as 
\begin{align}
    \frac{\expval{ \zeta_I(\q) \delta(\k_1,\eta_1) \delta(\k_2,\eta_2)}'}{P_{\zeta_I}(q)}
	 &=    \cI_{\mm/\rr}^{\rm inhom.}\!\left(\frac{q}{k_1},\frac{k_2}{k_1},k_1 \eta_1\right)\Dmr(k_2\eta_2) \Tmr^2(k_2)P_{\zeta_I}(k_2) +(1\leftrightarrow 2)\, \nonumber\\
 & \hskip -20pt +  \frac{ \Dmr(k_1\eta_1)  \Dmr(k_2\eta_2 )}{ \Dmr(k_1\eta_{\mm/\rr}) \Dmr(k_2\eta_{\mm/\rr})}{\cal H}_{\mm/\rr}(q,k_S,\mu)    P_{\delta}(k_S, \eta_{\mm/\rr},\eta_{\mm/\rr})\, .
 \label{eq:3pt_so}
\end{align}
Next, we re-express the wavenumbers $k_1,k_2$ in terms of $k_S$ and $\mu \equiv \hat q\cdot\hat{ k}_S$ using
\begin{align}
	k_1 = k_S\, \sqrt{1 - \frac{q}{k_S}\mu + \frac{1}{4} \left( \frac{q}{k_S} \right)^2}\,, \quad k_2 =  k_S\,\sqrt{1 + \frac{q}{k_S}\mu  + \frac{1}{4} \left( \frac{q}{k_S} \right)^2}\,,
\end{align}
which follows from $\q+\k_1+\k_2=\bz$.
We then expand the inhomogeneous term in small $q$ to obtain
\begin{align}
	&\lim_{\mathbf q \rightarrow \bz}\frac{\expval{ \zeta_I(\mathbf q) \delta(\mathbf k_1,\eta_1) \delta(\mathbf k_2,\eta_2)}'}{P_{\zeta_I}(q) \Tmr^2(k_S) P_{\zeta_I}(k_S) }
	= \sum^\infty_{s,t=0} {\mathcal B}_{\mm/\rr}^{[s,t]}(y_1,y_2) \left(\frac{q}{k_S}\right)^s \mu^t \nn
	&\qquad \qquad \qquad \quad +  \dblue{\frac{\Dmr(k_1\eta_1)  \Dmr(k_2\eta_2 )\Dmr^2(k_S\eta_{\mm/\rr})}{\Dmr(k_1\eta_{\mm/\rr})  \Dmr(k_2\eta_{\mm/\rr} )} }{\cal H}_{\mm/\rr}(q,k_S,\mu)  \,, \label{eq:3pt_with_ignorance}
 \end{align}
where $y_1 = k_S \eta_1$, $y_2 = k_S \eta_2$ here and throughout.
In the following, we will give the concrete expressions of $\mathcal B_{\mm/\rr}$ as appropriate.
Note that the $\O(q)$ consistency relation involves $s\le 1$.

\subsubsection{Matter-Dominated Era}

We first consider the MD era. 
The growth function of the first-order density perturbations is given by
\begin{align}
	\Dm(y) &=  \frac{y^2}{10}\,, \label{eq:trans_delta_mm_syn}
\end{align}
and the inhomogeneous kernel of the second-order density perturbations is given by~\cite{Tomita1967a,Bartolo:2005xa}
\begin{align}
\label{eq:inhomogeneousMD}
	\mathcal I_{\mm}^{\rm inhom.}(u,v,y) =\frac{u^4+v^4+12u^2v^2-2(u^2+v^2)+1}{700}\,  y^4\,,
\end{align}
where we have taken $\eta_\mm\rightarrow 0$ as described in Appendix \ref{app:secondorder} (see Eq.~(\ref{eq:inhomogeneousMDgeneral})). 
Plugging this into Eq.~\eqref{eq:3pt_so} and expanding in the squeezed limit,  
we find that the inhomogeneous contribution to the three-point function vanishes up to $\O(q)$, i.e.,~$\mathcal B^{[s,t]}_{\mm}=0$ for $s=0,1$ in Eq.~\eqref{eq:3pt_with_ignorance}. Since 
\begin{equation}
 \frac{ \Dm(k_1 \eta_1)  \Dm(k_2 \eta_2 )}{ \Dm(k_1 \eta_\mm) \Dm(k_2 \eta_\mm)}  P_\delta(k_S,\eta_\mm,\eta_\mm)
= \frac{\eta_1^2 \eta_2^2}{\eta_\mm^4} P_\delta(k_S,\eta_\mm,\eta_\mm)
=
P_\delta(k_S,\eta_1,\eta_2)\,,
\label{eq:p_delta}
\end{equation}
it immediately follows that for an initial non-Gaussianity or homogeneous term that satisfies the consistency relation Eq.~(\ref{eq:MDinitial}),  the late-time consistency relation of Eq.~(\ref{eq:general_gauge_cons_rel}) holds: 
\begin{align}
\lim_{\mathbf q \rightarrow \bz}\frac{\expval{ \zeta_I(\mathbf q) \delta(\mathbf k_1,\eta_1) \delta(\mathbf k_2,\eta_2)}'}{{P_{\zeta_I}(q)} P_\delta(k_S,\eta_1,\eta_2)}
	&= -  n_\delta(k_S,\eta_1,\eta_2) + 3f\big( \mu^2 n_\delta(k_S,\eta_1,\eta_2) + (1-3 \mu^2)\big)
\label{eq:ndelta_ks}
\end{align}
to $\mathcal O(q)$.   In particular at this order in $q$, the consistency relation for $f=0$ becomes a pure dilation of short-wavelength fluctuations in the long-wavelength mode, whereas for $f=1$ its angle average vanishes entirely,  as expected.

\subsubsection{Radiation-Dominated Era}
\label{subsubsec:3pt_rd}

Next, we summarize the expressions in the RD era.   
The growth function during the RD era is given by 
\begin{align}
	\Dr(y) &= -\frac{4(2-(2-\tilde y^2)\cos\tilde y - 2\tilde y\sin\tilde y)}{\tilde y^2}\, ,
 \label{eq:Tdrr}
\end{align}
where we have defined $\tilde y\equiv y/\sqrt 3$.   We will refer to the oscillatory features in this growth function as acoustic oscillations and the change from the initial power law behavior as the mode crossing the Jeans scale.
The expressions of the inhomogeneous kernel,  again taking the initial surface $\eta_\rr\rightarrow 0$, can be found in Appendix~\ref{app:secondorder}.
For brevity, we here directly go to the expressions of $\mathcal B^{[s,t]}_\rr$ whose nonzero values for $s\le 1$ are
\begin{align}
	\label{eq:cb_00_f}
	\cB_\rr^{[0	,0]} &=	-4 \left( \yy_2 \sin \yy_2 \Dr(y_1) + \yy_1 \sin \yy_1 \Dr(y_2) - 2 \Dr(y_1) \Dr(y_2) \right), \\
	\cB_\rr^{[0	,2]} &=	12f \left( \yy_2 \sin \yy_2 \Dr(y_1) + \yy_1 \sin \yy_1 \Dr(y_2) - 2 \Dr(y_1) \Dr(y_2) \right), \\
	\cB_\rr^{[1	,1]} &=	2(4 - 3f + (n_s-1)) \left( \yy_2 \sin \yy_2 \Dr(y_1) - \yy_1 \sin \yy_1 \Dr(y_2) \right), \\
	\label{eq:cb_13_f}	
	\cB_\rr^{[1	,3]} &=	-6f (1+ (n_s-1)) \left( \yy_2 \sin \yy_2 \Dr(y_1) - \yy_1 \sin \yy_1 \Dr(y_2) \right),
\end{align}
where $\tilde y_{1,2} \equiv y_{1,2}/\sqrt{3} = k_S \eta_{1,2}/\sqrt{3}$.
Unlike in the MD era, the inhomogeneous part contributes to the three-point function up to $\mathcal O(q)$.  This is because the RD growth function $\Dr$ depends on scale so that 
\begin{equation}
n_\delta(k,\eta_1,\eta_2) \equiv \frac{\dd\ln k^3 P_\delta(k,\eta_1,\eta_2) }{\dd\ln k} = \frac{\dd}{\dd\ln k} \ln \left[\frac{\Dr(k \eta_1)  \Dr(k \eta_2 )}{\Dr^2(k \eta_\rr)}\right]+
 n_\delta(k,\eta_\rr,\eta_\rr)\,.
\label{eq:n_deltaRD}
\end{equation}
For the consistency relation to hold at late times, it is not sufficient that the initial non-Gaussianity obeys the consistency relation Eq.~(\ref{eq:h_rr}): the second-order perturbations must evolve in the correct way such that the sum of the two pieces obeys dilation symmetry.

Let us check this preservation of the consistency relation explicitly.   At ${\cal O}(q)$, if we assume the initial piece obeys the consistency relation \eqref{eq:RDinitial}, then the total $s=0$ contribution  goes as 
\begin{align}
	\cB_\rr^{[0	,0]} -(n_s+3 -3f) \Dr(y_1) \Dr(y_2) 
	&= - \Dr(y_1)  \Dr(y_2) (n_\delta(k_S,\eta_1,\eta_2)-3f)\, , \\
	\cB_\rr^{[0	,2]} + 3f n_s\Dr(y_1) \Dr(y_2)  
	&= 3f \Dr(y_1)  \Dr(y_2) (n_\delta(k_S,\eta_1,\eta_2)-3)\,, 
 \end{align}
 whereas the total $s=1$ contribution vanishes:
 \begin{align}
 \cB_\rr^{[1,1]} -2 (n_s + 3 -3f) \left( \yy_2 \sin \yy_2 \Dr(y_1) - \yy_1 \sin \yy_1 \Dr(y_2) \right) &=0\, , \\
 \cB_\rr^{[1,3]} + 6f n_s \left( \yy_2 \sin \yy_2 \Dr(y_1) - \yy_1 \sin \yy_1 \Dr(y_2) \right) &=0\,. 
\end{align}
The sum then implies that the three-point function satisfies the consistency relation
\begin{align}
\lim_{\mathbf q \rightarrow \bz}\frac{\expval{\zeta_I(\mathbf q) \delta(\mathbf k_1,\eta_1) \delta(\mathbf k_2,\eta_2)}'}{P_{\zeta_I}(q)  P_\delta(k_S,\eta_1,\eta_2) }
	&= - n_\delta(k_S,\eta_1,\eta_2) + 3f\big( \mu^2 n_\delta(k_S,\eta_1,\eta_2) + (1-3 \mu^2)\big)\,,
\label{eq:ndelta_ks_rd}	
\end{align}
for all times $\eta_1,\eta_2 \ge \eta_\rr$.  Physically, the inhomogeneous piece dilates the acoustic scale of the oscillatory small-scale perturbations, while the homogeneous or initial piece dilates the initial scale dependence.  
Therefore in the subsequent MD era, an inflationary non-Gaussianity that obeys the consistency relation once evolved through RD  preserves the consistency relation for the ``initial conditions'' of the MD regime (\ref{eq:h_mm}), despite the change in $n_\delta(k_S,\eta_\mm,\eta_\mm) \ne n_s+3$ induced by the Jeans scale, or horizon at matter-radiation equality that is reflected in $T_\mm(k)$.  Moreover, even if the inflationary initial condition does not obey the consistency relation, e.g., in the local non-Gaussianity case, the dilation symmetry acting on $T_\mm(k)$ is enforced by the RD inhomogeneous evolution, which is independent of the initial non-Gaussianity.   We elaborate on these issues in Appendix \ref{app:NewtonianRD}.

\subsection{Separate Universe and Averaging}
\label{subsec:separate_universe}

In this section, we verify the consistency of the $\mathcal O(q^2)$ three-point correlation and the separate universe approach described in \S\ref{sec:SU}.
In the separate universe approach, we absorb the long-wavelength perturbation into the background, and follow the evolution of the short-wavelength perturbations in the modified background including the $\O(q^2)$ effects of $\zeta_I$ on the spatial curvature of the separate universe.  Since the separate universe is taken to be locally isotropic, this describes the angle-averaged three-point correlation~\cite{Sirko:2005uz,Gnedin:2011kj,Baldauf:2011bh,Li:2014sga,Wagner:2015gva,Hu:2016ssz}.

\subsubsection{Matter-Dominated Era}

First, let us consider the MD era case.  As discussed in \S \ref{sec:SU} (see \cite{Baldauf:2011bh,Li:2014jra} for details), there are in general two effects for the perturbation evolution in the separate universe approach.   
The first is that the changes in cosmological parameters of the local background, Eqs.~(\ref{eq:KL}) and (\ref{eq:HL}), {modify} the growth of the short-wavelength density fluctuations.
Converting $\zeta_I$ to the long-wavelength mode $\delta_L$ in real space using Eq.~(\ref{eq:delta_zeta_rel}), we find
\begin{equation}
\frac{\partial \ln \Dm(k_S \eta_{1,2})} {\partial \delta_L(\eta)} = \frac{13}{21} \frac{\eta^2_{1,2}}{\eta^2}\,.
\end{equation}
The second is that the local density fluctuation is measured with respect to the  local mean and that is perturbed as $1+\delta_L$, which adds 1 to the result, giving 34/21.
We can then write the short-wavelength mode in the presence of the long-wavelength mode, $\delta_L$, as 
\begin{align}
	\delta(\mathbf k)|_{\delta_L} = \delta(\mathbf k)|_0 + \frac{34}{21} \int \frac{\dd^3 q}{(2\pi)^3}\, \delta(\mathbf k-\mathbf q) |_0 \delta_L(\mathbf q)\,,
	\label{eq:d_k_on_d_L_md}
\end{align}
where $\delta(\mathbf k)$ is the density perturbation with respect to the global background  and $\delta(\mathbf k)|_0$ is the density perturbation without the long-wavelength mode.
Using this expression and scaling the evaluation time using  $\Dm$ from Eq.~(\ref{eq:trans_delta_mm_syn}), we  obtain
\begin{equation}
\frac{\partial \ln  P_{\delta}(k_S,\eta_1,\eta_2)}{\partial \delta_L(\eta)}= \frac{34}{21}  \frac{\eta_1^2 + \eta_2^2}{\eta^2}.
\label{eq:MDSUgrowth}
\end{equation}

Next, let us calculate the same quantity using the second-order perturbations. The inhomogeneous term contributes at  $\mathcal O(q^2)$ from the ${\mathcal B}^{[2,0]}_{{\rm m}}$ and  ${\mathcal B}^{[2,2]}_{{\rm m}}$ coefficients (see Eqs.~(\ref{eq:b_m_20_22}))
\begin{align}
	\lim_{\q \rightarrow \bz}\frac{\expval{ \zeta_I(\q) \delta(\k_1,\eta_1) \delta(\k_2, \eta_2)}'}{{P_{\zeta_I}(q)} {\Tm^2(k_S)}  P_{\zeta_I}(k_S)}\bigg|_{\O(q^2)}
	&= \frac{(5+2\mu^2) y_1^2 y_2^2(y_1^2 + y_2^2)}{3500} \left( \frac{q}{k_S}\right)^2 .
	\label{eq:q2_matter}
\end{align}
The homogeneous term carries the initial non-Gaussianity and the non-Gaussianity generated during the RD era.  
For single-field inflation, the former is of $\mathcal O(y_1^2 y_2^2 q^2/k_S^2)$ due to the $\mathcal O(q^2/k_S^2)$ corrections to ${\cal H}_\mm$ in Eq.~\eqref{eq:h_mm}, and the latter is of $\mathcal O(y^2_1 y^2_2 q^2  {\eta_\mm^2})$ due to the ${\cal O}(q^2\eta_\mm^2)$ corrections.
Both of these are negligible in comparison to the inhomogeneous contributions so long as the correlation is evaluated at late times $\eta_{1,2} \gg \eta_\mm$.
Converting the initial curvature and power spectra to density fluctuations using Eqs.~(\ref{eq:delta_zeta_rel}) and (\ref{eq:Pdelta}) we obtain
\begin{align}
\lim_{\q \rightarrow \bz,\,y_{1,2}\gg 1}\frac{\overline{\expval{ \delta(\q,\eta) \delta(\k_1,\eta_1) \delta(\k_2,\eta_2)}'}}{P_\delta(q,\eta,\eta) P_\delta(k_S,\eta_1,\eta_2)}\bigg|_{\O(q^2)}	
= \frac{34}{21}  \frac{\eta_1^2 + \eta_2^2}{\eta^2}  \,.
\label{eq:lhs_rd_qsq}
\end{align}
This is consistent with the separate universe approach, Eq.~(\ref{eq:MDSUgrowth}).

\subsubsection{Radiation-Dominated Era}
\label{sec:RD_SU}

During the RD era, the separate universe approach
again in principle gives a change in the growth function of small-scale perturbations $\Dr$ through a change in cosmological parameters~\cite{Zegeye:2021yml}.\footnote{Although the derivation of the expression in \cite{Zegeye:2021yml} was done in comoving gauge for the short-wavelength fluctuations, this becomes the same as the expression in synchronous gauge in the subhorizon limit.  On the other hand the long-wavelength comoving 
density perturbation $\delta_L|_{\rm com} = 4\delta_L|_{\rm synch}/3$ so that the short-wavelength response in Eq.~(\ref{eq:RDSUgrowth}) to $\delta_L|_{\rm com}$ would no longer take the simple separate universe form.}
However, gravitational growth of small-scale perturbations stabilizes at the Jeans scale and thereafter the growth of the long-wavelength density perturbation $\delta_L\propto \eta^2$ has no further impact on the amplitude of small-scale perturbations. In this case, the short-wavelength acoustic oscillations to leading order in $k_S\eta\gg1$ are given by 
\begin{align}
\lim_{\tilde y_S\gg 1}
\delta(\mathbf k_S)	&\simeq -4 (1 + \delta_L) \cos ( \tilde y_S [1+{\cal O}(\delta_L)])   \zeta_I(\mathbf k_S) \,,
	\label{eq:small_wave_pertb_rd}
\end{align}
where $\tilde y_S \equiv k_S \eta/\sqrt{3}$ and we have used $\Dr \rightarrow -4 \cos\tilde y_S$ 
from Eq.~(\ref{eq:Tdrr}) in Eq.~(\ref{eq:delta_zeta_rel}).

Note that the local cosmological parameters do change the local conformal time and hence the frequency of the oscillation by ${\cal O}(\delta_L)$~\cite{Sherwin:2012nh}.  If we were to expand this relation for a small $\delta_L$, it would appear that the phase shift provides the larger effect $\sin(\tilde y_S)\times {\cal O}(\tilde y_S\delta_L)$ since it continues to accumulate over time.   On the other hand, this shift is only observable by comparing modes at very different times. In addition, as discussed in \S \ref{sec:3point} the geometric effects induced by momentum conservation ${\bf q}+{\bf k}_1+{\bf k}_2=0$ also alter the frequencies for the three-point correlation and accumulate over time.

We  can cleanly extract the more relevant amplitude change $(1+\delta_L)$ due to the rescaling of the local background  by averaging over a cycle of the oscillation and comparing the short-wavelength modes at equal time in the three-point correlation.
We therefore get 
\begin{equation}
\frac{\partial \ln \overline{ P_{\delta}(k_S,\eta,\eta)}}{\partial \delta_L(\eta)}= 2\,,
\label{eq:RDSUgrowth}
\end{equation}
where the overline denotes the cycle average of the acoustic oscillation.

Next, let us see the expressions of the three-point function from the second-order perturbations.
In the subhorizon limit of the short-wavelength modes at equal times ($\eta_1 = \eta_2$), the $\mathcal O(q^2)$ contribution from the ${\mathcal B}^{[2,t]}_{{\rm r}}$ terms of Eqs.~(\ref{eq:cb_r_inhom20})--(\ref{eq:cb_r_inhom24}) \dblue{and the homogeneous contribution} give 
\begin{align}
	&\lim_{\mathbf q \to 0,\, y_1\gg 1}\frac{\expval{\zeta_I(\q) \delta(\k_1,\eta_1) \delta(\k_2,\eta_1)}'}{ P_{\zeta_I}(q) P_{\zeta_I}(k_S)}\bigg|_{\O(q^2)}\nonumber\\
&=  \dblue{-\bigg\{
	\bigg( \frac{4}{9} + \frac{16}{3} \mu^2 - 8f\mu^4\bigg)\sin(2 \tilde y_1)\tilde y_1^3  + 4f(3(n_s-1)-20)\cos(2\yy_1)\yy^2_1\mu^4}
	\\
	&\dblue{\quad\quad- \bigg[ \bigg( 4 + \bigg(-\frac{64}{3} +4(n_s-1)-24f\bigg) \cos( 2 \tilde y_1) \bigg) \mu^2} +\frac{4}{9} \bigg(33 +38 \cos( 2 \tilde y_1) \bigg) \bigg]\tilde y_1^2 
	 \bigg\} \bigg( \frac{q}{k_S} \bigg)^2.
\end{align}
Subleading terms in the subhorizon limit are given in Appendix~\ref{app:secondorder}.    
\dblue{We note that, unlike the MD case, the homogeneous term in Eq.~\eqref{eq:h_rr} contributes at $\mathcal O((q/k_S)^2 \tilde y^2_1)$ even for single-field inflation because the prefactor $D_\rr(k_1 \eta_1)D_\rr(k_2 \eta_1)$ for $\mathcal H_\rr$ gives a contribution in $\mathcal O((q/k_S)^2 \tilde y_1^2)$ in the squeezed limit.}
The oscillatory terms in brackets vanish under time averaging, leaving
\begin{align}
	\lim_{\q \rightarrow \bz,\, \yy_{1}\gg 1}\frac{\overline{\overline{\expval{ \zeta_I(\mathbf q) \delta(\mathbf k_1,\eta_1) \delta(\mathbf k_2,\eta_1)}'}}}{P_{\zeta_I}(q)P_{\zeta_I}(k_S)}\bigg|_{\mathcal O(q^2)}
	&=  
	16\tilde y_1^2 \left( \frac{q}{k_S} \right)^2  ,
	 \label{eq:time_angle_ave_3pt_r_syn}
\end{align}
where the double overline represents both angle and time averaging.  Finally we convert the initial curvature power spectrum to the time-averaged density power spectrum using Eq.~(\ref{eq:Pdelta}) to obtain
\begin{align}
\lim_{\q \rightarrow \bz,\, \yy_1\gg 1}\frac{\overline{\overline{\expval{ \delta(\mathbf q,\eta_1) \delta(\mathbf k_1,\eta_1) \delta(\mathbf k_2,\eta_1)}'}}}{P_\delta(q,\eta_1) \overline{P_\delta(k_S,\eta_1)}}\bigg|_{\O(q^2)}
= 2\, ,
\label{eq:lhs_rd_qsq_rd}
\end{align}
which is consistent with the leading-order separate universe expectation, Eq.~(\ref{eq:RDSUgrowth}).

\section{Conclusion}
\label{sec:conclusion}

In this paper, we have thoroughly analyzed the squeezed $N$-point function
to quadratic order in the wavenumber of the soft mode in synchronous gauge.  The equivalence principle implies that synchronous gauge is the natural coordinate system to analyze the impact of long-wavelength fluctuations on short-wavelength observables because it follows the free-fall frame of observers that are initially at rest with respect to the expansion.  It is then manifest that there are no dynamical effects of the gradient of the long-wavelength curvature perturbation  on short-wavelength modes in synchronous gauge. 
The entire impact of the long-wavelength mode at ${\cal O}(q)$ is a time-independent change of the local spatial coordinates, making the form of the consistency relation simple, even for late- and unequal-time correlation functions.   With the freedom to define initial spatial coordinates in synchronous gauge, for the three-point function we can make this effect as a pure dilation or remove it entirely through angle averaging. This single relation automatically includes the so-called Newtonian consistency relation that expresses the equivalence principle in Newtonian gauge.

At quadratic order $\mathcal O(q^2)$, this background wave method also provides the perturbation framework for the separate universe approach. In this approach, while synchronous time slicing still defines the constant-time surfaces of the local universe, the long-wavelength mode modifies the local 3-curvature and expansion rate. We have extended these separate universe methods for the RD universe, for which special care must be taken in the averaging of acoustic oscillations with respect to angle and time.

We have also analyzed the relationship between these kinematic and dynamical constraints on the squeezed three-point function and those predicted by second-order density perturbations. In doing so, we paid special attention to the role of the dynamical evolution sourced from first-order perturbations versus initial non-Gaussianity, which we refer to as inhomogeneous vs.~homogeneous contributions.  In particular, we have clarified the role of the evolution during the RD era  in providing the initial conditions for the MD era  for scales below the horizon during radiation domination.
During the RD era the initial second-order contributions are responsible for the dilation of the initial scale dependence of perturbations, whereas sourced evolution causes a dilation of the acoustic scale in the radiation and matter transfer functions.  Once this combination of effects is established in the prior RD epoch, it is maintained simply by the homogeneous evolution during the MD era.  We have clarified how this combination enters by explicitly keeping the matter transfer function $T_\mm$ throughout the calculations and the change in the tilt of the density power spectrum that it implies.

This decomposition of contributions to the consistency relation also has implications for the squeezed limit of the observable angular bispectrum of the CMB anisotropy~\cite{Creminelli:2011sq,Pajer:2013ana}.  In this case, the long-wavelength mode which traces the large angle temperature fluctuation dilates the angular coordinates of the small-scale temperature power spectrum, shifting their acoustic peaks from a projection of these $k$-space relations from the perspective of the observer today, which is not local in the plasma at recombination.  
One can likewise decompose its pieces into an initial piece that carries the initial tilt $n_s-1$ of the spectrum and a first-order sourced piece that  carries the acoustic scale.   Beyond single-field inflation where the former piece can break, the latter piece still contributes as we have shown explicitly for local non-Gaussianity in Appendix~\ref{app:cons_rel_newtonian} and \ref{app:new_to_syn}.  A breaking of this joint dilation of all scales would indicate a falsification of single-field inflation.

Beyond the consistency relation, the RD background wave and separate universe techniques introduced here are useful for the prediction of the dependence of any local quantity on the long-wavelength curvature perturbation, e.g., the formation of CMB spectral distortions \cite{Zegeye:2021yml} and the spatial dependence of the dark matter abundance relative to the radiation.  We leave these and other applications to future work.

\acknowledgments
We thank Austin Joyce for collaboration on the initial phase of this project and useful discussions throughout.
K.I. and H.L. were supported by the Kavli Institute
for Cosmological Physics at the University of Chicago
through an endowment from the Kavli Foundation and
its founder Fred Kavli. W.H. was supported by U.S.
Dept.\ of Energy contract DE-FG02-13ER41958 and the
Simons Foundation. 
We acknowledge the use of the
Mathematica package \texttt{xAct}~\cite{Pitrou:2013hga}.
We used OpenAI Codex (GPT-5; accessed July 2026) and Anthropic Claude (including models through Claude Opus 5) to find typos/errors in the previous version of this manuscript and their corrections are highlighted in \dblue{this color}.
All scientific judgments, derivations, calculations, interpretation, and responsibility for the manuscript remain solely with us.

\appendix

\section{Removal of the Delta Function}
\label{app:remove_delta}

In this appendix, we review the removal of the delta function from  the consistency relation in the case of $f=0$ (Eqs.~(\ref{eq:cons_rel_f1})--(\ref{eq:f_1_gauge_cons_rel})) and $f=1$ (Eqs.~(\ref{eq:cons_rel_zero_trace})--(\ref{eq:f_0_gauge_cons_rel})). See also Refs.~\cite{Creminelli:2012ed, Goldberger:2013rsa, Horn:2014rta}.

\subsection{Isotropic-Synchronous Gauge \texorpdfstring{($f=0$)}{(f=0)}}

We begin with the isotropic case. 
On the RHS of Eq.~(\ref{eq:cons_rel_f1}), the derivatives act on the delta function.
First, let us treat the $\mathcal O(q^0)$ term in the RHS of Eq.~(\ref{eq:cons_rel_f1}).
Using the relation
\begin{align}
	\sum_{a=1}^{N} k^i_a \partial_{k^i_a} \delta_D(\mathbf k_\tot) 
	=& \int \frac{\dd^3 x}{(2\pi)^3} x^i \partial_i \ee^{-i (\mathbf k_1 + \cdots + \mathbf k_N) \cdot \mathbf x} \nonumber \\
	=& -3 \delta_D(\mathbf k_\tot)\,,
\end{align}
with $\mathbf k_\tot \equiv \sum^N_{a=1} \mathbf k_a$, we can rewrite the consistency relation at $\mathcal O(q^0)$ as 
\begin{align}
	\lim_{\mathbf q \rightarrow \bz} &\expval{\zeta_I(\mathbf q) \delta({\mathbf k_1},\eta_1) \cdots \delta({\mathbf k_N},\eta_N)}'  \nn
 &\qquad\qquad
 =- P_{\zeta_I}(q)\left(3(N-1)+ \sum_{a=1}^{N} k^i_a \partial_{k^i_a}\right)\expval{\delta({\mathbf k_1},\eta_1) \cdots \delta({\mathbf k_N},\eta_N)}'\,,
	\label{eq:dilation_cons_rel_q0}
\end{align}
where note again the prime means the removal of the momentum-conserving delta function, $\expval{\cdots}= (2\pi)^3 \delta_D\left({\bf k}_{\rm tot}\right)\expval{\cdots}'$.

Next, let us examine the $\mathcal O(q)$ terms in Eq.~\eqref{eq:cons_rel_f1}.  The term $\sum_a 3 \mathbf q \cdot \partial_{\mathbf k_a} \delta_D(\mathbf k_\tot)$ can be rewritten as $3N \q \cdot \partial_{\k_\tot} \delta_D(\k_\tot)$, while the second derivative of the delta function can be expressed as 
\begin{align}
	\sum^N_{a=1}\left(q^i k^j_a \partial_{k^i_a} \partial_{k^j_a} - \frac{1}{2}\mathbf q \cdot \mathbf k_a \nabla_{k_a}^2 \right) \delta_D(\mathbf k_\tot) = -3 \mathbf q \cdot \partial_{\mathbf k_\tot} \delta_D(\mathbf k_\tot)\, .
\end{align}
In addition, Eq.~(\ref{eq:cons_rel_f1}) also has the first derivative of delta function multiplied by the first derivative of $\expval{\cdots}'$:
\begin{align}
	&\sum^N_{a=1}q^i  \Big(k^j_a(\partial_{k^i_a} \delta_D(\mathbf k_\tot)) \partial_{k^j_a}  + (\partial_{k^j_a} \delta_D(\mathbf k_\tot)) (k^j_a \partial_{k_a^i} -  k^i_a \partial_{k_a^j} )\Big)\expval{\cdots}' \nn
 &\quad\quad= \sum^N_{a=1}(q^i k^j_a (\partial_{k^i_a} \delta_D(\mathbf k_\tot)) \partial_{k^j_a} \expval{\cdots}' ,
\end{align}
where we have used the rotational invariance of the $N$-point function.
Using these expressions, we can rewrite the ${\mathcal O}(q)$ terms in the RHS of Eq.~(\ref{eq:cons_rel_f1}) as 
\begin{align}
	&(\ref{eq:cons_rel_f1})|_{\O(q)} = -P_{\zeta_I}(q) \bigg[(2\pi)^3 (\q \cdot \partial_{\mathbf k_\tot}  \delta_D(\mathbf k_\tot)) \left(3(N-1)+ \sum_{a=1}^{N} k^i_a \partial_{k^i_a}\right) \nonumber \\
	&\quad + (2\pi)^3 \delta_D(\mathbf q + \mathbf k_\tot)\sum_{a=1}^{N}q^i\bigg(3 \partial_{k^i_a} + k^j_a \partial_{k^j_a} \partial_{k^i_a} - \frac{1}{2} k^i_a \nabla^2_{k_a}\bigg) \bigg] \expval{\delta({\mathbf k_1},\eta_1) \cdots \delta({\mathbf k_N},\eta_N)}' \nonumber\\
  & \approx \expval{\zeta_I(\mathbf q) \delta({\mathbf k_1},\eta_1) \cdots \delta({\mathbf k_N},\eta_N)}'
(2\pi)^3(\q \cdot \partial_{\q}  \delta_D(\mathbf q + \mathbf k_\tot) )
  \label{eq:rhs_final_f1} \\
	&\quad - (2\pi)^3 \delta_D(\mathbf q + \mathbf k_\tot) P_{\zeta_I}(q)\sum_{a=1}^{N}q^i\bigg(3 \partial_{k^i_a} + k^j_a \partial_{k^j_a} \partial_{k^i_a} - \frac{1}{2} k^i_a \nabla^2_{k_a}\bigg)\expval{\delta({\mathbf k_1},\eta_1) \cdots \delta({\mathbf k_N},\eta_N)}',
\nonumber
\end{align}
where we have used the ${\cal O}(q^0)$ dilation consistency relation, Eq.~(\ref{eq:dilation_cons_rel_q0}), in the approximation.
Since this approximated term matches the expansion of the delta function on the LHS of Eq.~(\ref{eq:cons_rel_f1}) to ${\mathcal O}(q)$, we can rewrite the whole consistency relation without delta functions as Eq.~\eqref{eq:f_1_gauge_cons_rel}.

\subsection{Anisotropic-Synchronous Gauge \texorpdfstring{($f=1$)}{(f=1)}}

Next, we discuss the removal of the delta function in the anisotropic case, Eq.~(\ref{eq:cons_rel_zero_trace}).
At $\mathcal O(q^0)$, we can use the relation
\begin{align}
	\sum_{a=1}^{N}k^i_a \partial_{k^j_a}\delta_D(\mathbf k_\tot) 
	= -\delta_{ij} \delta_D(\mathbf k_\tot)\,.
\end{align}
From this, we can see that the term with the derivatives on the delta function in Eq.~(\ref{eq:cons_rel_zero_trace}) becomes zero due to the contraction with the traceless projector $ (\frac{q_i q_j}{q^2} - \frac{1}{3} \delta_{ij}) $.
We can then remove the delta function without changing the $\mathcal O(q^0)$ expression, giving
\begin{align}
&	\lim_{\mathbf q \rightarrow \bz} \expval{\zeta_I(\mathbf q) \delta({\mathbf k_1},\eta_1) \cdots \delta({\mathbf k_N},\eta_N)}'|_{\O(q^0)}\nn
&\qquad= 3P_{\zeta_I}(q)\sum_{a=1}^{N} \left( \frac{q_i q_j}{q^2} - \frac{1}{3} \delta_{ij} \right) k^i_a \partial_{k^j_a} \expval{\delta({\mathbf k_1},\eta_1) \cdots \delta({\mathbf k_N},\eta_N)}'.
    \label{eq:cons_rel_zero_trace_q0_wo_del}
\end{align}
At  $\mathcal O(q)$, the RHS of Eq.~(\ref{eq:cons_rel_zero_trace}) has the second derivative of the delta function, which can be expressed as 
\begin{align}
	\sum_{a=1}^{N} \left( \frac{q_i q_j}{q^2} - \frac{1}{3} \delta_{ij} \right) \left( k^i_a \partial_{k^j_a}(\partial_{\mathbf k_a} \cdot \mathbf q) - \frac{1}{2}(\mathbf k_a \cdot \mathbf q) \partial_{k^i_a} \partial_{k^j_a} \right)\delta_D(\mathbf k_\tot) = 0\, . 
\end{align}
Also, the RHS of Eq.~(\ref{eq:cons_rel_zero_trace}) has the first derivative of delta function multiplied by the first derivative of $\expval{\cdots}'$:
\begin{align}
	&\sum^N_{a=1}\left( \frac{q_i q_j}{q^2} - \frac{1}{3} \delta_{ij} \right) \left[q^l (\partial_{k^l_a} \delta_D(\mathbf k_\tot)) k^i_a \partial_{k^j_a}  + q^l (\partial_{k^j_a} \delta_D(\mathbf k_\tot)) ( k^i_a \partial_{k_a^l} -  k^l_a \partial_{k_a^i} ) \right]\expval{\cdots}' \nonumber\\
	&= \sum^N_{a=1}\left( \frac{q_i q_j}{q^2} - \frac{1}{3} \delta_{ij} \right) q^l (\partial_{k^l_a} \delta_D(\mathbf k_\tot)) k^i_a \partial_{k^j_a} \expval{\cdots}',
\end{align}
where we have used the rotational invariance of the $N$-point function.
Using these expressions, we can rewrite the RHS of Eq.~(\ref{eq:cons_rel_zero_trace}) as 
\begin{align}
    &(\ref{eq:cons_rel_zero_trace}) = 3P_{\zeta_I}(q) \left[ (2\pi)^3(\q \cdot \partial_{\k_\tot} \delta_D(\mathbf k_\tot)) \sum^N_{a=1} \left( \frac{q_i q_j}{q^2} - \frac{1}{3} \delta_{ij} \right) k^i_a \partial_{k^j_a} \right. \label{eq:rhs_final_syn_f0} \\
	&\left.  + (2\pi)^3 \delta_D(\mathbf k_\tot)\sum_{a=1}^{N} \left( \frac{q_i q_j}{q^2} - \frac{1}{3} \delta_{ij} \right) \left( k^i_a (\partial_{\mathbf k_a} \cdot \mathbf q) - \frac{1}{2} (\mathbf k_a \cdot \mathbf q) \partial_{k^i_a}  \right)\partial_{k^j_a}  \right]\expval{\delta({\mathbf k_1},\eta_1) \cdots \delta({\mathbf k_N},\eta_N)}'. \nonumber
\end{align}
As with the $f=0$ case, using the ${\cal O}(q^0)$ relation Eq.~(\ref{eq:cons_rel_zero_trace_q0_wo_del}), the first line of this equation is the same as the first-order expansion of the delta function on the LHS of Eq.~(\ref{eq:cons_rel_zero_trace}).
Then, we finally obtain the consistency relation without the delta function, Eq.~(\ref{eq:f_0_gauge_cons_rel}).

\section{Second-Order Relations in Synchronous Gauge}
\label{app:secondorder}

In this appendix, we provide details of the second-order calculation in synchronous gauge. We first present relevant second-order quantities in an arbitrary gauge in \S\ref{app:secondorder1}, and then solve the Einstein equations in the MD and RD eras for synchronous gauge in \S\ref{app:secondorder2}. 
We also provide explicit expressions for the three-point function in the squeezed limit at $\O(q^2)$.

\subsection{Second-Order Perturbation Theory}\label{app:secondorder1}
The line element perturbed around the flat FLRW metric can be written as
\begin{align}
	\dd s^2 = a^2(\eta)\Big[{-}(1+2\Phi)\dd\eta^2 + 2B_i \dd\eta \dd x^i +(\delta_{ij}+2C_{ij})\dd x^i\dd x^j\Big]\, .
	\label{eq:metric_def}
\end{align}
A perturbed quantity can be expanded as $Q_{ij\cdots} = \bar Q_{ij\cdots} + \sum_{n=1}^\infty \frac{1}{n!}Q^{(n)}_{ij\cdots}$, 
where the overbar indicates its background value. 
The inverse metric components up to second order are
\begin{align}
	g^{00} &= -a^{-2}\left(1-2\Phi+4\Phi^2-B_i^2\right) , \\
	g^{0i} &= a^{-2}\left(B_i-2\Phi B_i-2B_kC_{ki}\right) , \\
	g^{ij} &= a^{-2}\left(\delta_{ij}-2C_{ij}+4C_{ik}C_{kj} - B_i B_j \right) .
\end{align}
We remind the reader {of} our convention that the indices of 4-tensors are raised and lowered with the metric $g_{\mu\nu}$, while those of 3-tensors such as $B_i$ and $C_{ij}$ above are raised and lowered with $\delta_{ij}$. To minimize confusion, in this appendix we will lower all the 3-indices whenever possible and sum over the repeated lowered indices.

It is convenient to decompose the metric components into their {3}-scalar, vector, and tensor parts as
\begin{align}
	B_i &= \partial_i B + B_i^{\rm T}\, , \\
	C_{ij} &=  \frac{1}{3}C \delta_{ij} + D_{ij}\nabla^{-2} \hE + \partial_{(i}E_{j)}^{\rm T} + E_{ij}^{\rm T}\, ,
\end{align}
with $\partial_i B_i^{\rm T}=\partial_i E_i^{\rm T}=\partial_i E_{ij}^{\rm T}=0$, and we have defined
\begin{align}
	 C\equiv C_{kk} \, ,\quad	D_{ij} \equiv \partial_i\partial_j-\frac{1}{3}\delta_{ij}\nabla^2\, .
\end{align}
Up to second order, the Einstein tensor components take the form
\begin{align}
	a^2{G^{0}}_{0} &=-3\cH^2 -(2\cH \partial_\eta-\nabla^2) C -C_{kl,kl} + 6\cH^2\Phi + 2\cH B_{k,k} \nn[2pt]
	&\quad -\tfrac{1}{2}B_{k,k}^2+\tfrac{1}{2}B_{k,l}B_{(k,l)} +((3\cH^2+\tfrac{1}{2}\nabla^2) B_k-\tfrac{1}{2}B_{l,lk}) B_k \nn[2pt]
	&\quad -\tfrac{1}{2}(C'^2-C_{,k}^2-C_{kl}'^2+3C_{kl,m}^2) +C_{km,l}C_{kl,m}+2C_{km,k}C_{lm,l}- 2C_{,k}C_{kl,l}\nn[2pt]
	&\quad +2(2C_{km,lm}+2\cH C_{kl}'-\nabla^2 C_{kl} - C_{,kl})C_{kl}+4\cH( C'-B_{k,k}-3\cH\Phi)\Phi-2\cH \Phi_{,k}B_k\nn[2pt]
	&\quad + C'B_{k,k}-(C_{k,l}'+4\cH C_{kl})B_{k,l}+(C_{,k}'+2\cH C_{,k}-C_{kl,l}'-4\cH C_{kl,l})B_k\, , \\[5pt]
	a^2{G^{0}}_{i} &=  C_{,i}'-2\cH\Phi_{,i} - C_{ik,k}'+\tfrac{1}{2}\nabla^2B_i -\tfrac{1}{2}B_{k,ki} \nn[2pt]
	&\quad -2\cH B_k B_{k,i}+(2C_{kl,l}-C_{,k})C_{ik}'  + 2(C_{il,k}'-C_{kl,i}')C_{kl}-C_{kl}'C_{kl,i}\nn[2pt]
	&\quad +(B_{k,k}-C')\Phi_{,i}-(B_{(i,k)}-C_{ik}')\Phi_{,k} + (B_{k,ki}-\nabla^2 B_i+2C_{ik,k}' -2C_{,i}'+8\cH \Phi_{,i})\Phi \nn[2pt]
	&\quad +C_{,k}B_{[i,k]} + 2(C_{i[l,k]}B_{k,l} + C_{kl,l}B_{[k,i]}+ C_{kl}B_{[k,i]l}) \nn[2pt]
	&\quad + (C_{il,kl}+ C_{kl,il}-\nabla^2C_{ik} - C_{,ik} )B_k\, ,  \\[5pt]
	  a^2{G^{i}}_{j}&=  {\hat G} \delta_{ij} -\Phi_{,ij}-B_{(i,j)}'-2\cH B_{(i,j)}+C_{ij}''+2\cH C_{ij}'-\nabla^2 C_{ij} - C_{,ij}+2C_{k(i,j)k}\nn
	  &\quad -\tfrac{1}{2}(B_{i,k}B_{j,k}+B_{k,i}B_{k,j})+B_{(i,j)}B_{k,k}  +\tfrac{1}{2}(B_{k,kj}-\nabla^2B_j)B_i+(B_{(i,j)k}-B_{k,ij})B_k \nn
	  &\quad + C'C_{ij}'-C_{,k}C_{ij,k}-2C_{ik}'C_{jk}'+2C_{,k}C_{k(i,j)}+4C_{ik,l}C_{j[k,l]} + C_{kl,i}C_{kl,j}+2C_{ij,l}C_{kl,k}\nn
	  &\quad-4C_{kl,k}C_{l(i,j)}-2(C_{jk}''+2\cH C_{jk}'-\nabla^2C_{jk}-C_{,jk})C_{ik}+2(C_{kl,ij}+C_{ij,kl}-2C_{k(i,j)l})C_{kl}\nn
	&\quad -4C_{l(j,k)l}C_{ik}+\Phi_{,i}\Phi_{,j}+2\Phi\Phi_{,ij}+\Phi'B_{i,j} + 2(B_{i,j}'+2\cH B_{i,j})\Phi +2\cH \Phi_{,j} B_i\nn
	&\quad -\Phi'C_{ij}'+(2C_{k(i,j)}-C_{ij,k})\Phi_{,k} -2(C_{ij}''+2\cH C_{ij}')\Phi  + \Phi_{,jk}C_{ik}\nn
	&\quad -C'B_{(i,j)} - C_{ij}'B_{k,k}-C_{ij,k}B_k' + C_{ik}'B_{j,k} + 2C_{k(i,j)}B_{k}'+ C_{jk}'B_{i,k}\nn
	&\quad +(C_{jk,k}'-C_{,j}')B_i - 2(C_{ij,k}'+\cH C_{ij,k}-C_{k(i,j)}'-\cH C_{k(i,j)} )B_k\nn
	&\quad + 2(B_{(j,k)}'+2\cH B_{(j,k)})C_{ik}\, ,
\end{align}
with $\cH \equiv a'/a$, and the term multiplying the Kronecker delta in the spatial components ${G^i}_j$ is given by
\begin{align}
	 {\hat G}  &= -2(\cH^2+\cH')-(\partial_\eta^2+2\cH\partial_\eta-\nabla^2) C- C_{kl,kl}+(2\cH \partial_\eta+ 2\cH^2+4\cH'+\nabla^2)\Phi\nn[2pt]
	&\quad +(\partial_\eta+2\cH)B_{k,k}-\tfrac{1}{2}B_{k,k}^2+\tfrac{3}{4}B_{k,l}^2-\tfrac{1}{4}B_{\dblue{l,k}}^2+B_k((2\cH\partial_\eta+2\cH' -\cH^2 + \nabla^2 )B_k-B_{l,lk})\nn[2pt]
	&\quad-\tfrac{1}{2}(C'^2+C_{,k}^2+3C_{kl}'^2-3C_{kl,m}^2)+2(C_{kl}''+2\cH C_{kl}'-\nabla^2C_{kl}-C_{,kl}+2C_{km,lm})C_{kl}\nn
	&\quad +2(C_{km,m}-C_{,k})C_{kl,l}+C_{kl,m}C_{km,l}-\Phi_{,k}^2-2\Phi(4\cH\partial_\eta+4\cH'+2\cH^2+\nabla^2)\Phi\nn
	&\quad -\Phi'B_{k,k}-2\cH \Phi_{,k}B_k-2(B_{k,k}'+2\cH B_{k,k})\Phi+\Phi'C'+\Phi_{,k}C_{,k}-2\Phi_{,k}C_{kl,l}-2\Phi_{,kl}C_{kl} \nn
	&\quad +2(C''+2\cH C')\Phi+C_{,k}B_k'+C'B_{k,k}-B_{k,l}C_{kl}'-2B_k'C_{kl,l}\nn
	&\quad+2(C_k'+\cH C_{,k}-C_{kl,l}'-2\cH C_{kl,l})B_k -2(B_{k,l}'+2\cH B_{k,l})C_{kl}\, , \label{eq:hat_G}
\end{align}
where the comma $X_{,i}$ is the shorthand expression of $\partial_i X$. 
The energy-momentum tensor of a fluid with density $\rho$ and pressure $P$ is given by
\begin{align}
	T_{\mu\nu} = (\rho+P)u_\mu u_\nu +P g_{\mu\nu}+\Sigma_{\mu\nu}\, ,
\end{align}
where $u^\mu=dx^\mu/d\tau$ is the 4-velocity given by 
\begin{align}
	u^\mu   &=  a^{-1}\left(1-\Phi + \tfrac{3}{2}\Phi^2+\tfrac{1}{2}v_{k}(v_k+2B_k)\, ,\,  v_i\right),\\
	u_\mu  &= a\left(-1-\Phi+\tfrac{1}{2}\Phi^2-\tfrac{1}{2}v_i^2\, ,\, v_i+B_i- \Phi B_i + 2v_kC_{ki}\right),
\end{align}
up to second order, $v_i=\partial_i\hat v$ is the comoving 3-velocity with $\hat v$ the velocity potential, and $\Sigma_{\mu\nu}$ denotes the anisotropic stress, which can be non-vanishing in the presence of e.g.~free-streaming neutrinos. It is subject to the conditions ${\Sigma^\mu}_\mu=0$, $\Sigma_{\mu\nu}u^\nu=0$, and its spatial components can be decomposed as
\begin{align}
	\Sigma_{ij} = D_{ij}\sigma  + \partial_{(i}\Sigma_{j)}^{\rm T} + \Sigma_{ij}^{\rm T}\, ,
\end{align}
with $\partial_i\Sigma_{i}^{\rm T}=\partial_i\Sigma_{ij}^{\rm T}=0$. 
 Up to second order, we have
\begin{align}
	{T^0}_0 &=  -\rho - (\bar\rho+\bar P)v_k(v_k+B_k)\, , \\
	{T^0}_i &=(\rho+P)(v_i+B_i)+(\bar\rho+\bar P)(2v_kC_{ki}\dblue{-\Phi v_i} -\dblue{2}\Phi B_i) \dblue{+a^{-2} \Sigma_{ij}(v_j + B_j)}\,,\\
	{T^i}_j &= P \delta_{ij}+(\bar \rho+\bar P)v_i(v_j+B_j) + a^{-2}(D_{ij}-2C_{ik}D_{kj})\sigma\, ,
\end{align}
where $\bar\rho$ and $\bar P$ denote the background quantities obeying the Friedmann equations
\begin{align}
	\cH^2 = \frac{8\pi G}{3}a^2 \bar\rho\, , \quad \cH^2-\cH'=4\pi Ga^2(\bar\rho+\bar P)\, .
\end{align}

\subsection{Synchronous Gauge}\label{app:secondorder2}

We now specialize to synchronous gauge by setting $\Phi= B_i = 0$. 
We focus on scalar perturbations, and therefore set all vector and tensor perturbations to zero, and relabel the spatial trace of the metric as $\Psi=-\frac{1}{3}C$ to make contact with the notation used in the rest of the paper. This leaves $\Psi$ and $\hat E$ as the metric perturbations with the residual gauge freedom of choosing the time slicing and spatial threading of synchronous coordinates reflecting the choice of initial free-fall observers.
In synchronous gauge, the second-order Einstein equations take the form 
\begin{align}
	6\cH \Psi' -2\nabla^2\bigg(\Psi+\frac{1}{3}\hE\bigg)+3\cH^2 \delta &=  S_{00}\, ,\label{eq00}\\
	-2\partial_i\bigg(\Psi +\frac{1}{3} \hE\bigg)' -3(1+\bar P/\bar\rho)\cH^2\partial_i \hat v &=  S_{0i}\, ,\label{eq0i}\\
	\hE''+2\cH\hE'+ \nabla^2\bigg(\Psi+\frac{1}{3}\hE\bigg) -12\pi G a^2 \nabla^2\sigma &=  \hN_{ij}S_{ij}\, ,\label{eqij}\\
	6\Psi''+12\cH \Psi'-2\nabla^2\bigg(\Psi+\frac{1}{3}\hE\bigg)-9 \cH^2 (P-\bar P)/\bar\rho &=  S_{kk}\label{eqkk}\, ,
\end{align}
where $\hN_{ij}\equiv \frac{3}{2}\nabla^{-2}D_{ij}$ and
we have used the background equations and the second-order source terms ${S}_{\mu \nu}$ consist of first-order quantities
\begin{align}
	\label{eq:s00}
	S_{00}&= -3\cH^2(1+\bar P/\bar\rho)v_i^2+3(\Psi'^2+\Psi_{,k}^2)+ \frac{1}{6}\hE'^2 -\frac{5}{6}\hE_{,i}^2-\frac{4}{3}(9\cH \Psi'-6\nabla^2\Psi-2\nabla^2\hE)\Psi\nn
	&\quad - \frac{1}{2}E_{,ij}'^2 + \frac{1}{2}E_{,ijk}^2 -\frac{2}{3}(6\cH E'+3\Psi+\hE)_{,kl} E_{,kl}+\frac{2}{9}(6\cH\hE'+\nabla^2\hE+3\nabla^2\Psi) \hE \, ,\\[5pt]
 S_{0i}&= 3\cH^2\big[(\delta+(P-\bar P)/\bar\rho)v_i - 2(1+\bar P/\bar\rho)(\Psi v_i-E_{ij}v_j+\tfrac{1}{3} \hat Ev_i)\big] -E'_{ij}\Psi_{,j}+2E_{,ij}\Psi_{,j}' \nn
  &\quad +\frac{2}{3}\hat E'_{,j}E_{,ij}-\frac{4}{3}\hat E_{,j}E'_{,ij}+E_{,ijk}E'_{jk}- 4(\Psi'+\tfrac{1}{3}\hat E')_{,i}\Psi-4(\Psi+\tfrac{1}{3}\hat E)_{,i}\Psi'-\frac{1}{3}\hat E'\Psi_{,i}\nn
  &\quad+\frac{2}{3}\hat E\Psi'_{,i}-\frac{1}{9}\hat E'\hat E_{,i}+\frac{2}{9}\hat E\hat E'_{,i}\,,\\[3pt]
	 S_{ij} &= 3(1+\bar P/\bar\rho)\cH^2v_iv_j-E_{,ikl}E_{,jkl}+\frac{1}{3}E_{,ijk}(3\Psi+4\hE)_{,k}+2(2\cH E+E')_{,ik}E_{,jk}'\nn
	&\quad +2E_{,ik}E_{,jk}''- 2(\Psi+\tfrac{1}{3}\hE)E_{,ij}''-2(\Psi+\tfrac{1}{3}\hE)_{,ik}E_{,jk}-3(\Psi+\tfrac{1}{3}\hE)_{,i}(\Psi+\tfrac{1}{3}\hE)_{,j}\nn
	&\quad -2E_{,ij}(\partial_\eta^2+2\cH\partial_\eta-\nabla^2)(\Psi+\tfrac{1}{3}\hE)-(\Psi'+4\cH\Psi+\tfrac{4}{3}(\hE'+\cH\hE) ) E_{,ij}' \nn
 &\quad -2(2\Psi-\tfrac{1}{3}\hE) (\Psi+\tfrac{1}{3}\hE)_{,ij}+16\pi Ga^2 \Psi D_{ij}\sigma \, ,\\
	S_{kk} &= \dblue{3(1+\bar P/\bar \rho)\mathcal H^2 v_k v_k} + 3(\Psi_{,k}^2-\Psi'^2)+\frac{5}{6}(\hE'^2-\hE_{,k}^2)-\frac{5}{2}E_{,kl}'^2 +\frac{1}{2}E_{,klm}^2\nn
	& \quad -4(3\Psi''+6\cH \Psi' -2\nabla^2(\Psi+\tfrac{1}{3}\hE))\Psi -2(2E''+4\cH E'+\tfrac{1}{3}\hE+\Psi)_{,kl}E_{,kl}\nn 
    &\quad +\frac{2}{3}(2\hE''+4\cH \hE'+\nabla^2(\Psi+\tfrac{1}{3}\hE))\hE\, ,
\end{align}
with \dblue{$E\equiv \nabla^{-2}\hat E$}.
As in the main text, we hereafter assume a perfect fluid with a constant equation of state \dblue{$\bar P = w \bar \rho$} and vanishing anisotropic stress $\sigma=0$. The 00, trace, and trace-free equations are then three equations with three unknowns $\hE, \Psi,\delta$. The $0i$ equation then determines the final matter variable $\hat v$. 
We can combine the 00 and trace equations in two different ways to get
\begin{align}
	\Psi'' + \cH \Psi' - \frac{1}{2}(1+3w)\cH^2\delta  &=   \frac{1}{6}(S_{kk}-S_{00})\, ,\label{Cdeq}\\
		\Psi'' +(2+3w)\cH \Psi' -\frac{1}{9}(1+3w)\nabla^2(3\Psi+\hE) &= \frac{1}{6}(S_{kk} + 3wS_{00})\, .\label{CEeq1}
\end{align}
Combining \eqref{Cdeq}, \eqref{CEeq1}, and \eqref{eqij} in various ways,
 these equations can be decoupled into the ordinary differential equations
\begin{align}
	\cL_\O \O = S_\O\, \label{Oeq}
\end{align}
for the variables $\O\in \{\hE, \Psi,\delta\}$. The differential operators $\cL_\O$ take the form
\begin{align}
	\cL_{\hE} &\equiv \left(\partial_y^3+\frac{2(4+3w)}{(1+3w)y}\partial_y^2+\frac{w(1+3w)^2y^2+8}{(1+3w)^2y^2} \partial_y+\frac{2 w(1+3w)y^2-8}{(1+3w)^2y^3}\right)\partial_y\, , \label{LE} \\[3pt]
	\cL_\Psi &\equiv -\frac{1}{3}\cL_{\hE} +\frac{4w}{(1+3w)y^3}(y\partial_y-1)\partial_y, \label{LC}\\[3pt]
	\cL_\delta &\equiv \left(\partial_y^2-\frac{12w}{(1+3w)y^2}\left(y\partial_y+\frac{1-9w}{6w}\right)+w\right)\left(y\partial_y+\frac{3(1+w)}{1+3w}\right),\label{Ld}
\end{align}
in terms of the dimensionless variable $y\equiv k\eta$. The order of each differential operator indicates the number of pure gauge modes on top of the two physical (growing and decaying) solutions; there are two gauge modes for $\hE$ and $C$, and a single gauge mode for $\delta$. The source terms are given by
\begin{align}
	k^2S_{\hE} &\equiv \frac{w}{2}S_{00}+\frac{1+3w}{3}\hN_{ij}S_{ij}+\frac{1}{6}S_{kk}+\bigg(\partial_y+\frac{2(2+3w)}{(1+3w)y} \bigg)\partial_y\hN_{ij}S_{ij}\, ,\label{SE}\\
	k^2 S_\Psi &\equiv  -\frac{w}{6} S_{00} -\frac{1+3w}{9}\hN_{ij}S_{ij} -\frac{1}{18}S_{kk} +\frac{1}{6y}\bigg(\partial_y + \frac{4}{(1+3w)y}\bigg)\partial_y(S_{kk}+3wS_{00})\, , \label{SC}\\
	k^2S_\delta &\equiv \frac{1}{12}\Big[24(1+w)+(1+3w)(11+9w)y\partial_y+(1+3w)^2y^2\partial_y^2\Big]\partial_yS_{00}-\frac{2}{3}\partial_y S_{kk}\nn[2pt]
	&\quad -\frac{1+3w}{36}y\Big[\big((1+3w)y\partial_y+(5+9w)\big)(2\hN_{ij}S_{ij}+S_{kk})+6\partial_y^2S_{kk}\Big]\, .\label{Sd}
\end{align}
Since the source terms start at second order, the first-order perturbations obey the  homogeneous equations \eqref{Oeq}, which admit analytic solutions for a constant $w$ in terms of the generalized hypergeometric functions. At early times, the independent solutions have the following power-law behavior: 
 \begin{align}
	\lim_{\eta\to 0}\hE^{(1)}(\k,\eta)\ &\supset \ y^2 ,\,  y^{1-\frac{4}{1+3w}},\, y^{1-\frac{2}{1+3w}} ,\, y^0\, ,\label{E1w}\\
	\lim_{\eta\to 0}\Psi^{(1)}(\k,\eta)\ &\supset \  y^2  ,\, y^{3-\frac{4}{1+3w}}  ,\, y^{-1-\frac{2}{1+3w}}
	,\, y^0 \, ,\\
	\lim_{\eta\to 0}\delta^{(1)}(\k,\eta)\ &\supset\  y^2,\, y^{3-\frac{4}{1+3w}} ,\, y^{-1-\frac{2}{1+3w}}\, .\label{d1w}
 \end{align}
 For the two metric variables, the constant mode in the
 curvature perturbation $\Psi^{(1)} +\frac{1}{3}\hat E^{(1)}$ corresponds to the growing mode in the density solution, which scales as $y^2$.  As discussed in \S \ref{sec:synchfreefall} (see Eq.~\eqref{eq:f_def}), there is a residual spatial gauge freedom that is fixed by the choice of initial observers as to how to assign the curvature to the two metric perturbations individually.  In the cases where the constant curvature is assigned to one or the other entirely, $f=0$ or $f=1$ in the main text, then the other will scale as $y^2$.  The other two power-law solutions correspond to the decaying mode and the residual gauge freedom in setting the initial time surface.
We shall make these assignments more concrete in the MD and RD cases below, but in general we set them to zero.

We will be particularly interested in solving the equation for the second-order density perturbation in the MD and RD eras with $w=0$ and $w=1/3$. 
We will express the first- and second-order solutions in terms of the initial first-order curvature perturbation as
\begin{align}
	\delta^{(1)}(\k,\eta) &=  \Dmr(y) \Tmr(k) \zeta_I^{(1)}(\k)\,, \label{delta1st}\\[3pt]
 \label{delta2nd}
 	\delta^\so (\mathbf k,\eta) &= 
\int \frac{\dd^3 p_1\dd^3 p_2}{(2\pi)^3}\delta_D(\k-\p_1-\p_2) \cI_{\mm/\rr}(u,v,y) T_{\mm/\rr}(p_1)T_{\mm/\rr}(p_2)\zeta^{(1)}_I(\p_1)\zeta^{(1)}_I(\p_2)\,,
\end{align}
 with
 \begin{align}
	{\cal I}_{\text{m/r}}(u,v,y) &= {\cal I}_{\text{m/r}}^{\rm inhom.}(u,v,y)+{\cal I}_{\text{m/r}}^{\rm hom.}(u,v,y)\,.
 \end{align} 
The inhomogeneous kernel ${\cal I}_{\text{m/r}}^{\rm inhom.}$ captures the part that is generated by the dynamical evolution from the initial time surface to late times, and the homogeneous part ${\cal I}_{\text{m/r}}^{\rm hom.}$ is fixed by the initial condition at early times.
In the following, we will solve the inhomogeneous solution using the Green's function, and leave the homogeneous solution to be specified by the type of initial non-Gaussianity.

\subsubsection{Matter-Dominated Era}

We start by solving Eqs.~\eqref{LE}--\eqref{Ld} at first order. For $w=0$, the solutions are simple power-law functions. From \eqref{E1w}--\eqref{d1w}, it follows that\footnote{Note that the restriction $\eta\to 0$ only applies to the gauge mode associated with $e_\Psi$, since each of the other modes is given by a single power-law function.}
\begin{align}
	\lim_{\eta\to 0}\hE^{(1)}(\k,\eta) &= c_{\hE} y^2  + d_{\hE} y^{-3}  +e_{\hE} y^{-1}+f_{\hat E}\, , \label{E1}\\
    \lim_{\eta\to 0}\Psi^{(1)}(\k,\eta) &=  c_\Psi y^2  + d_\Psi y^{-1}  +e_\Psi y^{-3} 
	+f_{\Psi} \, ,\label{C1}\\
	\lim_{\eta\to 0}\delta^{(1)}(\k,\eta) &= c_\delta y^2+d_\delta y^{-1} + e_\delta y^{-3}\, ,\label{d1}
\end{align}
where  $c_\O$, $d_\O$, $e_\O$, $f_\O$ depend on $\k$.
We set the decaying mode of adiabatic perturbations to zero by setting $d_\O=0$. In terms of the choice of synchronous observers, $e_\O$ corresponds to the initial time slicing of synchronization, and our choice of $e_\O=0$ specifies that the observers are at rest with respect to the background expansion.   
The time-independent gauge freedom $f_\O$ then corresponds to their initial spatial coordinates, which we parameterize according to \eqref{eq:f_def} as
\begin{align}
	f_{\hat E} = -3f \Tm(k)\zeta^{(1)}_I(\k) \,,\quad f_{\Psi} = (f-1)\Tm(k)\zeta^{(1)}_I(\k)\,.\label{f1st}
\end{align}
Physically, this condition arises from the fact that the 3-curvature on our   $e_\O=0$ synchronous slicing coincides with that of comoving slicing outside the sound horizon or maximal Jeans scale  for any $f$. For first order perturbations this implies 
$\zeta= -\Psi -\frac{1}{3}\hat E$ and $\Tm(k)$ transfers the initial $\zeta$ to the MD epoch.

The constraint equations between the variables and the initial condition on the growing mode fix the $c_\O$ coefficients to be
\begin{align}
	c_{\hE}= -3c_\Psi= - c_\delta = -\frac{1}{10} \Tm(k)\zeta_I^{(1)}(\k)\, .
\end{align}
Matching with \eqref{delta1st}, we find 
\begin{align}
	  \Dm(y)=\frac{y^2}{10}\,,
	  \label{eq:t_delta_m}
\end{align}
which gives \eqref{eq:delta_zeta_rel} and \eqref{eq:trans_delta_mm_syn} in the main text. Finally, the $0i$ Einstein equation \eqref{eq0i} implies $v_i^{(1)}=0$. This completes the specification of the first-order solutions.

The second-order perturbations obey the inhomogeneous equations \eqref{Oeq}. To solve these, we use the Green's functions obeying $\cL_\O G_\O(\eta,\eta')=\delta_{\rm D}(\eta-\eta')$; these are
\begin{align}
	 G_{\hE}(\eta,\eta')=-\frac{1}{3}G_{\Psi}(\eta,\eta') &=\frac{\eta^2\eta'}{30}\left(1-\frac{\eta'}{\eta}\right)^3\left(1+\frac{3\eta'}{\eta}+\frac{\eta'^2}{\eta^2}\right)\theta(\eta-\eta')\,,\\[3pt]
	G_{\delta}(\eta,\eta')&= \frac{\eta^2}{30}\left(2-\frac{5\eta'^3}{\eta^3}+\frac{3\eta'^5}{\eta^5}\right)\theta(\eta-\eta')\, .
\end{align}
Focusing on the second-order density perturbation, we integrate the Green's function $G_\delta$ over the source from $\eta_\mm$ to $\eta$ to obtain 
\begin{align}
{\cal I}^{\rm inhom.}_{\text{m}}(u,v,y)      =\frac{u^4+v^4+12u^2v^2-2(u^2+v^2)+1}{2100y^3}  (3y^7-7y^5y_\mm^2+7y^2y_\mm^5-3y_\mm^7)\,,
\label{eq:inhomogeneousMDgeneral}
\end{align}
where $y_\mm \equiv k \eta_\mm$.
Notice that the lower boundary at $\eta_\mm$ (or $y_\mm$) contributes to terms that correspond to the
growing mode $c_\delta$, decaying mode $d_\delta$, and temporal gauge mode $e_\delta$ of linear theory (\ref{d1}) at $\eta>\eta_\mm$.  These homogeneously evolving contributions represent the evolution of the impact of the  sources at the boundary.  Conversely, with a model where there are sources at $\eta<\eta_\mm$ there will also be homogeneous contributions to each of these modes at the boundary.  At $\eta\gg \eta_\mm$ only the growing mode component of contributions from $\eta_\mm$ or before survive.  Moreover if we take $\eta_\mm\rightarrow 0$ these terms formally disappear, the inhomogeneous term includes all sourced contributions, and Eq.~(\ref{eq:inhomogeneousMDgeneral}) reduces to Eq.~(\ref{eq:inhomogeneousMD}).  The homogeneous term then represents the ``true'' unsourced initial conditions, which we assume are initially in the growing mode.  Hereafter we implicitly assume this approach of sending the initial surface to zero time.

Interestingly, the inhomogeneous solution is independent of the purely spatial gauge freedom parameterized by $f$, whereas the consistency condition depends on $f$.  As discussed in the main text and shown explicitly below, the MD consistency relation comes directly from non-Gaussianity at the initial surface and not from the sourced second-order evolution.

With $\eta_\mm \rightarrow 0$, the homogeneous part ${\cal I}^{\rm hom.}_{\delta,\text{m}}$ provides
a growing mode solution
$c_\delta$,  a time-independent function of $u$ and $v$, that depends on the initial second-order perturbation.
While this can be fully fixed with a full specification of this solution, its implications for the  three-point function at late times can be fixed by specifying the initial three-point function  as given by Eq.~(\ref{eq:MDinitial}) through ${\cal H}_\mm$.

We here summarize the expressions of the $\mathcal O(q^2)$ inhomogeneous terms of the three-point function, parameterized as
\begin{align}
	\frac{\expval{ \zeta_I(\mathbf q) \delta(\mathbf k_1,\eta_1) \delta(\mathbf k_2,\eta_2)}'}{P_{\zeta_I}(q)\Tm^2(k_S) P_{\zeta_I}(k_S)}
	&=   \sum_{s,t=0}^\infty {\mathcal B}_{\mm}^{[s,t]}(y_1,y_2) \left(\frac{q}{k_S}\right)^s \mu^t + {\cal H}_\mm(k_S,\mu,\eta_\mm) \frac{\eta_1^2\eta_2^2}{\eta_\mm^4}\,,
\end{align}
where ${\mathcal B}_{\mm}$ denotes the contribution from the inhomogeneous solution. The nonzero coefficients for $s=2$ are 
\begin{align}
	\cB_\mm^{[2,0]}&= \frac{y_1^2y_2^2(y_1^2+y_2^2) 
 }{700}\,,
 \quad \cB_\mm^{[2,2]}= \frac{y_1^2y_2^2(y_1^2+y_2^2)
}
 {1750}\,, \label{eq:b_m_20_22}
\end{align}
Notice that since the inhomogeneous contributions have no $s=0,1$ components, the validity of the consistency relation relies entirely on the initial condition.   If ${\cal H}_\mm$ satisfies the consistency condition initially, as it does when this initial non-Gaussianity arises from single-field inflation, then it is trivially preserved by matter-dominated evolution.  This holds because matter-dominated evolution has no intrinsic scale and the consistency relation in synchronous gauge is then a pure remapping  of initial scales.

\subsubsection{Radiation-Dominated Era}

The first-order solutions in the RD era take the form
\begin{align}
	\label{eq:hat_e_r}
	\hE^{(1)}(\k,\eta) &= c_{\hE}\frac{\sin{\tilde y}-{\tilde y}\, \widetilde\Ci({\tilde y})}{{\tilde y}}+d_{\hE}\frac{\cos{\tilde y}+{\tilde y}\, \Si({\tilde y})}{{\tilde y}} + e_{\hE}\log {\tilde y}+f_{\hE}\, , \\
	\Psi^{(1)}(\k,\eta) &=  c_{\Psi}\frac{\cos{\tilde y}+{\tilde y}\sin{\tilde y}-{\tilde y}^2( \widetilde\Ci({\tilde y})-\tilde y^{-2})}{{\tilde y}^2}+d_{\Psi}\frac{\sin{\tilde y}-{\tilde y}\cos{\tilde y}-{\tilde y}^2\, \Si({\tilde y})}{{\tilde y}^2} \nn
 &\quad+ e_{\Psi}({\tilde y}^{-2}-\log {\tilde y})+f_{\Psi}\, , \\
	\delta^{(1)}(\k,\eta) &= c_\delta\frac{-2+(2-{\tilde y}^2)\cos{\tilde y}+2\tilde y \sin{\tilde y}}{{\tilde y}^2}+d_\delta\frac{(2-{\tilde y}^2)\sin{\tilde y}-2\tilde y \cos{\tilde y}}{{\tilde y}^2} + e_\delta {\tilde y}^{-2}\, ,
\end{align}
where $\tilde y\equiv y/\sqrt 3$, and we have defined $\widetilde\Ci({\tilde y})\equiv \Ci({\tilde y})-\log\tilde y-\gamma_{\rm E}+1$ with the property $\widetilde\Ci(0)=1$. We have parameterized the solutions so that the $\k$-dependent coefficients $c_\O$, $d_\O$, $e_\O$, and $f_\O$ have the same meanings as in the MD case. The growing mode condition again sets $d_\O=e_\O=0$. The constraint equations together with the initial condition fix
\begin{align}
	 c_{\hE}=-\frac{1}{3}c_\Psi =-\frac{3}{2}c_\delta = -6\zeta_I^{(1)}(\k)\, , 
\end{align}
while $f_\O$ are fixed according to \eqref{f1st}. The transfer function in \eqref{delta1st} is then given by 
\begin{equation}
    \Dr(y) = -\frac{4(2-(2-\tilde y^2)\cos\tilde y - 2\tilde y\sin\tilde y)}{\tilde y^2}\,,
    \label{eq:t_dr_syn_app}
\end{equation}
which was shown in \eqref{eq:Tdrr}. This scales as $y^2$ in the $y\to 0$ limit, as expected, and reduces to $\Dr(y)\sim -4\cos\yy$ at late times. The $0i$ Einstein equation finally gives
\begin{align}
    \hat v^{(1)} &= \frac{\sqrt{3}(2\cos\yy+\yy\sin\yy-2)}{k\yy} \zeta^\fo_I(\k)\,.
\end{align}

Moving on to the second-order solution, the Green's functions for the equation \eqref{Oeq} in the RD era are given by
\begin{align}
	 G_{\hE}(k,\eta,\eta') &= \frac{\tilde y'}{4\tilde k^3}\bigg[\frac{\sin(\tilde y-\tilde y')+\tilde y'\cos(\tilde y-\tilde y')}{\tilde y}+\log(\tilde y/\tilde y ') -1\nn
	 &\hskip -60pt +(\cos\tilde y'+\tilde y'\sin\tilde y')({\rm Ci}(\tilde y')-{\rm Ci}(\tilde y))+(\sin\tilde y'-\tilde y'\cos\tilde y')({\rm Si}(\tilde y')-{\rm Si}(\tilde y))\bigg]\theta(\eta-\eta')\, ,\\
	 G_{\Psi}(k,\eta,\eta') &= \frac{-3\tilde y'}{8\tilde k^3}\bigg[\frac{(2+\tilde y\tilde y')\cos(\tilde y-\tilde y')+(2\tilde y-\tilde y')\sin(\tilde y-\tilde y')}{\tilde y^2}+2(\log(\tilde y/\tilde y ')-\tilde y^{-2}) -1 \nn
	 &\hskip -60pt  +(2\cos\tilde y'+\tilde y'\sin\tilde y')({\rm Ci}(\tilde y')-{\rm Ci}(\tilde y))+(2\sin\tilde y'-\tilde y'\cos\tilde y')({\rm Si}(\tilde y')-{\rm Si}(\tilde y))\bigg]\theta(\eta-\eta')\, ,\\
	G_\delta(k,\eta,\eta') &= \frac{2\tilde y \sin(\tilde y-\tilde y')+(2-\tilde y^2)\cos(\tilde y-\tilde y')-2+\tilde y'^2}{3(\tilde k \tilde y)^2}\theta(\eta-\eta')\, ,
\end{align}
with $\tilde k\equiv k/\sqrt 3$.

Integrating the Green's function $G_\delta$ over the source from $\eta_\rr$ to $\eta$ and then sending $\eta_\rr \rightarrow 0$ gives the inhomogeneous solution for the second-order density perturbation, which can be expressed in the squeezed limit as 
\begin{align}
	{\cal I}^{\rm inhom.}_{\text{r}}(u,v,y)= \frac{1}{2-\yy^2}\Big[
		\Dr(y) ,\, {\tilde y}\sin{\tilde y} ,\, \yy^2\Big]\cdot M_\delta \cdot \begin{bmatrix}
		 u^{-2}(1-v)^2\\u^{-2}(1-v)^3\\u^{-2}(1-v)^4\\1\\ 1-v\\ (1-v)^2 \\ u^2
	\end{bmatrix} + {\cal O}(u^3)\, ,
	\label{eq:i_r_inhom}
\end{align}
where 
\begin{align}
	M_\delta\equiv \begin{bmatrix}
	-12f(2-\yy^2) & 12f(2-\yy^2) & 0 \\
	-\frac{3}{2}f\yy^4 & 6f(2+\yy^2) & -12f  \\
	-\frac{1}{4}f(72-36\yy^2+\yy^4) & 2f(10-6\yy^2+\yy^4) & -2f  \\
	4(2-\yy^2) & -4(2-\yy^2) & 0\\
	-\frac{1}{2}(24-12\yy^2-\yy^4) +6f(2-\yy^2) & 8(1-\yy^2)-6f(2-\yy^2) & 4 \\
	-5(2-\yy^2+\frac{7}{60}\yy^4) + \frac{3}{4}f(8-4\yy^2+\yy^4) & \frac{4}{3}(11-9f-\yy^4) & \frac{2}{3}(9f-7) \\
	14-6\yy^2-\frac{31}{36}\yy^4 & -\frac{1}{9}(100-78\yy^2+\yy^4) & -\frac{26}{9} 
	\end{bmatrix}^T\, .
\end{align}
Since $\Dr(k\eta)\sim \eta^2$ at early times, this way of organizing various terms manifests that the second-order density kernel vanishes as we approach the initial time surface at $\eta_\rr\approx 0$. More precisely, the above solution scales as $\eta^4$ as $\eta\to 0$.

 Unlike the MD case, the inhomogeneous solution in RD has a nonzero contribution at $\O(q^0)$ at late times. 
The parameterization of the three-point function is given by 
\begin{equation}
	\frac{\expval{ \zeta_I(\mathbf q) \delta(\mathbf k_1,\eta_1) \delta(\mathbf k_2,\eta_2)}'}{P_{\zeta_I}(q) P_{\zeta_I}(k_S)}
	=  \sum^\infty_{s,t=0} {\mathcal B}_{\rr}^{[s,t]}(y_1,y_2) \left(\frac{q}{k_S}\right)^s \mu^t+ {\cal H}_\rr(k_S,\mu,\eta_\rr)  \frac{\Dr(y_1) \Dr(y_2)}{\Dr^2(k_S\eta_\rr)}\,,
\end{equation}
where the $s=0,1$ components were shown in \eqref{eq:cb_00_f}--\eqref{eq:cb_13_f}, and the nonzero components for $s=2$ are 
\begin{align}
	\cB_\rr^{[2,0]}
	&=
	\left(
	360+
	\frac{\dblue{71}\yy_1^4-144}{\yy_1^2-2}
	+\frac{\dblue{71}\yy_2^4-144}{\yy_2^2-2}
	\right)\frac{D_1D_2}{72}
	\nn
	&\quad
	+\bigg(
	\frac{
	35\yy_2^2+
	(\yy_2^4-33\yy_2^2-8)S_2
	}{
	9(\yy_2^2-2)
	}D_1
	+\onetwo
	\bigg)
	-4S_1S_2
	\nn
	&\quad
	+(n_s-1)
	\bigg(
	D_1D_2-\frac{D_1S_2+\onetwo}{2}
	\bigg)\, ,
	\label{eq:cb_r_inhom20}
	\\[5pt]
	\cB_\rr^{[2,2]}
	&=
	\bigg[
	1+45f
	-\bigg(
	\frac{(25+9f)\yy_1^4}{24(\yy_1^2-2)}
	+1\leftrightarrow2
	\bigg)
	\bigg]D_1D_2
	-\dblue{
	\bigg(
	12(1-3f)
	+\frac{8}{\yy_1^2-2}
	+\frac{8}{\yy_2^2-2}
	\bigg)S_1S_2
	}
	\nn
	&\quad
	+\bigg[
	\bigg(
	6-39f
	+\frac{\yy_1^4}{2(\yy_1^2-2)}
	+\frac{80+36f+5\yy_2^4}{6(\yy_2^2-2)}
	\bigg)D_1S_2
	\nn
	&\qquad
	-\frac{(25+9f)\yy_2^2}{3(\yy_2^2-2)}D_1
	+\dblue{\frac{4\yy_2^2}{\yy_2^2-2}S_1}
	+\onetwo
	\bigg]
	\nn
	&\quad
	-(n_s-1)
	\bigg(
	\dblue{(2+9f)}D_1D_2
	+8S_1S_2
	-\frac{(10+9f)D_1S_2+\onetwo}{2}
	\bigg)
	\nn
	&\quad
	+\dblue{
	\Big((n_s-1)^2+k_Sn_s'\Big)
	\bigg(
	D_1D_2-\frac{D_1S_2+\onetwo}{2}
	\bigg)
	}\, ,
	\\[5pt]
 \label{eq:cb_r_inhom24}
	\cB_\rr^{[2,4]}
	&=
	f\bigg\{
	\bigg(
	\frac{23\yy_1^4}{\yy_1^2-2}
	+\frac{\dblue{23\yy_2^4}}{\yy_2^2-2}
	-744
	\bigg)\frac{D_1D_2}{8}
	+12\bigg(
	\frac{2}{\dblue{\yy_1^2-2}}
	+\frac{2}{\dblue{\yy_2^2-2}}
	-3
	\bigg)S_1S_2
	\nn
	&\quad
	+\bigg[
	\bigg(
	120
	-\frac{3\yy_1^4}{\yy_1^2-2}
	-\frac{3\yy_2^4}{\yy_2^2-2}
	+\dblue{\frac{2(\yy_2^4-44)}{\yy_2^2-2}}
	\bigg)\frac{D_1S_2}{2}
	+\frac{23\yy_2^2}{\yy_2^2-2}D_1
	-\dblue{\frac{12\yy_2^2}{\yy_2^2-2}S_1}
	+\onetwo
	\bigg]
	\nn
	&\quad
	+24(n_s-1)(D_1-S_1)(D_2-S_2)
	-3\Big((n_s-1)^2+k_Sn_s'\Big)
	\bigg(
	D_1D_2-\frac{D_1S_2+\onetwo}{2}
	\bigg)
	\bigg\}\,,
\end{align}
where $D_i\equiv D_\rr(y_i)$, $S_i\equiv \yy_i\sin\yy_i$, $n_s$ is evaluated at $k_S$, and $n_s' = \dd n_s(k_S)/\dd k_S$.
Note that the terms proportional to $\mu^4$ vanish for the $f=0$ gauge choice.

\section{Relations in Newtonian Gauge}
\label{app:cons_rel_newtonian}

In this appendix, we focus on the three-point function in Newtonian gauge in parallel to that in synchronous gauge, discussed in the main text and Appendix~\ref{app:secondorder}.
We discuss the consistency relation up to $\mathcal O(q)$ and the second-order density perturbations up to $\mathcal O(q^2)$ in Newtonian gauge.
For the second-order perturbations, we use the results in \cite{Inomata:2020cck}, which takes the non-Gaussianity to be spatially local in uniform density gauge. 
Unlike in the main text and Appendix~\ref{app:secondorder}, we do not explicitly split the contributions into the homogeneous and the inhomogeneous ones in this appendix. 
Instead, we use the total kernel for the second-order density perturbation.
Note that, even if we separate the homogeneous and inhomogeneous contributions from the total kernel in Newtonian gauge, this does not correspond to the homogeneous and the inhomogeneous contributions in synchronous gauge.
This is because there is no time-slicing invariant way to split the homogeneous and the inhomogeneous contributions.
Since our synchronous condition sets its  time slicing to be the same as comoving and uniform density gauge as $\eta\rightarrow 0$, that split has a well-defined meaning  for local non-Gaussianity (see Appendix~\ref{app:summary_of_gauge_trans}).

We take the following metric notation in Newtonian gauge:
\begin{align}
	\dd s^2 = a^2(\eta)\Big[{-}(1+2\Phi)\dd\eta^2 +(1-2\Psi) \delta_{ij}\dd x^i\dd x^j\Big]\, ,
	\label{eq:metric_def_new}
\end{align}
with $\Phi = \Phi^\fo + \frac{1}{2} \Phi^\so + \cdots$ and $\Psi = \Psi^\fo + \frac{1}{2} \Psi^\so + \cdots$, where the dots mean higher order in perturbation theory.

\subsection{Dilation and Newtonian Consistency Relation}

Unlike in synchronous gauge, the consistency relation in Newtonian gauge requires not only a time-independent or initial spatial coordinate transformation but also time-dependent space and time transformation, complicating both its form and an interpretation~\cite{Weinberg:2008zzc, Creminelli:2013mca, Horn:2014rta}. 

Requiring the coordinate transformation to maintain the gauge condition $g_{0i} =0$ and $g_{ij}|_{i\neq j} =0$, we obtain:
\begin{align}
		\label{eq:eta_trans_new}
		\tilde \eta &= \eta + \epsilon(\eta) + \mathbf x \cdot \mathbf \xi(\eta)\,, \\
		\label{eq:x_trans_new}		
		\tilde x^i & = x^i(1+ \lambda + 2 \mathbf b \cdot \mathbf x) -  x^2 b^i + \int^\eta \xi^i\, \dd \eta\,.
\end{align}
Under this coordinate transformation, $\Phi$ and $\Psi$ in Eq.~(\ref{eq:metric_def_new}) transforms as 
\begin{align}
	\tilde \Phi &= \Phi - \epsilon' - \mathbf x \cdot \mathbf \xi' - \mathcal H (\epsilon + \mathbf x \cdot \mathbf \xi)\,, \\
	\tilde \Psi &= \Psi + \lambda + 2 \mathbf x \cdot \mathbf b + \mathcal H (\epsilon + \mathbf x \cdot \mathbf \xi)\,.
\end{align}
By choosing $\tilde \Phi = \tilde \Psi =0$, the new coordinate absorbs the long-wavelength modes of $\Phi$ and $\Psi$ up to $\mathcal O(q)$ if
\begin{align}
	\label{eq:phi_l_new_gauge}
	\Phi &= \epsilon' + \mathbf x \cdot \mathbf \xi' + \mathcal H (\epsilon + \mathbf x \cdot \mathbf \xi)\,, \\
	\label{eq:psi_l_new_gauge}	
	\Psi &= - \lambda - 2 \mathbf x \cdot \mathbf b - \mathcal H (\epsilon + \mathbf x \cdot \mathbf \xi)\,.
\end{align}
However, for the new coordinates to absorb the physical mode, the parameters $\lambda$, $\epsilon$ and $\xi$ must satisfy the adiabatic condition~\cite{Weinberg:2003sw}, which requires the long-wavelength mode to satisfy the Einstein equations with a finite wavenumber $q$.  These conditions can be expressed as
\begin{align}
	(\mathcal H' - \mathcal H^2) \hat v = (\Psi' + \mathcal H \Phi)\,, 
	\label{eq:first_adi_cond}
\end{align}
where recall that $\hat v$ is the velocity potential  
and 
\begin{align}
	\Phi = \Psi - 8\pi G \sigma\,,
	\label{eq:second_adi_cond}	
\end{align}
where $\sigma$ is the anisotropic stress.
Substituting Eqs.~(\ref{eq:phi_l_new_gauge}) and (\ref{eq:psi_l_new_gauge}) into Eq.~(\ref{eq:first_adi_cond}), we obtain 
\begin{align}
	\hat v = - (\epsilon + \mathbf x \cdot \mathbf \xi)\,.
\end{align}
Also, the long-wavelength comoving curvature ($\zeta = -\Psi + \mathcal H \hat v$) can be expressed as 
\begin{align}
	 \zeta_I = \lambda + 2 \mathbf x \cdot \mathbf b\,.
\end{align}
From this, we can see $\lambda = \zeta_{I}(\mathbf x=\bz)$ and $2b^i = \partial_i \zeta_{I}(\mathbf x=\bz)$.
Using Eq.~(\ref{eq:second_adi_cond}), we obtain 
\begin{align}
	\label{eq:ep_eom}
	\epsilon' + 2 \mathcal H \epsilon  &= - \lambda - 8 \pi G \sigma\, , \\
	\label{eq:xi_eom}	
	{\xi^i}' + 2 \mathcal H \xi^i  &= - 2 b^i - 8\pi G \partial_i  \sigma\,.
\end{align}
Under these conditions, the coordinate change in Eqs.~(\ref{eq:eta_trans_new}) and (\ref{eq:x_trans_new}) can erase the physical long-wavelength mode up to $\mathcal O(q)$.

In the absence of anisotropic stress ($\sigma = 0$), we can solve for $\epsilon$ and $\xi^i$ in terms of the growth function of $\hat v$, which satisfies
\begin{align}
		V'' + 2 \mathcal H V' - 1 = 0\,.
\end{align}
Namely, Eqs.~(\ref{eq:ep_eom}) and (\ref{eq:xi_eom}) lead to $\epsilon = -V'(\eta) \zeta_{I}$ and $\xi^i = -V'(\eta) \partial_i \zeta_{I}$.
Then following~\cite{Creminelli:2013mca, Horn:2014rta}, the consistency relation in Newtonian gauge under the perfect fluid condition is given by
\begin{align}
	\lim_{\mathbf q \rightarrow \bz} \frac{\expval{\zeta_I(\mathbf q) \delta({\mathbf k_1},\eta_1) \cdots \delta({\mathbf k_N},\eta_N)}'}{ P_{\zeta_I}(q)} =& -( \cD_{\rm N} + q^i \cK^i_{\rm N})\expval{\delta({\mathbf k_1},\eta_1) \cdots \delta({\mathbf k_N},\eta_N)}'\,,
	\label{eq:dilation_cons_rel_new_final}
\end{align}
where\footnote{Note that the last term in Eq.~(\ref{eq:sct_new}) is omitted in \cite{Horn:2014rta} (their Eq.~(92)).}
\begin{align}
	\cD_{\rm N}  &= \cD+ \sum_{a=1}^{N} V'(\eta_a)\! \left( \partial_{\eta_a} +\frac{\bar\rho'(\eta_a)}{\bar \rho(\eta_a)}\right) , \\
	\cK_{\rm N}^i  &= \cK^i + \sum_{a=1}^N\! \left[  V'(\eta_a)\! \left( \partial_{\eta_a} + \frac{\bar \rho'(\eta_a)}{\bar \rho(\eta_a)}\right)\! \partial_{k^i_a} + V(\eta_a) k^i_a 
	+4 V'(\eta_a)\frac{\bar \rho'(\eta_a)}{\bar \rho(\eta_a)}
	    \frac{K_{\Phi}(k_a\eta_a)}{ K_{\delta}(k_a\eta_a)  }\frac{k_a^i}{k_a^2}
	 \right],
	 \label{eq:sct_new}
\end{align}
and recall that $\cD$ and $\cK^i$ were defined in \eqref{Dop} and \eqref{SCTop}. 
Throughout, we put the subscript ``N'' when we would like to emphasize that the Newtonian gauge expressions are different from the synchronous gauge expressions. 
Although $V$ has an arbitrary additive, temporally-constant term, it does not appear in the final results because of the momentum conservation $\sum_a k^i_a = 0$. 
We here use $K$ instead of $D$ for the growth function to clarify that it is for Newtonian gauge, which are defined by
\begin{align}	
	\delta^\fo (\mathbf k, \eta) &=  \, K_{\delta,\mm/\rr}(x) \Tmr(k) \zeta^\fo_I(\mathbf k) \,, \\
	\Phi^\fo (\mathbf k, \eta) &= \,K_{\Phi,\mm/\rr}(x) \Tmr(k)\zeta^\fo_I(\mathbf k)\,.
\end{align}
Under the perfect fluid condition $\Phi^\fo = \Psi^\fo$, the growth functions during the MD and RD eras become~\cite{Inomata:2020cck} 
\begin{align}
	K_{\delta,\mm}(y) &= \frac{6}{5} + \frac{y^2}{10\,}\,,\quad
	\hskip-20pt &&K_{\delta,\rr}(y) = \frac{4 ( \tilde y(2 - \tilde y^2) \cos \tilde y - 2 (1 - \tilde y^2) \sin \tilde y )}{\tilde y^3}\,, \label{eq:trans_delta_mm} 
 \\
	K_{\Phi,\mm}(y) &= -\frac{3}{5} \,,\quad  \hskip-20pt &&K_{\Phi,\rr}(y) = \frac{2 (\tilde y \cos \tilde y-\sin \tilde y)}{\tilde y^3}\,.\label{eq:trans_phi_mm}
\end{align}
Given $V'(\eta) = \eta/5$ during the MD era, $\cD_{\rm N}$ and $\cK_{\rm N}$ become 
\begin{align}
	\cD_{\rm N,\mm} &= \cD + \sum^N_{a=1}  \frac{\eta_a}{5} \left(  \partial_{\eta_a}-\frac{6}{\eta_a}  \right) , \\
	\cK^i_{\rm N,\mm} &=\cK^i +  \sum_{a=1}^N \left[  \frac{\eta_a}{5} \left(  \partial_{\eta_a} -\frac{6}{\eta_a}\right) \partial_{k^i_a} + \frac{\eta_a^2}{10} k^i_a 
	- \frac{24}{5}
	  \frac{k_a^i}{k_a^2}  \frac{K_{\Phi,\mm}(k_a\eta_a)}{ K_{\delta,\mm}(k_a\eta_a)  }
	 \right],
\end{align}
while during the RD era we have $V'(\eta) = \eta/3$ and 
\begin{align}
	\cD_{\rm N,\rr} &=  \cD + \sum^N_{a=1} \frac{\eta_a}{3} \left( \partial_{\eta_a}-\frac{4}{\eta_a}  \right) , \\
	\cK^i_{\rm N,\rr} &= \cK^i + \sum_{a=1}^N \left[  \frac{\eta_a}{3} \left( \partial_{\eta_a}  -\frac{4}{\eta_a} \right) \partial_{k^i_a} + \frac{\eta_a^2}{6} k^i_a 
	- \frac{16}{3}
	  \frac{k_a^i}{k_a^2}  \frac{K_{\Phi,\rr}(k_a\eta_a)}{ K_{\delta,\rr}(k_a\eta_a)  }
	 \right].
\end{align}
For the MD era, the squeezed three-point function then takes the form
\begin{align}
    &\lim_{\mathbf q \rightarrow \bz} \frac{\expval{ \zeta_I(\mathbf q) \delta(\mathbf k_1,\eta_1) \delta(\mathbf k_2,\eta_2)}'}{P_{\zeta_I}(q)} 
    = -(\cD_{\rm N,\mm} + q^i \cK_{\rm N,\mm}^i )	K_{\delta,\mm}(k_S \eta_1) K_{\delta,\mm}(k_S \eta_2 )T_\mm^2(k_S) P_{\zeta_I}( k_S) \,,\nn[3pt]
    & \qquad= 
	\bigg[\frac{3(144 - y_1^2 y_2^2)}{125} - \frac{(12 + y_1^2) (12 + y_2^2) (\tilde n_s-1) }{100} \label{eq:3pt_new_q01}\\
	&\qquad\qquad	+  \frac{q\mu}{k_S} \frac{y_1^2 - y_2^2}{1000} (168 - 12(y_1^2 + y_2^2) - y_1^2 y_2^2-24(\tilde n_s-1)) \bigg]\Tm^2(k_S) P_{\zeta_I}(k_S) \,,\nonumber
\end{align}
with
\begin{equation}\tilde n_s(k)-1 \equiv \frac{\dd \ln k^3 \Tm^2(k)P_{\zeta_I}(k)}{\dd \ln k}=
n_s -1 + 2 \frac{\dd \ln \Tm}{\dd\ln k}\,.
\label{eq:nstns}
\end{equation}
Here we follow the convention that $n_s$ denotes the tilt of the initial curvature power spectrum $P_{\zeta_I}$ and $\tilde n_s(k)$ denotes its tilt in the MD era after evolution through the RD era.   The correspondence to the tilt of the synchronous gauge density power spectrum in the main text is $\tilde n_s(k) = n_\delta(k) -3$
for $\eta > \eta_\mm$.

The ${\cal O}(q)$ piece of
Eq.~(\ref{eq:3pt_new_q01}) is the so-called Newtonian consistency relation during MD. 
We note that this equation is exact rather than leading order in $y_{1,2}\gg 1$ as in Ref.~\cite{Horn:2014rta} due to our inclusion of the last term in Eq.~(\ref{eq:sct_new}), and corrects typographical errors in  Ref.~\cite{Creminelli:2013mca} (their Eq.~(62)).  It also generalizes the results in both \cite{Horn:2014rta,Creminelli:2013mca} to scales below the horizon at matter-radiation equality by including $T_\mm$ and the modified tilt $\tilde n_s$.

We remark that $\mathcal O(q)$ terms are nonzero when $\eta_1 \neq \eta_2$, while those in synchronous gauge are always zero.  
This highlights one of the advantages of synchronous gauge. 
In synchronous gauge, the time coordinate is always the proper time of  free-falling observers at each spatial point.
This leaves only the initial  time-independent spatial coordinate choice to distinguish local and synchronous coordinates, even when comparing perturbations at different times. 
On the other hand, in Newtonian gauge the consistency relation is associated with the time-dependent coordinate transformation, which leads to time-dependent operators in the consistency relation. 
These time-dependent operators give nonzero $\mathcal O(q)$ terms for $\eta_1 \neq \eta_2$ even in the $k_S$ reference convention.

Similarly, the $\mathcal O(q^0)$ and $\mathcal O(q)$ terms during the RD era are given by 
\begin{align}
  \lim_{\mathbf q \rightarrow \bz} &\frac{\expval{ \zeta_I(\mathbf q) \delta(\mathbf k_1,\eta_1) \delta(\mathbf k_2,\eta_2)}'}{P_{\zeta_I}(q)}=-(\cD_{\rm N,\rr} + q^i \cK_{\rm N,\rr}^i )	K_{\delta,\rr}(k_S \eta_1) K_{\delta,\rr}(k_S \eta_2 )P_{\zeta_I}( k_S) \nn
  &=\bigg\{\bigg(1-n_s+\frac{8(16-7(\yy_1^2+\yy_2^2)+3\yy_1^2\yy_2^2)}{3(2-\yy_1^2)(2-\yy_2^2)}\bigg)K_{\delta,\rr}(y_1)K_{\delta,\rr}(y_2) \nn
  &\qquad- \bigg(\frac{16(2+2\yy_1^2-\yy_1^4)}{3\yy_1(2-\yy_1^2)}\sin\yy_1 K_{\delta,\rr}(y_2) + 1\leftrightarrow2\bigg)\nn[3pt]
    &\qquad+\frac{q\mu}{k_S}\bigg[\frac{(\yy_1^2-\yy_2^2)(n_s-1-\yy_1^2\yy_2^2+2(\yy_1^2+\yy_2^2)-6)}{3(2-\yy_1^2)(2-\yy_2^2)}K_{\delta,\rr}(y_1)K_{\delta,\rr}(y_2) \label{eq:new_cons_rd}\\
     &\ \qquad\qquad+\frac{2}{3}\bigg(\frac{(n_s-3)\yy_1^4-2(n_s-5)\yy_1^2-2(n_s+1)}{\yy_1(2-\yy_1^2)}\sin\yy_1 K_{\delta,\rr}(y_2) - 1\leftrightarrow2\bigg)\bigg]\bigg\}P_{\zeta_I}(k_S)\,,\nonumber
\end{align}
where recall $\tilde y = y/\sqrt{3}$. 
Here we have again expressed the results in terms of the tilt of the initial curvature spectrum, which is related to the tilt of the synchronous gauge density power spectrum at $\eta_\rr$ in the main text as $n_s = n_\delta(k,\eta_\rr,\eta_\rr)-3$. 
Using the above expressions, we can express the three-point function in the large $\tilde y_1$ and $\tilde y_2$ limit up to $\mathcal O(q)$ as 
\begin{align}
	\lim_{\mathbf q \rightarrow \bz,\, y_{1,2}\gg 1} \frac{\expval{ \zeta_I(\mathbf q) \delta(\mathbf k_1,\eta_1) \delta(\mathbf k_2,\eta_2)}'}{P_{\zeta_I}(q) P_{\zeta_I}(k_S)}
	&= \frac{64(\yy_2 \cos \yy_1 \sin \yy_2 + \yy_1 \sin \yy_1 \cos \yy_2)}{3}   \nonumber \\
	&\quad\dblue{-\frac{q\mu}{k_S} \frac{16(\yy_1^2 - \yy_2^2) \cos \yy_1 \cos \yy_2}{3}}    + \mathcal O(q^2)\,.
	\label{eq:3pt_new_q01_rd}
\end{align}
As with the MD epoch, this Newtonian consistency relation in the RD era has an unequal-time contribution at ${\cal O}(q)$.

\subsection{Three-Point Function from Local Non-Gaussianity}
\label{app:NewtonianlocalNG}

Next, we show the expressions of the second-order density perturbations in Newtonian gauge. 
Unlike in Appendix~\ref{app:secondorder}, we take a specific choice of initial non-Gaussianity in this appendix, which amounts to specifying the homogeneous term of second-order density perturbations (see Appendix~\ref{app:new_to_syn}).
In particular, we follow \cite{Inomata:2020cck} in choosing  the local-type non-Gaussianity where  the second-order curvature in uniform density gauge $\zeta_\delta^\so$ at $\eta \rightarrow 0$ is parameterized as~\cite{Bartolo:2006fj}
\begin{align}
	\zeta_\delta^\so({\bf x}) = 2(b_\NL + 1) [\zeta_\delta^\fo({\bf x})]^2\,.
	\label{eq:bnl}
\end{align}
Note that $b_\NL=0$ corresponds to the single-field, slow-roll, inflationary prediction.
Strictly speaking, for this to apply in the MD epoch given an initial local non-Gaussianity from inflation, all modes should be well outside the horizon at matter-radiation equality.  We shall see that applying this ansatz directly in the MD era on smaller scales, without evolving second-order perturbations through the RD era, will miss the full impact of dilation on scales where $T_{\rm m}(k) \ne 1$ and technically violate the consistency relation for any $b_\NL$.

We express the second-order perturbation as 
\begin{align}
 \delta^\so_\New (\mathbf k,\eta)=\int \frac{\dd^3 p_1\dd^3 p_2}{(2\pi)^3}\delta_D(\k-\p_1-\p_2) \cJ_{\mm/\rr}(u,v,y) T_{\mm/\rr}(p_1)T_{\mm/\rr}(p_2)\zeta^{(1)}_I(\p_1)\zeta^{(1)}_I(\p_2)\,,\label{eq:delta_so_final_exp_newton}
\end{align}
where we have used the symbol $\mathcal J$ for the kernel, as opposed to $\mathcal I$ in synchronous gauge, to clarify the difference of the gauge. 
This leads to the three-point function
\begin{equation}
\hskip-8pt	\frac{\expval{ \zeta_I(\mathbf q) \delta(\mathbf k_1,\eta_1) \delta(\mathbf k_2,\eta_2)}'}{P_{\zeta_I}(q)} 
	= K_{\delta, \mm/\rr}(k_2\eta_2) \cJ_{\mm/\rr}\!\left(\frac{q}{k_1},\frac{k_2}{k_1},k_1 \eta_1\right)\!\Tmr^2(k_2)  P_{\zeta_I}(k_2)  + 1 \leftrightarrow 2\,,
	\label{eq:integrated_form_new}
\end{equation}
with kernels given below (see Eqs.~(\ref{eq:calJ_m}) and (\ref{eq:calJ_r})). 
In general we can again expand this expression in the squeezed limit in powers of $q$ as 
\begin{align}
	\lim_{\mathbf q \rightarrow \bz}\frac{\expval{ \zeta_I(\mathbf q) \delta(\mathbf k_1,\eta_1) \delta(\mathbf k_2,\eta_2)}'}{P_{\zeta_I}(q)\Tmr^2(k_S) P_{\zeta_I}(k_S)} 
	= \sum_{s,t}\tilde {\mathcal C}^{[s,t]}_{\mm/\rr}(y_1,y_2)\left(\frac{q}{k_S}\right)^s\mu^t \,,
	\label{eq:3pt_func_so_cal_new}
\end{align}
where we used the notation $\tilde{\cal C}$ to parameterize the total squeezed bispectrum for local non-Gaussianity in Newtonian gauge, as opposed to ${\cal B}$ used in synchronous gauge, which just includes the inhomogeneous part but applies to any type of initial non-Gaussianity.
In Appendix \ref{app:new_to_syn}, we will also see that the results in this appendix once gauge transformed to synchronous gauge  are consistent with those in Appendix~\ref{app:secondorder} for the homogeneous contributions of local non-Gaussianity.

\subsubsection{Matter Era}
\label{sec:NewtonianMD}

In the MD era, the second-order Newtonian-gauge kernel $\mathcal J$ is given by~\cite{Bartolo:2005xa,Inomata:2020cck}\footnote{The second-order kernel $\mathcal J$ in this paper corresponds to $A^2 uv I$ in \cite{Inomata:2020cck}, where $A = 3/5$ for the MD era and $A = 2/3$ for the RD era.} 
\begin{align}
  \cJ_\mm (u,v,y) &=  \frac{3(15 + 20 b_\NL + 6 (u^2 + v^2) - 9(u^4 + v^4) + 18 u^2 v^2)}{25}  \nonumber \\
	&\quad 
	+\frac{ 59 + 140 b_\NL - 125(u^2 + v^2) - 18(u^4 + v^4) +36 u^2 v^2}{700}y^2   \nonumber \\
	& 
	\quad + \frac{2 +3 (u^2 + v^2) - 5(u^4 + v^4) + 10 u^2 v^2}{1400} y^4 \,
	 .
	 \label{eq:calJ_m}
\end{align}
Note that this form ignores any previous epoch of radiation domination by sending $\eta_\mm \rightarrow 0$, so that all modes start above the horizon with local non-Gaussianity from Eq.~(\ref{eq:bnl}).  We shall return to this point  below when discussing the apparent violation of the consistency relation.
In the squeezed limit with $u \ll 1$ and $|v-1| \simeq \mathcal O(u)$, $\mathcal J_\mm$ can be expressed as
\begin{align}
    \cJ_\mm(u,v,y) &=  \frac{5 b_\NL (y^2+12)-3 y^2+36}{25} -\frac{(y^2+46) y^2+288}{100}(v-1) \label{eq:curlj_m} \\
   &\quad -\frac{\left(27y^4+466 y^2+8064\right)}{1400}(v-1)^2 + \frac{\left(13 y^4-178 y^2+4032\right)}{1400} u^2 + \mathcal O(u^3)\,.\nonumber 
\end{align}
The nonzero coefficients in \eqref{eq:3pt_func_so_cal_new} up to $\O(q)$ are then given by
\begin{align}
	\tilde {\mathcal C}^{[0,0]}_{\mm}(y_1,y_2)&= \frac{3(144-y_1^2 y_2^2)}{125} + \frac{b_\NL (12+y_1^2)(12+y_2^2)}{25} \,, \\
	\tilde {\mathcal C}^{[1,1]}_{\mm}(y_1,y_2)&= -\frac{y_1^2-y_2^2}{1000} \left(480 b_\NL+ 12 (y_1^2+y_2^2) -168 + 144 (\tilde n_s - 1) + y_1^2y_2^2\right).
\end{align}
Using the above expressions, we  can reassemble the three-point function up to $\O(q)$ as 
\begin{align}
	&\lim_{\mathbf q \rightarrow \bz} \frac{ \expval{ \zeta_I(\mathbf q) \delta(\mathbf k_1,\eta_1) \delta(\mathbf k_2,\eta_2)}'}{P_{\zeta_I}(q)\Tm^2(k_S) P_{\zeta_I}(k_S)}
	= 
	\frac{3(144-y_1^2 y_2^2)}{125} + \frac{b_\NL (12+y_1^2)(12+y_2^2)}{25} \label{eq:3pt_new_q01bNL} \\
	&\quad\qquad	-  \frac{q\mu}{k_S} \frac{y_1^2-y_2^2}{1000} \left(480 b_\NL+ 12 (y_1^2+y_2^2) -168 + 144 (\tilde n_s - 1) + y_1^2y_2^2\right)+ \mathcal O(q^2)\,.
	\nonumber
\end{align}
In general, satisfying the consistency relation (\ref{eq:3pt_new_q01}) would require
 $b_\NL \rightarrow (1- \tilde n_s(k_S))/4$ whereas $b_\NL=$\,const.  If we assume local non-Gaussianity is the only source of non-Gaussianity in the MD era as we do here, then the consistency relation is violated for any $b_\NL$ and any scales with $T_\mm(k_S) \ne 1$, i.e.,~those that approach or are within the horizon at matter-radiation equality.  As discussed in  the main text for synchronous gauge, this is because the inhomogeneous contributions to the
second-order density perturbation during the RD era are responsible for carrying the dilation of the Jeans or acoustic scale.  We return to how consistency is restored by radiation-dominated evolution in Newtonian gauge below (see discussion following Eq.~(\ref{eq:tC01})).

We now derive the analogue of the separate universe constraints in Newtonian gauge.
Expanding the three-point function up to $\O(q^2)$ gives
\begin{align}
	\label{eq:c_m_20}
	\tilde {\mathcal C}^{[2,0]}_{\mm}(y_1,y_2) =&\ 
	\frac{13 y_1^2 y_2^2 (y_1^2 + y_2^2)+156(y_1^4+y_2^4)-398y_1^2y_2^2+1896(y_1^2+y_2^2)+78624}{14000} 
 \nonumber \\
   &+ \frac{b_\NL ( y_1^2 y_2^2- 12 (y_1^2+y_2^2) - 432 )}{200} \nonumber  \\
   & + (\tilde n_s - 1)\frac{432-3y_1^2y_2^2+5b_\NL (12+y_1^2)(12+y_2^2)}{1000}\,, \\
	\tilde{\mathcal C}^{[2,2]}_{\mm}(y_1,y_2) =&\  
	 \frac{y_1^2 y_2^2(y_1^2 + y_2^2)}{14000}+\frac{31 y_1^2 y_2^2}{7000}+\frac{9 (y_1^4+y_2^4)}{700}-\frac{531 (y_1^2+y_2^2)}{1750}-\frac{54}{125} \nonumber \\
	 &+ b_\NL \left(\frac{7}{200} y_1^2 y_2^2+\frac{9 (y_1^2+y_2^2)}{10} +\frac{54}{5} \right) -(\tilde n_s -1) \bigg(\frac{ b_\NL}{25}(12+y_1^2)(12+y_2^2)\nonumber \\
   & +\frac{y_1^2 y_2^2(y_1^2 + y_2^2)+12(y_1^4+y_2^4)+44y_1^2y_2^2+552(y_1^2+y_2^2)+13824}{2000}\bigg)\nonumber \\
   &+ \left( (\tilde n_s -1)^2 + k_S \tilde n_s'\right) \frac{432-3y_1^2y_2^2+5b_\NL (12+y_1^2)(12+y_2^2)}{1000}\,.
\end{align}
Taking the angle average in the limit of $y_1 \gg 1$ and $y_2 \gg 1$, we obtain 
\begin{equation}
	\lim_{\substack{\mathbf q \rightarrow \bz,\, y_{1,2}\gg 1}}\frac{\overline{\expval{ \zeta_I(\mathbf q) \delta(\mathbf k_1,\eta_1) \delta(\mathbf k_2,\eta_2)}'}}{P_{\zeta_I}(q) \Tm^2(k_S) P_{\zeta_I}(k_S)}\bigg|_{\mathcal O(q^2)}
	\! = \left(\frac{34}{21} - \frac{n_\delta(k_S,\eta_1,\eta_2)}{6}\right)  \frac{y_1^2 y_2^2(y_1^2 + y_2^2)}{1000}\! \left( \frac{q}{k_S} \right)^2 ,
	\label{eq:3pt_new_q2}
\end{equation}
where we have used $n_\delta (k,\eta_1,\eta_2) = \tilde n_s(k) + 3 + \O(1/y^2_{1,2})$ in Newtonian gauge.
Unlike in synchronous gauge, the growth function in Newtonian gauge depends on the horizon scale and therefore the tilt of the density power spectrum depends on $\eta_1$ and $\eta_2$, though its time-dependent parts are subdominant in $y_{1,2} \gg 1$.
In the case of a subhorizon long-wavelength mode $q\eta \gg 1$, we can rewrite this expression as 
\begin{equation}
\lim_{\q \rightarrow \bz,\,y_{1,2}\gg 1,\, q\eta \gg 1}\frac{\overline{\expval{ \delta(\q,\eta) \delta(\k_1,\eta_1) \delta(\k_2,\eta_2)}'}}{P_\delta(q,\eta,\eta) P_\delta(k_S,\eta_1,\eta_2)}\bigg|_{\O(q^2)}	
= \left( \frac{34}{21} - \frac{n_\delta(k_S,\eta_1,\eta_2)}{6} \right) \frac{\eta_1^2 + \eta_2^2}{\eta^2}  .
\label{eq:lhs_rd_qsq_new}
\end{equation}
Note that, except for the term proportional to $n_\delta$, this expression is the same as Eq.~(\ref{eq:lhs_rd_qsq}) for synchronous gauge.
The $n_\delta$ term denotes the effect of the dilation, which comes from the difference between the spatial coordinates in Newtonian and synchronous gauges, whereas the time slicing shift causes only a small effect in the MD era due to the subhorizon growth of density fluctuations.  
The $n_\delta$ term comes from the combination of the $\mathcal O(y^4 (v-1))$ term in Eq.~(\ref{eq:curlj_m}) and the expansion of $K_{\delta,\mm}(k \eta)\Tm^2(k)P_{\zeta_I}(k)$ with respect to $k_S$.
We will concretely see that the $\mathcal O(y^4(v-1))$ term is cancelled by the spatial gauge shift from Newtonian to synchronous gauge in Appendix~\ref{app:new_to_syn}.

\subsubsection{Radiation Era}
\label{app:NewtonianRD}

The homogeneous and inhomogeneous kernels during the RD era in the squeezed limit ($u ,|v-1| \ll 1$) are given by 
\begin{align}
	\mathcal J_\rr (u,v,y) & = -\frac{16 (\tilde y^4-3 (b_\NL+2) \tilde y^2+3 b_\NL+8) \sin \tilde y+8 \tilde y (3 (b_\NL+2)
   \tilde y^2-2 (3 b_\NL+8)) \cos \tilde y}{3 \tilde y^3}\nonumber \\
   & \quad
   -\frac{4 \left((\tilde y^4+20 \tilde y^2-34) \sin
   \tilde y+\tilde y \left(\tilde y^4-6 \tilde y^2+34\right) \cos \tilde y\right)}{3 \tilde y^3}(v-1) \nonumber \\
   &\quad
   -\frac{4 \left((-4 \tilde y^4+3 \tilde y^2+34) \sin \tilde y+\tilde y (2 \tilde y^4-9
   \tilde y^2-34) \cos \tilde y\right)}{3 \tilde y^3}(v-1)^2\label{eq:calJ_r} \\
   &\quad
   +\frac{4 (2 \tilde y^6-49 \tilde y^4+333
   \tilde y^2-300) \sin \tilde y-4 \tilde y (23 \tilde y^4+193 \tilde y^2-300) \cos \tilde y}{45
   \tilde y^3}u^2 + \O(u^3)\,,\nonumber
   \end{align}
where note again $\tilde y = y/\sqrt{3}$.
See \cite{Inomata:2020cck} for the full expression of $\mathcal J_{\rr}$. The nonzero coefficients in \eqref{eq:3pt_func_so_cal_new} up to $\O(q)$ are then given by
\begin{align}
    \tilde\cC^{[0,0]}_\rr(y_1,y_2) &=\left(4b_\NL+ \frac{8(3\yy_1^2\yy_2^2-7(\yy_1^2+\yy_2^2)+16)}{3(2-\yy_1^2)(2-\yy_2^2)}\right)K_{\delta,\rr}(y_1)K_{\delta,\rr}(y_2)\,,\nn
    &\quad +\left(\frac{16(\yy_1^4-2\yy_1^2-2)}{3\yy_1(2-\yy_1^2)}\sin\yy_1 K_{\delta,\rr}(y_2) +1\leftrightarrow2 \right), \label{eq:tC00}\\
    \tilde\cC^{[1,1]}_\rr(y_1,y_2) &=\frac{(\yy_1^2-\yy_2^2)(4(b_\NL+\frac{n_s-1}{3})-\frac{1}{3}(\yy_1^2\yy_2^2-2(\yy_1^2+\yy_2^2)+6))}{(2-\yy_1^2)(2-\yy_2^2)}K_{\delta,\rr}(y_1)K_{\delta,\rr}(y_2)\nn
    &\hskip -35pt+\left(\frac{8(b_\NL+\frac{n_s-1}{3})(\yy_1^4-2\yy_1^2-2)-\frac{4}{3}(\yy_1^4-4\yy_1^2+2)}{\yy_1(2-\yy_1^2)} \sin\yy_1 K_{\delta,\rr}(y_2) - 1\leftrightarrow2\right).\label{eq:tC01}
\end{align}
 Notice that unlike specifying local non-Gaussianity directly in MD, here there is a self-consistent choice that verifies the 
 consistency relation: $b_{\rm NL}= (1- n_s)/4$.  
 Substituting this into Eqs.~(\ref{eq:tC00}) and (\ref{eq:tC01}) reproduces Eq.~(\ref{eq:new_cons_rd}).
 This initial consistency condition is then dynamically preserved for any later $\eta_1,\eta_2$ in the RD era.  For this value of the initial local non-Gaussianity, evolution through the RD era then restores consistency in the MD era as well by supplying the missing conversion of $b_{\rm NL} \rightarrow (1-\tilde n_s(k_S))/4$ that includes the dilation of $T_{\mm}(k_S)$ discussed below Eq.~(\ref{eq:3pt_new_q01bNL}).

Using the above expressions, we can express the three-point function in the large $y_1$ and $y_2$ limit up to $\mathcal O(q)$ as 
\begin{align}
	\lim_{\mathbf q \rightarrow \bz,\, y_{1,2}\gg 1} \frac{\expval{ \zeta_I(\mathbf q) \delta(\mathbf k_1,\eta_1) \delta(\mathbf k_2,\eta_2)}'}{P_{\zeta_I}(q) P_{\zeta_I}(k_S)}
	&= \frac{64(\yy_2 \cos \yy_1 \sin \yy_2 + \yy_1 \sin \yy_1 \cos \yy_2)}{3}   \nonumber \\
	&\quad\dblue{-  \frac{q\mu}{k_S} \frac{16(\yy_1^2 - \yy_2^2) \cos \yy_1 \cos \yy_2}{3}}     + \mathcal O(q^2)\,,
	\label{eq:3pt_new_q01_rd_2}
\end{align}
which is the same as Eq.~(\ref{eq:3pt_new_q01_rd}).  Notice that the initial consistency relation plays no role in this $y_{1,2} \gg 1$ limit.
As discussed at the end of \S\ref{subsec:3pt_cr}, in the RD era the effect of dilation on the scale of the sinusoidal acoustic oscillations dominates over that of the initial power spectrum, though both pieces are explicitly verified if the consistency condition holds initially using the exact coefficients above.

Expanding the coefficients at $\mathcal O(q^2)$ as
\begin{align}
	\tilde\cC^{[2,t]}_{\rr}(y_1, y_2) &= \sum^\infty_{\ell} \tilde{\mathcal C}^{[2,t]}_{\rr,\ell}(y_1, y_2) (n_s-1)^\ell\,,
\end{align}
the nonzero components for $\ell=0$ are
\dblue{
\begin{align}
	\tilde\cC^{[2,0]}_{\rr,0} (y_1,y_2) &=\bigg(\frac{(4(77+180 b_\NL) \yy_1^2- 122 \yy_1^4 + 61 \yy_1^4 \yy_2^2) + 1\leftrightarrow 2) + (142 - 315 b_\NL)\yy_1^2 \yy_2^2 }{90 (2-\yy_1^2)(2-\yy_2^2)}\nn
	&\quad-\frac{2(32 + 27 b_\NL)}{3 (2-\yy_1^2)(2-\yy_2^2)}\bigg) K_1K_2+\bigg[\bigg(-\frac{2 b_\NL(\yy_1^4-2\yy_1^2-2)}{\yy_1(2-\yy_1^2)}\nn
 &\quad- 2\frac{-760 - 356 \yy_1^2 - 96 \yy_1^4 + 8 \yy_1^6 + 2(160 + 59 \yy_1^2 + 39 \yy_1^4 - 2 \yy_1^6)\yy_2^2}{45 \yy_1(2-\yy_1^2)(2-\yy_2^2)}\bigg) \tilde S_1K_2+1\leftrightarrow 2\bigg]\nn
 &\quad-\frac{16(\yy_1^4-2\yy_1^2-2)(\yy_2^4-2\yy_2^2-2)}{3\yy_1\yy_2(2-\yy_1^2)(2-\yy_2^2)}\tilde S_1\tilde S_2\,,\\[10pt]
		\tilde\cC^{[2,2]}_{\rr,0} (y_1,y_2) 
&=\bigg[
b_\NL\bigg(
\frac{1}{2}k_Sn_s'
-\frac{
\big(
\yy_1^2(\yy_2^4+52-2\yy_1^2)
+1\leftrightarrow2
\big)
-23\yy_1^2\yy_2^2-108
}{
2(2-\yy_1^2)(2-\yy_2^2)
}
\bigg)
\nn
&\qquad
+k_Sn_s'
\frac{
3\yy_1^2\yy_2^2
-7(\yy_1^2+\yy_2^2)+16
}{
3(2-\yy_1^2)(2-\yy_2^2)
}
\nn
&\qquad
+\frac{
\big[
\yy_1^2
(2\yy_2^4+57\yy_2^2-442-4\yy_1^2)
+1\leftrightarrow2
\big]+1408
}{
6(2-\yy_1^2)(2-\yy_2^2)
}
\bigg]K_1K_2
\nn
&\quad
+\bigg[
\bigg(
k_Sn_s'
\frac{
2(\yy_1^4-2\yy_1^2-2)
}{
3\yy_1(2-\yy_1^2)
}
\nn
&\qquad
+2b_\NL
\frac{
\yy_2^2(3\yy_1^4-2\yy_1^2+2)
-10\yy_1^4+12\yy_1^2+4
}{
\yy_1(2-\yy_1^2)(2-\yy_2^2)
}
\nn
&\qquad
-\frac{
2\big[
\yy_2^2(29\yy_1^4-102\yy_1^2-118)
-90\yy_1^4+280\yy_1^2+308
\big]
}{
3\yy_1(2-\yy_1^2)(2-\yy_2^2)
}
\bigg)\tilde S_1K_2
+1\leftrightarrow2
\bigg]
\nn
&\quad
+16\bigg[
\frac{
\big[
\yy_1^4(-11\yy_2^2-10)+26\yy_2^2
+1\leftrightarrow2
\big]
+4\yy_1^4\yy_2^4
+4(7\yy_1^2\yy_2^2+6)
}{
3\yy_1\yy_2
(2-\yy_1^2)(2-\yy_2^2)
}
\nn
&\qquad
-b_\NL
\frac{
(\yy_1^4-2\yy_1^2-2)
(\yy_2^4-2\yy_2^2-2)
}{
\yy_1\yy_2
(2-\yy_1^2)(2-\yy_2^2)
}
\bigg]\tilde S_1\tilde S_2,
\end{align}
}
and the nonzero components for $\ell > 0$ are
\begin{align}
	\tilde\cC^{[2,0]}_{\rr,1} (y_1,y_2) &=\tilde\cC^{[2,2]}_{\rr,2} (y_1,y_2) \nn[3pt]
	&\hskip-50pt =\bigg(\frac{b_\NL}{2}+\frac{3\yy_1^2\yy_2^2-7(\yy_1^2+\yy_2^2)+16}{3(2-\yy_1^2)(2-\yy_2^2)}\bigg)K_1K_2+\frac{2}{3}\bigg(\frac{\yy_1^4-2\yy_1^2-2}{\yy_1(2-\yy_1^2)}\tilde S_1K_2+1 \leftrightarrow2\bigg)\,,\\[10pt]
	\tilde\cC^{[2,2]}_{\rr,1} (y_1,y_2)&=-\bigg(4b_\NL+ \frac{(\yy_1^2(\yy_2^4+60\yy_2^2-154-2\yy_1^2)+1\leftrightarrow2)-60\yy_1^2\yy_2^2+392}{\dblue{6}(2-\yy_1^2)(2-\yy_2^2)} \bigg)K_1K_2\nn
	&\quad+\bigg(\frac{2(\yy_2^2(15\yy_1^4-32\yy_1^2-26)-38\yy_1^4+80\yy_1^2+68)}{3\yy_1(2-\yy_1^2)(2-\yy_2^2)} \tilde S_1K_2+1\leftrightarrow 2\bigg)\nn
	&\quad-\frac{32(\yy_1^4-2\yy_1^2-2)(\yy_2^4-2\yy_2^2-2)}{3\yy_1\yy_2(2-\yy_1^2)(2-\yy_2^2)}\tilde S_1\tilde S_2\,,
\end{align}
where $K_i\equiv K_{\delta,\rr}(y_i)$ and $\tilde S_i\equiv \sin\yy_i$.
Then, the time-angle averaged correlation function with $\eta_1 = \eta_2$ becomes 
\begin{align}
	\lim_{\mathbf q \rightarrow \bz,\,y_{1}\gg 1}\frac{\overline{\overline{\expval{ \zeta_I(\mathbf q) \delta(\mathbf k_1,\eta_1) \delta(\mathbf k_2,\eta_1)}'}}}{P_{\zeta_I}(q) P_{\zeta_I}(k_S)}\bigg|_{\mathcal O(q^2)}
	&= \dblue{\left(\frac{104}{45} - 3032 b_\NL - \frac{8}{3}(n_s-1) \right) \left( \frac{q}{k_S} \right)^2 \yy_1^2 \,,}
	 \label{eq:time_angle_ave_3pt_q2_new}
\end{align}
where the double overline means both the time and angle average.
Even if we substitute the single-field condition $b_\NL = (1-n_s)/4$, the expression still has an $(n_s-1)$-dependent term as 
\begin{align}
	\lim_{\mathbf q \rightarrow \bz,\,y_1\gg 1}\frac{\overline{\overline{\expval{ \zeta_I(\mathbf q) \delta(\mathbf k_1,\eta_1) \delta(\mathbf k_2,\eta_1)}'}}}{P_{\zeta_I}(q)P_{\zeta_I}(k_S) }\bigg|_{\mathcal O(q^2)}
	&= \dblue{\left(\frac{104}{45} + \frac{2266}{3}(n_s-1) \right) \left( \frac{q}{k_S} \right)^2 \tilde y_1^2 \,.}
	 \label{eq:time_angle_ave_3pt_q2_new2}
\end{align}
Similar to the MD era case, the origin of the $(n_s-1)$-dependent term is the $\mathcal O((v-1) \tilde y^2 \cos \tilde y)$ term in Eq.~(\ref{eq:calJ_r}).
Apart from that, the $(n_s-1)$-independent term is also different from that in synchronous gauge, Eq.~(\ref{eq:time_angle_ave_3pt_r_syn}).
In Appendix~\ref{app:new_to_syn}, we will also see that the $(n_s-1)$ dependent or independent terms in the RD era are modified with the spatial and time shift, respectively, in the transformation from Newtonian to synchronous gauge.

\section{Relations Between Gauges}
\label{app:new_to_syn}

In this appendix, we show how the various consistency relations and separate universe expressions are related in different gauges via a second-order gauge transformation. We begin in \S \ref{app:summary_of_gauge_trans} with the active view of gauge transformations at second order and discuss their interpretation as a change in time slicing and spatial threading associated with the chosen gauge constraints.  In \S \ref{sec:LNG}, we then transform the second-order Newtonian gauge treatment  for local non-Gaussianity in Appendix~\ref{app:cons_rel_newtonian} to synchronous gauge.
By doing so, we  gain geometric intuition as to why the Newtonian gauge consistency relations are simplified in synchronous gauge.  In addition, this provides a cross-check for the general synchronous gauge derivation in  Appendix~\ref{app:secondorder}, which is valid for any type of initial non-Gaussianity.   Finally in \S \ref{sec:3ricci}, we relate the  curvature perturbation to the 3-geometry of the time slicing and show how initial non-Gaussianity in uniform density gauge and the various spatial coordinatizations of synchronous gauge are related.

\subsection{Second-Order Gauge Transformations}
\label{app:summary_of_gauge_trans}

Let us first summarize the gauge transformation properties of the quantities that we use (see Ref.~\cite{Malik:2008im} for a detailed derivation). 
Unlike in the main text, we take the active approach for interpreting $\xi^\mu$ as the generator of the gauge transformation. A (tensorial) quantity $\textbf{T}$ transforms as
\begin{align}
	\widetilde {\textbf{T}} = \ee^{\mathsterling_{\xi}} \textbf{T}\,,\label{eq:gt}
\end{align}
where the tilde denotes a gauge-transformed quantity and $\mathsterling_{\xi}$ is the Lie derivative with respect to $\xi^\mu$. 
Hereafter, we neglect third-order perturbations. 
Then, the perturbation of the quantity is given by
\begin{align}
	\textbf{T} = \bar {\textbf{T}} + \textbf{T}^\fo + \frac{1}{2} \textbf{T}^\so.
\end{align}
Similarly, the generating vector $\xi^\mu=(\alpha,\beta_{,i})$ can be expanded as
\begin{align}
	\xi^\mu = {\xi^\mu}^\fo + \frac{1}{2} {\xi^\mu}^\so = \left(\alpha^\fo + \frac{1}{2}\alpha^\so,\ \beta^\fo_{,i} + \frac{1}{2}\beta^\so_{,i} \right).
	\label{eq:xi_decompose}
\end{align}
Up to second order, perturbations transform under \eqref{eq:gt} as
\begin{align}
\widetilde {\overline{\textbf{T}}} & = {\overline{\textbf{T}}}\, , \\
	\widetilde {\textbf{T}}^\fo &= {\textbf{T}}^\fo + \mathsterling_{\xi^\fo} {\overline{\textbf{T}}}\,, \\	
	\widetilde {\textbf{T}}^\so &= {\textbf{T}}^\so + \mathsterling_{\xi^\so} {\overline{\textbf{T}}} + \mathsterling_{\xi^\fo}^2 {\overline{\textbf{T}}} + 2\mathsterling_{\xi^\fo} {\textbf{T}}^\fo\,.
\end{align}
The Lie derivatives with respect to $\xi^\mu$ of a scalar $\varphi$, a vector $v_\mu$, and a tensor $t_{\mu \nu}$ are given by 
\begin{align}
		\mathsterling_\xi \varphi &= \xi^\lambda \varphi_{,\lambda}\,, \\
		\mathsterling_\xi v_\mu &= v_{\mu,\alpha} \xi^\alpha + v_{\alpha} \xi^\alpha_{, \ \mu}\,, \\
		\mathsterling_\xi t_{\mu\nu} &= t_{\mu\nu,\lambda} \xi^\lambda + t_{\mu\lambda} \xi^\lambda_{, \ \nu} +  t_{\lambda\nu} \xi^\lambda_{, \ \mu}\,.
\end{align}
Since the energy density $\rho$ is a scalar and the background $\bar\rho$ does not transform, the density contrast $\delta=\rho/\bar\rho$ obeys the following scalar transformation rule:
\begin{align}
 	\widetilde{\delta} &= \delta + \frac{{\bar\rho}'}{\bar \rho}\alpha  + \alpha\left( \frac{1}{2}\frac{{\bar\rho}''}{\bar\rho} \alpha + \frac{1}{2}\frac{{\bar\rho}'}{\bar \rho}\alpha' +  \delta' +  \frac{\bar \rho'}{\bar \rho} \delta \right) + \beta_{,k}\left( \delta_{,k} + \frac{1}{2}\frac{{\bar\rho}'}{\bar \rho} \alpha_{,k}\right) . \label{eq:rho_so_trans}
\end{align}
The 4-velocity vector $u^\mu$ obeys the vector transformation rule. 
From the $i$-component, we can obtain the transformation of the velocity potential, 
\begin{align}
	\widetilde {\hat v}^\fo + \widetilde{B}^\fo = \hat v^\fo + B^\fo - \alpha^\fo.
\end{align}
The metric obeys the tensor transformation rule.
The metric perturbations are given by $g_{\mu \nu} = \bar g_{\mu \nu} + \delta g_{\mu \nu}$. Then, we obtain
\begin{align}
    \widetilde {\delta g_{\mu \nu}} =&\  \delta g_{\mu \nu} + \bar g_{\mu\nu,\lambda} {\xi^\lambda} + \bar g_{\mu\lambda} \xi^\lambda_{,\nu} +  \bar g_{\lambda\nu} \xi^\lambda_{,\mu} +   \delta g_{\mu\nu,\lambda} {\xi^\lambda} + \delta g_{\mu\lambda} {\xi^\lambda}_{,\nu} +  \delta g_{\lambda\nu} \xi^\lambda_{,\mu}  \nonumber \\
	& + \frac{1}{2}\bar g_{\mu \nu, \lambda \alpha} {\xi^\lambda} {\xi^\alpha} +\frac{1}{2} \bar g_{\mu \nu, \lambda } \xi^\lambda_{,\alpha} {\xi^\alpha} +  \bar g_{\mu \lambda, \alpha } {\xi^\alpha} \xi^\lambda_{, \nu} + \bar g_{\lambda\nu, \alpha} {\xi^\alpha} \xi^\lambda_{, \mu} +  \bar g_{\lambda \alpha } \xi^\lambda_{,\mu} \xi^\alpha_{, \nu}  \nonumber \\
	& +\frac{1}{2} \bar g_{\mu \lambda} \left(\xi^\lambda_{,\nu \alpha} {\xi^\alpha} + \xi^\lambda_{,\alpha} \xi^\alpha_{,\nu} \right) +\frac{1}{2} \bar g_{\lambda \nu} \left(\xi^\lambda_{, \mu \alpha} {\xi^\alpha} + \xi^\lambda_{, \alpha} \xi^\alpha_{,\mu} \right). \label{eq:g_mn_trans_so}
\end{align}
From the first-order part of the $0i$ and $ij$ components, we get
\begin{align}
	\widetilde B^{(1)} &= B^{(1)} - \alpha^{(1)} + {\beta^{(1)}}'\,, \label{eq:b_fo_trans} \\
	\widetilde{\hat E}^{(1)} & = \hat{E}^{(1)} + \nabla^{2}\beta^{(1)}\,, \label{eq:ehat_fo_trans}\\
	\widetilde \Psi^{(1)} & = \Psi^{(1)} - \mathcal H \alpha^{(1)} - \frac{1}{3}\nabla^2 \beta^{(1)}\,,\label{eq:psi_fo_trans}
\end{align}
and the first and the second order parts of the $00$ components give
\begin{align}
	\widetilde \Phi =&\ \Phi + \mathcal H \alpha + \alpha' + \frac{1}{2}\alpha \big({\alpha}'' + 5 \mathcal H {\alpha}' + (\mathcal H' + 2 \mathcal H^2) \alpha + 4 \mathcal H \Phi + 2 {\Phi}' \big) \nonumber \\
	& +  {\alpha}' ({\alpha}' + 2 \Phi) + \frac{1}{2}\beta_{,k} \left({\alpha}' + \mathcal H \alpha + 2 \Phi \right)_{,k} + \frac{1}{2}\beta_{,k}' \big({\alpha}_{,k} - 2 B_k - {\beta_{,k}'} \big)\,. \label{eq:phi_so_trans}
\end{align}
To go from Newtonian to synchronous gauge, we must erase $\Phi$ while keeping $B_{,i} = 0$. From Eq.~(\ref{eq:phi_so_trans}), we can then relate $\Phi$ in Newtonian gauge to the gauge transformation parameter $\alpha$ as
\begin{align}
	0 =&\ \Phi + \mathcal H \alpha + \alpha' + \frac{1}{2}\alpha \big({\alpha}'' + 5 \mathcal H {\alpha}' + (\mathcal H' + 2 \mathcal H^2) \alpha + 4 \mathcal H \Phi + 2 {\Phi}' \big) \nonumber \\
	& +  {\alpha}' ({\alpha}' + 2 \Phi) + \frac{1}{2}\beta_{,k} \left({\alpha}' + \mathcal H \alpha + 2 \Phi \right)^{,k} + \frac{1}{2}\beta_{,k}' \big({\alpha}_{,k} - {\beta_{,k}'} \big)\,. \label{eq:alpha_new_to_syn}
\end{align}
By imposing $B_{,i} = 0$ in both sides of Eq.~(\ref{eq:b_fo_trans}), we can also relate $\alpha$ to $\beta$ at first order as 
\begin{align}
	\beta^\fo = \int \dd \eta\, \alpha^\fo + C_e\,, \label{eq:beta_new_to_syn}
\end{align}
where $C_e$ is an integration constant.

Before going to the concrete gauge transformation from Newtonian to synchronous, let us briefly comment on uniform density and comoving gauges here. 
The uniform density gauge condition at first order is given by 
\begin{align}
	\delta^\fo = 0, \ \hat E^\fo = 0\,.
\end{align}
Meanwhile, the comoving gauge condition at first order is given by 
\begin{align}	
	\hat v^\fo + B^\fo = 0, \ \hat E^\fo = 0\,.
\end{align}
To go from synchronous to uniform density gauge, the first-order gauge transformation parameters need to satisfy
\begin{align}
	\alpha^\fo = -\frac{\bar \rho}{\bar \rho'} \delta^\fo_\Syn, \quad \beta^\fo = -\nabla^{-2}\hat E^\fo_\Syn\,.
\end{align}
where the subscript ``S'' denotes quantities in synchronous gauge.
 Because $\delta_\Syn \propto (q\eta)^2$ for the growing mode outside the horizon, we have $\alpha^\fo=0$ up to $\mathcal O(q)$.
Using \eqref{eq:rho_so_trans} we can then see  that $\alpha^\so=0$ to $\O(q)$ as well, for the synchronous $f=0$ gauge where $\hat E^\fo_\Syn=\O((q\eta)^2)$.
 Since the time slicing is independent of the induced density perturbation from the $\beta^\fo$ change in the spatial threading between surfaces, the time slicings of synchronous gauge for any $f$ and uniform density gauge coincide in second-order perturbation theory in this limit as well.  We shall see in Eqs.~(\ref{eq:curvpertrelation}) and (\ref{eq:zeta_so})  that this also has the consequence that $\zeta_\delta = -\Psi$ for $f=0$, since the intrinsic 3-geometries of the same time slice must agree.  The equivalence of uniform density and comoving gauge slicing and $\zeta_\delta=\zeta$ follows the same logic because the comoving gauge density perturbation scales as  $(q\eta)^2$ as well.

Synchronous gauge differs from uniform density and comoving gauges in the spatial threading of the time slices.
To move from synchronous gauge to comoving gauge, the first-order gauge transformation parameter should satisfy
\begin{align}
	\alpha^\fo = \hat v^\fo_\Syn, \quad \beta^\fo = -\nabla^{-2}\hat E^\fo_\Syn\,.
 \label{eq:CotoS}
\end{align}
We can again see that the time slicing remains the same to leading order in the superhorizon limit if $\hat v^\fo_\Syn =0$ initially.
For both  uniform density and  comoving gauges, the spatial coordinates differ due to the isotropic condition $\hat E^\fo=0$.
Since $\hat E_\Syn$ depends on $\nabla^2 \sigma$ from Eq.~(\ref{eqij}), the above gauge parameter $\beta$ depends on $\sigma$ itself.
This indicates that, in order to obtain the isotropic threading of spatial coordinates in uniform density or comoving gauges in the presence of the time-dependent anisotropic stress by a coordinate change, we need to perform the time-dependent spatial coordinate change $\beta^{(1)}$ that depends on $\sigma$ at leading $\mathcal O(q^0)$. This complicates the derivation and interpretation of the consistency relation in uniform density or comoving gauges for unequal-time correlators (see also the discussion in the beginning of \S\ref{sec:3point}).

\subsection{Second-Order Kernel from Local Non-Gaussianity}
\label{sec:LNG}

We now derive second-order results for local non-Gaussianity in synchronous gauge by using the results of  Newtonian gauge and performing the gauge transformation between the two.  By doing so we also extract the inhomogeneous kernel in synchronous gauge which is independent of the initial non-Gaussianity. 

\subsubsection{Matter-Dominated Era}

We first discuss the gauge transformation in the MD era.
The synchronous gauge conditions, Eqs.~(\ref{eq:alpha_new_to_syn}) and (\ref{eq:beta_new_to_syn}), leave two free temporal integration constants (free spatial functions) that correspond to the initial choice of synchronous time slicing and spatial coordinates. 
Throughout, we fix the initial time slicing by imposing that $\lim_{\eta\rightarrow 0} \alpha = 0$, which corresponds to $e_\O=0$ in Appendix~\ref{app:secondorder}. 
For the spatial coordinate gauge fixing, we first take $\lim_{\eta\rightarrow 0} \beta = 0$ which corresponds to taking  $f_{\hat E} = 0$ in Appendix~\ref{app:secondorder} or equivalently $f=0$.\footnote{From this fully gauge-fixed $f=0$ synchronous gauge, we can reach any other $f$ via a further time-independent spatial transformation $\beta$ as we shall explicitly show at the end of this section.}
For this case, we obtain 
from Eqs.~(\ref{eq:alpha_new_to_syn}) and (\ref{eq:beta_new_to_syn}),
\begin{align}
	\alpha^\fo(\k,\eta) =&\, \frac{\eta}{5} \Tm(k) \zeta^\fo_I(\k)\,, \\
	{\beta^\fo}(\k,\eta) =&\,  \frac{\eta^2}{10} \Tm(k) \zeta^\fo_I(\k)\,.
\end{align}
Also, we are neglecting the decaying mode in Newtonian gauge, which sets
 $d_\O = 0$ in synchronous gauge and leaves only the growing mode.
During the MD era, we can rewrite Eq.~(\ref{eq:rho_so_trans}) as 
\begin{align}
	\widetilde{\delta} &= \delta - \frac{6}{\eta} \alpha + \alpha \left(\frac{21}{\eta^2} \alpha - \frac{3}{\eta} {\alpha}' +  {\delta}' - \frac{6}{\eta} \delta \right) +  \beta_{,k}\left( \delta_{,k} - \frac{3}{\eta} \alpha_{,k} \right).
	\label{eq:delta_rho_trans_re_md}
\end{align}
At first order, this gives
\begin{align}
	\delta^\fo_\Syn(\mathbf k,\eta) = \frac{y^2}{10}\Tm(k)\zeta_I^\fo(\mathbf k)\,, 
\end{align}	
where we here denote the quantity in synchronous gauge by the subscript ``S'' and the above expression is consistent with Eq.~(\ref{eq:t_delta_m}). 
At  second order, we can express Eq.~(\ref{eq:delta_rho_trans_re_md}) in Fourier space as 
\begin{equation}
	\delta^\so_{\Syn}(\k)
	= \delta^\so_{\New}(\k) - \frac{6}{\eta} \alpha^\so(\k) + \convimpl{ (\mathcal U_{\mm}(u,v,y) + \mathcal V_{\mm}(u,v,y))\conv}\,,
\label{eq:delta_rho_trans_re_md_f}
\end{equation}
with $u=p_1/k$, $v=p_2/k$.
Note that we have omitted the time arguments, and we will also sometimes omit the wavenumber arguments when they are not important.
Here and in the rest of this appendix, we use the shorthand convention
\begin{equation}
\convimpl{ F(\k,\p_1,\p_2) } \equiv
\int \frac{\dd^3 p_1\dd^3 p_2}{(2\pi)^{3}}\delta_D(\k-\p_1-\p_2) F(\k,\p_1,\p_2)\,,
\label{eq:convimpl}
\end{equation}
to denote implicit convolution. 
The $\mathcal U$ and $\mathcal V$  terms correspond to the terms proportional to $\alpha^\fo$ and $\beta^\fo$ in Eq.~(\ref{eq:delta_rho_trans_re_md}), respectively, and are explicitly
given by
\begin{align}
	\mathcal U_{\mm}(u,v,y) &\equiv \frac{36}{25} + \frac{2}{10}\Big( u y K_{\delta,\mm}'(u y)  - 6 K_{\delta,\mm}(u y) +(u\leftrightarrow v)\Big) \, \nn
	 &= -\frac{2}{25}(18 + y^2) - \frac{4}{25}y^2(v-1) - \frac{2}{25}y^2 (v-1)^2 - \frac{2}{25}y^2 u^2 + \O(u^3)\,, \\[3pt]
	\mathcal V_{\mm}(u,v,y) &\equiv  \frac{u^2 + v^2-1}{20} \left( K_{\delta,\mm}(u y) + K_{\delta,\mm}(v y) - \frac{6}{5} \right)y^2\,\nn
	 &= \frac{y^2(12+y^2)}{100} (v-1) + \frac{y^2(12 + 5 y^2)}{200} (v-1)^2 + \frac{y^2(12+y^2)}{200} u^2 + \O(u^3)\,.\label{eq:i_s2_m_n}
\end{align}
Let us here determine $\alpha^\so$. Using Eq.~(\ref{eq:alpha_new_to_syn}), we can obtain
\begin{align}
	0 &= \Phi^\so + \mathcal H \alpha^\so + {\alpha^\so}' + \convimpl{\mathcal W_{\mm}(u,v,y)\conv}\,, \label{eq:alpha_so_new_syn2}
\end{align}
where 
\begin{align}
	\mathcal W_{\mm}(u,v,y) \equiv -\frac{18}{25} - \frac{3(u^2 + v^2-1)}{100} y^2\,.
\end{align}
In Newtonian gauge, $\Phi^\so$ is related to $\Psi^\so$ as~\cite{Inomata:2020cck} 
\begin{align}
	\Phi^\so &= \Psi^\so + 4 (\Phi^\fo)^2 - F_{\mm}^\so\,,\quad {F^\so_{\mm}} = \frac{20}{3}  \nabla^{-2} \hat N_{ij} (\Phi^\fo_{,i} \Phi^\fo_{,j})\,.
\end{align}
In Fourier space, we have
\begin{align}
     F^\so_{\mm}(\mathbf k)  = \convimpl{\frac{20}{3}  \frac{1 + 2(u^2 + v^2) -3(u^2-v^2)^2}{8} \left( \frac{3}{5} \right)^2\conv}\,,
\end{align}
and $\Psi^\so$ during the MD era is given by~\cite{Inomata:2020cck}
\begin{align}
 \Psi^\so (\mathbf k,\eta) &= \convimpl{\cJ_{\Psi,\mm}(u,v,y)\conv}\,,\\[4pt]
 	\mathcal J_{\Psi,\mm}(u,v,y) &\equiv \frac{9(2(u^2+v^2) - 3(u^2 - v^2)^2 - 5 - 10 b_\NL) }{75} - \frac{3(2 + 3 (u^2 +v^2) - 5 (u^2 -v^2)^2)}{700} y^2\,.\nonumber
\end{align}
Combining these expressions, we get 
\begin{align}
	\Phi^\so (\mathbf k,\eta) &= \convimpl{\cJ_{\Phi,\mm}(u,v,y)\conv}\,, \\
	\mathcal J_{\Phi,\mm}(u,v,y) &\equiv \mathcal J_{\Psi,\mm}(u,v,y) + \frac{9}{25} \left(4 - \frac{5}{3} \frac{1 + 2(u^2 + v^2) -3(u^2-v^2)^2}{2} \right).
\end{align}
Plugging these into Eq.~(\ref{eq:alpha_so_new_syn2}), we get the following equation for $\alpha^\so$:
\begin{align}
	{\alpha^\so}' + \frac{2}{\eta} \alpha^\so= -  \convimpl{\left( \mathcal J_{\Phi,\mm}(u,v,y) + \mathcal W_{\mm}(u,v,y) \right) \conv}\,,
\end{align}
whose solution is given by
\begin{align}
	{\alpha^\so} &= k^{-1} \convimpl{\cJ_{\alpha,\mm}(u,v,y) \conv}\,, \\
	\mathcal J_{\alpha,\mm}(u,v,y) &\equiv - \frac{1}{y^2} \int^y_0 \dd \bar y\, \bar y^2 \left( \mathcal J_{\Phi,\mm}(u,v,\bar y) + \mathcal W_{\mm}(u,v,\bar y) \right)  \\
	&= \frac{9}{25} y\left[\frac{10}{9} b_\NL - \frac{4}{3} (v-1) - \frac{56 + y^2}{21}(v-1)^2 + \frac{28 + y^2}{21}u^2 \right] + \mathcal O(u^3)\,,\nonumber
\end{align}
where we have set the integration constant to zero, consistent with our choice of initial time slicing for synchronous observers.
From this expression, we can see that $\mathcal J_{\alpha,\mm}$ is subdominant in the limit $y \gg 1$ compared to $\mathcal V_{\mm}$, which has $\mathcal O(y^4)$ terms (see Eq.~(\ref{eq:i_s2_m_n})).
Combining these expressions, we finally obtain the kernel of $\delta_\Syn^\so$ for $f=0$ as
\begin{align}
	\delta^\so_\Syn(\mathbf k) &= \convimpl{\cI_{\mm}(u,v,y) \conv}\,, \\ 
	\cI^{f=0}_{\mm}(u,v,y) &= -\frac{6}{y} \mathcal J_{\alpha,\mm}(u,v,y) + \mathcal J_{\mm}(u,v,y) + \mathcal U_{\mm}(u,v,y) + \mathcal V_{\mm}(u,v,y) \label{eq:i_m_l_ng} \nn
	&= \frac{1 +4b_\NL - 5(u^2 + v^2)}{20}y^2 + \frac{1 - 2(u^2 + v^2) + 12 u^2 v^2 + u^4 + v^4}{700}y^4  \\
	&= y^2\left[ \frac{b_\NL -1}{5} - \frac{1}{2}(v-1) - \frac{175-4y^2}{700}(v-1)^2 -\frac{35-2y^2}{140} u^2\right] + \O(u^3)\,.\nonumber
\end{align}  
The $\mathcal O(y^4)$ terms, which determine the $\mathcal O(q^2)$ expressions of the three-point function in the late time limit ($y\gg 1$), appear only in $\mathcal J_\mm$ and $\mathcal V_{\mm}$ (see Eq.~(\ref{eq:3pt_new_q2}) for $\mathcal J_\mm$).
In particular, the $\mathcal O(y^4(v-1))$ term in $\mathcal V_{\mm}$ cancels the $\mathcal O(y^4(v-1))$ in $\mathcal J_{\mm}$.  More specifically, the $n_\delta$ terms in the three-point function in the Newtonian gauge at $\mathcal O(q^2)$ in Eq.~(\ref{eq:lhs_rd_qsq_new}) are cancelled by the gauge transformation term coming from $\beta_{,k}\delta_{,k} $ in Eq.~(\ref{eq:delta_rho_trans_re_md}).
In other words, the appearance of the $\mathcal O(q^2)$ term as the dilation effect in Newtonian gauge originates from the difference between the spatial coordinates in Newtonian and synchronous gauges induced by the first-order long-wavelength mode, as anticipated below Eq.~(\ref{eq:lhs_rd_qsq_new}).

We have so far derived the kernel in synchronous gauge with $f=0$ by the gauge transformation from Newtonian gauge. 
We here perform the gauge transformation from $f=0$ to a general $f$ in synchronous gauge, which leads to $\hat E =0 \to -3 f \zeta_I$ in the superhorizon limit. 
From Eq.~(\ref{eq:ehat_fo_trans}), this gauge transformation is just the spatial shift given by 
\begin{align}
	\beta(\mathbf x) = \int \frac{\dd^3 q}{(2\pi)^3} \frac{3 f \, \ee^{i\mathbf q \cdot \mathbf x}}{q^2} \zeta_I(\mathbf q)\,.
	\label{eq:f_gauge_trans}
\end{align}
From Eq.~(\ref{eq:rho_so_trans}), we see that the second-order density is transformed as 
\begin{align}
	\tilde \delta^\so = \delta^\so + 2 \beta^{\fo}_{,i}\delta^\fo_{,i}\,.
	\label{eq:gauge_trans_fneq}
\end{align}
In Fourier space, we can approximate the gauge shift between synchronous gauges  as 
\begin{align}
	\left[2 \beta_{,i}^\fo\delta_{,i}^\fo \right](\mathbf k_1, \eta_1) &\simeq - 6f\convimpl{ \Dm(k_2 \eta_1)\, \frac{\mathbf k_2 \cdot \mathbf q}{q^2} \Tm(k_2)\zeta_I^\fo(\mathbf q) \zeta_I^\fo(\mathbf k_2) }\nonumber \\
	&= - 6f  \convimpl{\Dm(v k_1 \eta_1) \frac{1-v^2 - u^2}{2 u^2} \Tm(k_2)\zeta_I^\fo(\mathbf q) \zeta_I^\fo(\mathbf k_2)}\,, \label{eq:gauge_term}
\end{align}
where we have focused on the kernel for $\expval{\zeta_I(\mathbf q)\delta^\so(\mathbf k_1) \delta^\fo(\mathbf k_2)}$ with $\beta(\mathbf q)$, $u= q/k_1$, and $v=k_2/k_1$.
The kernel from the gauge term then becomes
\begin{align}
	&\mathcal F_{\mm}(u,v,y)= - 3f \left( \Dm(v y) \frac{1-v^2 - u^2}{2 u^2} + (u \leftrightarrow v) \right) \label{eq:gauge_kernel_md} \\ 
	&=3 f \frac{\Dm(y)}{2} \left(1 + 2(v-1) + (v-1)^2 + 2 \frac{(v-1) + \frac{5}{2}(v-1)^2 +2 (v-1)^3 + \frac{1}{2} (v-1)^4}{u^2} \right) \,,
	\nonumber
\end{align}
up to $\O(u^2)$, where we have symmetrized the kernel with respect to $u$ and $v$ in the first line.
The second-order kernel for general $f$ is then given by
\begin{align}
	\cI_\mm(u,v,y) = \cI^{f=0}_{\mm}(u,v,y) + \mathcal F_\mm(u,v,y)\,.
	\label{eq:i_m_general_f}
\end{align}
Using this kernel, we finally obtain the following expression for the local non-Gaussianity ansatz {in synchronous gauge}:
\begin{align}
	\frac{\expval{ \zeta_I(\mathbf q) \delta(\mathbf k_1,\eta_1) \delta(\mathbf k_2,\eta_2)}'}{P_{\zeta_I}(q)\Tm^2(k_S) P_{\zeta_I}(k_S)}
	&=   \sum_{s,t=0}^\infty \tilde{\mathcal Q}_{\mm}^{[s,t]}(y_1,y_2)\left(\frac{q}{k_S}\right)^s \mu^t \,.
	\label{eq:3ptmd_syn_from_new}
\end{align}
The nonzero coefficients of $\tilde{\mathcal Q}^{[s,t]}$ in $s \leq 2$ are 
\begin{align}
	\tilde{\mathcal Q}^{[0,0]}_\mm =&\, \left(\frac{(b_\NL -1) }{25} + \frac{3f}{100} \right)y_1^2 y_2^2\,,\nonumber \\
	\tilde{\mathcal Q}^{[0,2]}_\mm =&\, \frac{3f}{100}( n_\delta-3)\, y_1^2 y_2^2\,,\nonumber\\
	\tilde{\mathcal Q}^{[2,0]}_\mm =&\, \left(\frac{(-7 - 3b_\NL  + (b_\NL-1) n_\delta)}{200} + \frac{3f}{800} (n_\delta - 3)\right)y_1^2 y_2^2 + \frac{(y_1^2 + y_2^2)y_1^2 y_2^2}{700}\,,\nonumber \\	
	\tilde{\mathcal Q}^{[2,2]}_\mm =&\, \left(\frac{(-5 + 55 b_\NL + (6 - 16b_\NL) n_\delta + (b_\NL -1) (n_\delta^2 + k_S n_\delta'))}{200} \right. \nonumber \\
	&\left. \quad
	+ \frac{3f}{400}  (15 - 8 n_\delta + n_\delta^2 + k_S n_\delta') \right)y_1^2 y_2^2 + \frac{y_1^2y_2^2(y_1^2+y_2^2)}{1750} \,,\nonumber\\
	\tilde{\mathcal Q}^{[2,4]}_{\mm} =&\, \dblue{\frac{f}{800} (-105 -15 n_\delta^2 + n_\delta^3 -14 k_S n_\delta' + n_\delta(71 + 3 k_S n_\delta') + k_S^2 n_\delta'') y_1^2 y_2^2}\,.
 \label{eq:tQMD}
\end{align}
Notice again that if we mimic the single-field initial condition by replacing $b_\NL \rightarrow (4-n_\delta(k_S))/4$, then $\tilde{\mathcal Q}^{[0,t]}$ terms become consistent with Eq.~(\ref{eq:general_gauge_cons_rel}), though $b_\NL$ is constant in $k_S$ in the local non-Gaussianity model (\ref{eq:bnl}).
For a more general value of $b_\NL$, there remains a $\O(q^0)$ piece that does not simply reflect the relationship between local and synchronous initial spatial coordinates.  Instead, the squeezed bispectrum represents a modulation of the amplitude of the local short-wavelength power spectrum in the long-wavelength mode, and produces locally observable effects such as the scale dependence of halo bias \cite{Dalal:2007cu}. Moreover, there are no inhomogeneous contributions up to $\mathcal O(q)$ in the MD era, so all effects are determined by the choice of an initial non-Gaussianity.

Finally, with these full expressions for local non-Gaussianity in synchronous gauge via gauge transformation from Newtonian gauge, we can compare the results to  Appendix~\ref{app:secondorder}, for the same initial non-Gaussianity.    To make contact with those results, we first need to extract the initial non-Gaussianity and define the homogeneous term at some initial surface $\eta_\mm$.
Using Eq.~(\ref{eq:i_m_general_f}) and imposing $\mathcal I^{\rm inhom.}_\mm =0$ at $\eta_\mm \rightarrow 0$, we can find the homogeneous part as
\begin{align}
	&\frac{\mathcal I^{\rm hom.}_\mm(u,v,y)}{\Dm(y)} ={} \frac{1 +4b_\NL - 5(u^2 + v^2)}{2}  
	-3f \frac{(1-v^2 - u^2)(u^4+v^4)}{2u^2v^2}\,.
\end{align}
Notice that, in $\eta_\mm\rightarrow 0$, the homogeneous piece is purely in the growing mode $\Dm$ since all of the sourced terms are grouped into the inhomogeneous term.
We have also confirmed that this expression of the homogeneous kernel can be obtained directly by using the expressions of the second-order uniform-density curvature in synchronous gauge.
Specifically, $c_\delta^\so(u,v)$ in Eq.~(\ref{eq:zeta_so}) corresponds to $\mathcal I^{\rm hom.}_\mm(u,v,y)/(10 D_\mm(y))$ and 
\begin{align}
\lim_{y\to 0}\zeta^\so_\delta 
=&{} \lim_{y\to 0}\zeta^\so=- \lim_{y\to 0} \Psi^\so|_{f=0} \nonumber\\
=&{} 2(b_\NL +1) \convimpl{\conv}\,,
\label{eq:curvpertrelation}
\end{align}
where the first line comes from the equivalence of the time slicing between these gauges and that the spatial coordinates are chosen in the same way to make the spatial metric isotropic, i.e.,~isotropic threading.

Subtracting the homogeneous piece from Eq.~(\ref{eq:i_m_general_f})  returns the inhomogeneous result from  Appendix~\ref{app:secondorder},
\begin{align}
			{\cal I}^{\rm inhom.}_{\text{m}}(u,v,y) &=\frac{u^4+v^4+12u^2v^2-2(u^2+v^2)+1}{700}\,  y^4\, ,
\end{align}	
and demonstrates the consistency of these second-order relations between synchronous gauge and Newtonian gauge via gauge transformation.  
Finally, the explicit homogeneous bispectrum terms, or equivalently ${\cal H}_\mm$,  can be extracted from  $\tilde {\mathcal Q} - \mathcal B$ for local non-Gaussianity
\begin{equation}
{\cal H}_\mm(q,k_S,\mu) \dblue{ \frac{y_1^2 y_2^2}{100}}
= \sum_{s,t=0}^{\infty} (\tilde {\mathcal Q} - \mathcal B)^{[s,t]}_\mm(y_1,y_2) 
\left( \frac{q}{k_S} \right)^s \mu^t\,,
\end{equation}
 where \dblue{we have used the relation Eq.~(\ref{eq:3pt_with_ignorance}) and} the concrete expressions of $\mathcal B$ are already shown in Eq.~(\ref{eq:b_m_20_22}) and $\tilde {\mathcal Q}$ in Eq.~(\ref{eq:tQMD}).

\subsubsection{Radiation-Dominated Era}

Next, we discuss the gauge transformation in the RD era. 
Similar to the MD case, we take this in two steps.
We first perform the gauge transformation to the synchronous gauge with $f=0$, and then to the one with a general $f$.
From Eqs.~(\ref{eq:alpha_new_to_syn}) and (\ref{eq:beta_new_to_syn}), we can obtain 
\begin{align}
	\alpha^\fo(\k,\eta) &= k^{-1}K_{\alpha,\rr}(y) \zeta^\fo_I(\mathbf k)\,,\quad \hskip -50pt&&K_{\alpha, \rr}(y) \equiv \frac{2 \sqrt{3} (\tilde y- \sin \tilde y )}{\tilde y^2}\,, \\
	{\beta^\fo}(\k,\eta) &= k^{-2}K_{\beta,\rr}(y) \zeta^\fo_I(\mathbf k)\,,\quad \hskip -50pt&&K_{\beta,\rr}(y)  \equiv 6\left( \frac{\sin \tilde y}{\tilde y}-\tilde{\Ci}(\tilde y) \right)\,.
\end{align}
We have imposed $\alpha, \beta \rightarrow 0$ in the limit of $\eta \rightarrow 0$ to determine the integration constants, similar to the MD era case.
Note again these conditions correspond to taking $d_\O=e_\O=f=0$ in Appendix~\ref{app:secondorder}.
During the RD era, we can rewrite Eq.~(\ref{eq:rho_so_trans}) as 
\begin{align}
	\widetilde{\delta} &= \delta - \frac{4}{\eta}\alpha + \alpha \left( \frac{10}{\eta^2} \alpha - \frac{2}{\eta} {\alpha}' +  {\delta}' - \frac{4}{\eta} \delta \right) + \beta_{ ,k}\left( \delta_{,k} -\frac{2}{\eta} \alpha_{,k} \right) .
	\label{eq:delta_rho_trans_re_rd}
\end{align}
From the first-order equation, we obtain 
\begin{align}
	\delta^\fo_\Syn(\mathbf k) = -\frac{4(2-(2-\tilde y^2)\cos \tilde y - 2\tilde y\sin \tilde y)}{\tilde y^2} \zeta^\fo_I(\k)\,,
\end{align}	
which is consistent with Eq.~(\ref{eq:t_dr_syn_app}).
At second order, we can express Eq.~(\ref{eq:delta_rho_trans_re_rd}) in Fourier space as 
\begin{align}
	\delta^\so_\Syn
	&= \delta^\so_\New - \frac{4}{\eta} \alpha^\so + \convimpl{(\mathcal U_{\rr}(u,v,y) + \mathcal V_{\rr}(u,v,y))\convr}\,,
	\label{eq:delta_rho_trans_re_rd2}
\end{align}
where $\mathcal U$ and $\mathcal V$ correspond to the terms proportional to $\alpha^\fo$ and $\beta^\fo$ in Eq.~(\ref{eq:delta_rho_trans_re_rd}), respectively, and are given by 
\begin{align}
 	\mathcal U_{\rr}(u,v,y) &\equiv \frac{K_{\alpha,\rr}(u y)}{uvy^2} \Big( 10K_{\alpha,\rr}(v y) - 2vy  K_{\alpha,\rr}'(vy) +  v^2y^2 K_{\delta,\rr}'(vy) - 4vy K_{\delta,\rr}(vy)\Big) 
	 + (u \leftrightarrow v) \,, \nn
	\mathcal V_{\rr}(u,v,y) &\equiv  \frac{u^2 + v^2-1}{4}   \left[ \frac{K_{\beta,\rr}(uy)}{u^2}  \left( 2 K_{\delta,\rr}(v y) -\frac{4}{vy} K_{\alpha,\rr}(v y) \right) + (u \leftrightarrow v)  \right] . \label{eq:i_s2_r_n}
\end{align}
We here see the expression of $\alpha^\so$ in Eq.~(\ref{eq:delta_rho_trans_re_rd2}).
Using Eq.~(\ref{eq:alpha_new_to_syn}), we can obtain
\begin{align}
	0 = \Phi^\so + \mathcal H \alpha^\so + {\alpha^\so}'+  \convimpl{\mathcal W_{\rr}(u,v,y)\convr}\,,
	\label{eq:phi_alpha_rr}
\end{align}
where
\begin{align}
	\mathcal W_{\rr}(u,v,y) \equiv&\,  \bigg[ \frac{K_{\alpha,\rr}(uy)}{2uy}\left( vy K_{\alpha,\rr}''(v y) + 5 K_{\alpha,\rr}'(v y)+ \frac{K_{\alpha,\rr}(v y)}{v y} + 4K_{\Phi,\rr}(v y) + 2vy K_{\Phi,\rr}'(vy) \right)  \nonumber \\
	& \qquad\qquad\quad
	+ K_{\alpha,\rr}'(uy)\left(  K_{\alpha,\rr}'(vy) + 2 K_{\Phi,\rr}(v y) \right)  + (u \leftrightarrow v) \bigg] \nonumber \\
	&\quad + \frac{u^2 + v^2-1}{4} \left[\frac{K_{\beta,\rr}(uy)}{u^2} \left( K_{\alpha,\rr}'(vy) + \frac{1}{vy} K_{\alpha,\rr}(vy) + 2 K_{\Phi,\rr}(v y) \right)\right. \nonumber \\
	& \left. \qquad\qquad\qquad\quad + \frac{K_{\beta,\rr}'(uy)}{uv} \left( K_{\alpha,\rr}(vy) -  K_{\beta,\rr}'(vy)\right) + (u \leftrightarrow v)
	 \right]. 
\end{align}
In the RD era, $\Phi^\so$ is related to $\Psi^\so$ as~\cite{Inomata:2020cck}
\begin{align}
	\Phi^\so &= \Psi^\so + 4 (\Phi^\fo)^2 -  F^\so_{\rr}\,,\\
 	F^\so_{\rr} &= \nabla^{-2} \hat N_{ij}\Big[6 \Phi^{\fo}_{,i} \Phi^\fo_{ ,j} + 2\eta \left(\Phi^{\fo}_{,i} \Phi^\fo_{ ,j} \right)' + 2\eta^2 {\Phi^{\fo}_{,i}}' {\Phi^\fo_{,j}}'\Big]\,,
\end{align}
with the corresponding  Fourier space expression
\begin{align}
    F_{\rr}^\so(\mathbf k) &=  \convimpl{\mathcal Z_{\rr}(u,v,y)  \convr}\,,\\[3pt]
	\mathcal Z_{\rr}(u,v,y) &\equiv \frac{1 + 2(u^2 + v^2) -3(u^2-v^2)^2}{16}   \Big[ 6 K_{\Phi,\rr}(uy) K_{\Phi,\rr}(v y)  \nn
 &\quad + 4 u y K_{\Phi,\rr}'(uy) K_{\Phi,\rr}(v y)+ 2 uv y^2 K_{\Phi,\rr}'(uy)K_{\Phi,\rr}'(vy) + (u \leftrightarrow v)  \Big]\,,\nonumber
\end{align}
For $\Psi^\so$, we have
\begin{align}
	\Psi^\so = \convimpl{\cJ_{\Psi,\rr}(u,v,y)\convr}\,, 
\end{align}
where $\mathcal J_{\Psi,\rr}(u,v,y)$ in the squeezed limit ($u\ll 1, |v-1| \simeq \mathcal O(u)$) is given by~\cite{Inomata:2020cck}
\begin{align}
	\mathcal J_{\Psi,\rr}(u,v,y) 
	\simeq & \ \frac{4\left(\left(-8-3 b_\NL + 2 \tilde y^2\right) \sin \tilde y+ (8 + 3 b_\NL) \tilde y \cos \tilde y \right)}{3 \tilde y^3}
	\nn
 &+\frac{2 \left(\left(17-8 \tilde y^2\right) \sin \tilde y+ \tilde y \left(\tilde y^2-17\right) \cos \tilde y \right)}{3 \tilde y^3}(v-1) \nonumber \\
	&-\frac{2 \left(\left(17-7 \tilde y^2\right) \sin \tilde y+ \tilde y \left(4 \tilde y^2-17\right) \cos \tilde y\right)}{3 \tilde y^3} (v-1)^2 \nonumber \\
	&+\frac{2 \left(\left(-2 \tilde y^4+57 \tilde y^2-150\right) \sin \tilde y+ \tilde y \left(13 \tilde y^2+150\right) \cos \tilde y\right)}{45 \tilde y^3}u^2
	 + \mathcal O(u^3)\,.
\end{align}
Combining these expressions, we get 
\begin{align}
	\Phi^\so (\mathbf k,\eta) &= \convimpl{\mathcal J_{\Phi,\rr}(u,v,y)  \convr}\,, \\
	\mathcal J_{\Phi,\rr}(u,v,y) &\equiv \mathcal J_{\Psi,\rr}(u,v,y) + 4 K_{\Phi,\rr}(uy) K_{\Phi,\rr}(vy) - \mathcal Z_{\rr}(u,v,y)\,.
\end{align}
This allows us to rewrite Eq.~(\ref{eq:phi_alpha_rr}) as 
\begin{align}
	{\alpha^\so}' + \frac{1}{\eta} \alpha^\so  = -  \convimpl{\left( \mathcal J_{\Phi,\rr}(u,v,y) + \mathcal W_{\rr}(u,v,y) \right) \convr}\,.
\end{align}
Solving this equation, we obtain 
\begin{align}
	{\alpha^\so} &= k^{-1} \convimpl{\cJ_{\alpha,\rr}(u,v,y) \convr}\,, \\
	\mathcal J_{\alpha,\rr}(u,v,y) &\equiv - \frac{1}{y} \int^y_0 \dd \bar y\, \bar y \left( \mathcal J_{\Phi,\rr}(u,v,\bar y) + \mathcal W_{\rr}(u,v,\bar y) \right) \nonumber \\
	&\hskip -30pt 	= \frac{\sqrt{3} (\tilde y (14 +12 b_\NL +7 \cos \tilde y)-3 (7 + 4 b_\NL) \sin \tilde y)}{3 \tilde y^2} \nonumber \\
		&\hskip -30pt +\frac{\sqrt{3} (\tilde y (-6 \tilde y^2 \tilde{\Ci}(\tilde y) -3
   \tilde y^2-6 \cos \tilde y+4)+(\tilde y^2+2) \sin \tilde y)}{6 \tilde y^2} (v-1)\nonumber \\
   &\hskip -30pt -\frac{\sqrt{3}
    (3\tilde y (-6 \tilde y^2 \tilde{\Ci}(\tilde y) -7
   \tilde y^2+52)+(21 \tilde y^2-128) \sin \tilde y+4 \tilde y (\tilde y^2-7) \cos \tilde y)}{12 \tilde y^2} (v-1)^2\nonumber \\
   &\hskip -30pt
   	-\frac{\sqrt{3}
   (90 \tilde y^2 (\tilde y \tilde{\Ci}(\tilde y) +\sin \tilde y)+3 (19\tilde y^3-540 \tilde y+400 \sin \tilde y)+\left(13 \tilde y^2+420\right) \tilde y \cos \tilde y)}{180
   \tilde y^2}u^2 \nonumber \\
   &\hskip -30pt + \O(u^3)\,,
\end{align}
where we have again set the integration constant to zero to focus on the growing mode.
Combining these expressions, we finally obtain the kernel of $\delta_\Syn^\so$ for $f=0$:
\begin{align}
	\delta^\so_\Syn(\k,\eta) &= \convimpl{\cI_{\rr}(u,v,y)\convr}\,, \\[3pt]
	\mathcal I^{f=0}_{\rr}(u,v,y) &\equiv -\frac{4}{y} \mathcal J_{\alpha,\rr}(u,v,y) + \mathcal J_{\rr}(u,v,y) + \mathcal U_{\rr}(u,v,y) + \mathcal V_{\rr}(u,v,y) \nonumber \\
	& = -\frac{4}{\tilde y^2}\bigg[4 (b_\NL + 1) -2(b_\NL + 1)(2-\tilde y^2)\cos\tilde y-(4(b_\NL + 1)-\tilde y^2)\tilde y\sin\tilde y \nn[-2pt]
	&  \quad+ \bigg(2-\frac{4-2{\tilde y}^2-{\tilde y}^4}{2}\cos{\tilde y}-(2-{\tilde y}^2){\tilde y}\sin{\tilde y}\bigg)(v-1)\label{eq:i_r_f_0}\\
	& \quad -\bigg(15-\frac{180-90{\tilde y}^2+7{\tilde y}^4}{12}\cos{\tilde y}-\frac{180-22{\tilde y}^2-4{\tilde y}^4}{12}{\tilde y}\sin{\tilde y}\bigg)(v-1)^2\nn
	&\quad + \bigg(9+{\tilde y}^2-\frac{324-126{\tilde y}^2-31{\tilde y}^4}{36}\cos{\tilde y}-\frac{324-14{\tilde y}^2+{\tilde y}^4}{36}{\tilde y}\sin{\tilde y}\bigg)u^2\bigg]+{\cal O}(u^3)\,. \nonumber
\end{align}
For the time- and angle-averaged three-point function $\langle\overline{\overline{\zeta \delta \delta}}\rangle$ at $\mathcal O(q^2)$ in the limit of $\tilde y \gg 1$, only the $-\frac{4}{\eta}\alpha \delta$ in Eq.~(\ref{eq:delta_rho_trans_re_rd}) modifies the term independent of $(n_s-1)$ in the Newtonian result Eq.~(\ref{eq:time_angle_ave_3pt_q2_new}), and the $\delta_{,k} \beta^{,k}$ in Eq.~(\ref{eq:delta_rho_trans_re_rd}) cancels the $(n_s-1)$ term in Eq.~(\ref{eq:time_angle_ave_3pt_q2_new}).
This can be easily checked by calculating the time- and angle-averaged three-point function only with the two terms, $-\frac{4}{\eta} \alpha\delta$ and $\delta_{,k} \beta^{,k}$.

Similar to Eq.~(\ref{eq:gauge_kernel_md}), the kernel for the gauge transformation to a general $f$ in the RD era is given by 
\begin{align}
	\mathcal F_{\rr}(u,v,y)= - 3f \left(\Dr(v y) \frac{1-v^2 - u^2}{2 u^2} + (u \leftrightarrow v)\right).
	\label{eq:gauge_kernel_rd}
\end{align}
The kernel for a general $f$ is then given by
\begin{align}
	\mathcal I_\rr(u,v,y) = \mathcal I^{f=0}_{\rr}(u,v,y) + \mathcal F_\rr(u,v,y)\,.
	\label{eq:i_r_general_f}
\end{align}	
With this kernel,  the local non-Gaussianity ansatz in synchronous gauge becomes
\begin{align}
	\frac{\expval{ \zeta_I(\mathbf q) \delta(\mathbf k_1,\eta_1) \delta(\mathbf k_2,\eta_2)}'}{P_{\zeta_I}(q) P_{\zeta_I}(k_S)}
	&=   \sum_{s,t=0}^\infty {\tilde {\mathcal Q}_{\rr}^{[s,t]}}(y_1,y_2) \left(\frac{q}{k_S}\right)^s \mu^t \,.
\end{align}
The nonzero $\tilde{\mathcal Q}$ for $s \leq 2$ are
\begin{align}
	\label{eq:cal_q_r_00}
	&\tilde{\mathcal Q}^{[0,0]}_\rr = (4b_\NL +4 + 3f) D_1 D_2 -4 (S_2 D_1 + S_1 D_2) \,,\\
	&\tilde{\mathcal Q}^{[0,2]}_\rr = 3f((n_s-1)-7) D_1 D_2 + 12 f(S_2 D_1 + S_1 D_2)\,,\\	
	&\tilde{\mathcal Q}^{[1,1]}_\rr = (8b_\NL + 2 (n_s-1))(S_2 D_1 - S_1 D_2)\,,\\
	&\tilde{\mathcal Q}^{[2,0]}_\rr = - 4 S_1S_2 + \frac{1}{72 (\yy_1^2-2)
   (\yy_2^2-2)} \bigg(\big(252 \yy_1^2 \yy_2^2 (1 - b_\NL) + 144 (11-7 b_\NL) \big)D_1 D_2  \nonumber \\
   &\quad +\big[8 S_2 D_1 \left(\yy_1^2-2\right) \left(18
   b_\NL \left(\yy_2^2-2\right)+\yy_2^4-51 \yy_2^2+28\right) +  1 \leftrightarrow 2 \big] \\
   &
   \quad+ \big[\yy_2^2 D_1\left(280 \yy_1^2  +8 (9 (7 b_\NL-9) D_2-70)\right) + 71(\yy_2^2-2)\yy_1^4 D_1 D_2 + 1 \leftrightarrow 2\big] \bigg)  \nn
   &\quad- \frac{3f(7 D_1 D_2-4 S_1 D_2 -4 S_2 D_1)}{8}+ \frac{n_s-1}{2} \left( \frac{4(b_\NL + 1)+3f}{4}  D_1 D_2-  (S_1 D_2 + S_2 D_1)\right) , \nn
&\tilde{\mathcal Q}^{[2,2]}_\rr =-\frac{ 4S_1S_2 \left[ \yy_1^2\yy_2^2(4b_\NL-1)+4(4b_\NL-3) \right] }{ (\yy_1^2-2)(\yy_2^2-2) } +\frac{ (69b_\NL-33)(\yy_1^2\yy_2^2+4) }{ 6(\yy_1^2-2)(\yy_2^2-2) }D_1D_2 \nonumber\\ &\quad +\left[ \left( \frac{ 384\yy_2^2(2b_\NL-1) +192(4b_\NL-2)\yy_1^2 }{ 24(\yy_1^2-2)(\yy_2^2-2) }S_2 +\frac{4\yy_2^2}{\yy_2^2-2} \right)S_1 +1\leftrightarrow2 \right] \nonumber\\ &\quad +\left[ \left( \frac{S_1D_2}{ 24(\yy_1^2-2)(\yy_2^2-2) } \Big( 4\yy_2^2 \left( 36b_\NL\yy_1^2-24b_\NL +5\yy_1^4+32 \right) -288b_\NL\yy_1^2 \right.\right. \nonumber\\ &\left.\left. \hspace{75pt} -64(4-3b_\NL) +12(\yy_1^2-2)\yy_2^4 -40\yy_1^4 \Big) -\frac{(12b_\NL+13)\yy_1^2}{ 3(\yy_1^2-2) } \right) +1\leftrightarrow2 \right] \nonumber\\ &\quad +\left[ \frac{D_1D_2}{ 24(\yy_1^2-2)(\yy_2^2-2) } \left[ (-12b_\NL-13) (\yy_1^2-2)\yy_2^4 +8(33-69b_\NL)\yy_2^2 \right] +1\leftrightarrow2 \right] \nonumber\\ &\quad +\dblue{ \frac{1}{ 3(\yy_1^2-2)(\yy_2^2-2) } \bigg\{ (12b_\NL+13) \left[ D_1\yy_2^2(2-\yy_1^2) +D_2\yy_1^2(2-\yy_2^2) \right] } \nonumber\\ &\dblue{ \hspace{70pt} +48(1-2b_\NL) (\yy_1^2+\yy_2^2)S_1S_2 +(24b_\NL+26) \left( \yy_1^2\yy_2^2-\yy_1^2-\yy_2^2 \right) \bigg\} } \color{black} \nonumber\\ &\quad +\frac{3f}{4} \bigg\{ 32S_1S_2+59D_1D_2 +\left[ \left( \frac{4(4S_2-D_2-4)}{\yy_2^2-2} -\left(44S_2+D_2\yy_1^2+8\right) \right)D_1 +1\leftrightarrow2 \right] \bigg\} \nonumber\\ &\quad +\frac{ k_sn_s'+(n_s-1)^2 }{2} \left[ \left( b_\NL+1+\frac{3f}{2} \right)D_1D_2 -\left(S_1D_2+S_2D_1\right) \right] \nonumber\\ &\quad -(n_s-1) \bigg[ (4b_\NL+3)D_1D_2 -f\left( 6S_1D_2+6S_2D_1-12D_1D_2 \right) +8S_1S_2 -5S_1D_2-5S_2D_1 \bigg]\,
\, , \nn
	\label{eq:cal_q_r_24}   
   &\tilde{\mathcal Q}^{[2,4]}_\rr =f \left\{
   \frac{-693D_1 D_2 (\yy_1^2 \yy_2^2+4)}{8 (\yy_1^2-2) (\yy_2^2-2)}
   +\left[ D_1
   \left(\frac{D_2 \left(20\yy_1^4 \yy_2^2-40 \yy_1^4+1386 \yy_1^2\right)}{8 (\yy_1^2-2) (\yy_2^2-2)}+\frac{20
   \yy_2^2}{\yy_2^2-2}\right. \right. \right.\nonumber \\
   &\left. \left. \left. \quad
   +\frac{S_2 \left(-3 \yy_1^4
   \yy_2^2-\yy_1^2 \yy_2^4+129 \yy_1^2 \yy_2^2+6 \yy_1^4-334 \yy_1^2+2 \yy_2^4-258
   \yy_2^2+668\right)}{2 (\yy_1^2-2) (\yy_2^2-2)}\right) + 1 \leftrightarrow 2\right] \right. \nonumber \\
   &\left.
   \quad -\frac{48 S_1S_2 (\yy_1^2 \yy_2^2+6)}{(\yy_1^2-2)
   (\yy_2^2-2)} + \left[ \left(-\frac{12 \yy_2^2 }{\yy_2^2-2}+\frac{120
   \yy_1^2 }{(\yy_1^2-2)
   (\yy_2^2-2)}S_2\right)S_1 + 1 \leftrightarrow 2\right] \right. \nonumber \\
   &\quad +\left. (n_s-1) \left(\frac{3 k_s n_s' D_1 D_2}{8}+12 S_1S_2 + \frac{239D_1D_2 ( \yy_1^2 \yy_2^2+4)}{8
   (\yy_1^2-2) (\yy_2^2-2)}\right. \right.  \\
   &\left.  \left.
   \quad + \left[ D_2 \left(-\frac{3 S_1
   \left(13 \yy_1^2-30\right)}{2 \left(\yy_1^2-2\right)}-\frac{3 \yy_1^2}{\yy_1^2-2}
   + 
   \frac{D_1 \left(-3 \yy_1^4 \yy_2^2 +6 \yy_1^4-478 \yy_1^2\right)}{8
   (\yy_1^2-2) (\yy_2^2-2)}\right) + 1 \leftrightarrow 2\right]\right.\right. \nonumber \\
   &\left.
   \quad +(n_s-1)^2 \left(\frac{3 S_1 D_2}{2}+D_1 \left(\frac{3
   S_2}{2}-\frac{27 D_2}{8}\right)\right)+\frac{1}{8} (n_s-1)^3 D_1 D_2\right.  \nn
   &\left.
 \quad   +k_s n_s' \left(\frac{3
   S_1 D_2}{2}+D_1 \left(\frac{3 S_2}{2}-\frac{13 D_2}{4}\right)\right) 
   + k_s^2 n_s'' \frac{D_1D_2 ( \yy_1^2 \yy_2^2-2 \yy_1^2-2 \yy_2^2+4)}{8 (\yy_1^2-2) (\yy_2^2-2)} \right\}\, ,\nonumber
\end{align}
where $D_i\equiv D_\rr(y_i)$, $S_i\equiv \yy_i\sin\yy_i$, $n_s'' = \dd^2 n_s(k_S)/\dd k_S^2$, and we have considered the limit of $\eta_\rr/\eta_{1,2} \ll 1$ for simplicity, as we did in \S\ref{sec:second_order}.

To see the correspondence to Appendix~\ref{app:secondorder}, similar to the MD era case, we here split the kernel as $\mathcal I_\rr = \mathcal I^{\rm hom.}_\rr + \mathcal I^{\rm inhom.}_\rr$.
Imposing $\mathcal I^{\rm inhom.} = 0$ at $\eta_\rr \rightarrow 0$ in Eqs.~(\ref{eq:i_r_f_0})--(\ref{eq:i_r_general_f}), we obtain the following expression of the homogeneous kernel:
\begin{align}
	&\mathcal I^{\rm hom.}_{\rr}(u,v,y) =  \Dr(y)\bigg[\frac{1}{2} \left( 4(b_\NL -1) - 10 (v-1) - 5 (v-1)^2 - 5 u^2 \right)   \\
	&
	+ 3f \left( \frac{1}{2} + (v-1) + \frac{1}{2} (v-1)^2 + \frac{(v-1) + \frac{5}{2}(v-1)^2 + 2(v-1)^3 + \frac{1}{2}(v-1)^4}{u^2} \right) + \O(y_I^2)\bigg]\,,\nonumber
\end{align}
where we have also confirmed that this expression of the homogeneous kernel can be obtained directly by using the expressions of the second-order uniform-density curvature in synchronous gauge, similar to the MD era case. 
Also, after some calculations, one can find that the inhomogeneous part ($\mathcal I^{\rm inhom.}_\rr = \mathcal I_\rr - \mathcal I^{\rm hom.}_\rr$) is the same as  Eq.~(\ref{eq:i_r_inhom}) in the same limit.

Similar to the MD era case, we can also extract the explicit homogeneous bispectrum terms or equivalently ${\cal H}_\rr$  from  $\tilde {\mathcal Q} - \mathcal B$ for local non-Gaussianity
\begin{equation}
{\cal H}_\rr(k_S,\mu,\eta_\rr) \dblue{\frac{\Dr(k_1 \eta_1) \Dr(k_2 \eta_2) \Dmr^2(k_S\eta_{\mm/\rr})}{\Dmr(k_1\eta_{\mm/\rr})  \Dmr(k_2\eta_{\mm/\rr} )} } 
= \sum_{s,t=0}^{\infty} (\tilde {\mathcal Q} - \mathcal B)^{[s,t]}_\rr(y_1,y_2) 
\left( \frac{q}{k_S} \right)^s \mu^t,
\end{equation}
 where \dblue{we have used the relation Eq.~(\ref{eq:3pt_with_ignorance}) and} the concrete expressions of $\mathcal B$ are already shown in Eqs.~(\ref{eq:cb_r_inhom20})--(\ref{eq:cb_r_inhom24}) and $\tilde {\mathcal Q}$ in Eqs.~(\ref{eq:cal_q_r_00})--(\ref{eq:cal_q_r_24}).

\subsection{Second-Order Curvature and 3-Geometry }
\label{sec:3ricci}

In this section, we clarify the relation between the intrinsic spatial curvature of the constant-time slice $R_3$ and the curvature perturbation $\zeta_\delta$ at second order, and examine the spatial gauge dependence of these quantities.  Note that the intrinsic 3-geometry depends only on the choice of time slicing, which we have shown is initially the same for synchronous, comoving, and uniform-density gauge as $\eta\rightarrow 0$, but its spatial coordinatization and representation in terms of the trace and tracefree part of the spatial metric can differ.

At second order, the intrinsic spatial curvature is given by~\cite{Malik:2008im}\footnote{This fixes the relative normalization factor of 2 between the second-order and first-order-squared quantities in Eq.~(C.13) of \cite{Malik:2008im}.} 
\begin{align}
    a^2R_3 &=4\nabla^2(\Psi+\tfrac{1}{3}\hat E) -4C_{km,m}C_{kn,n} + 3C_{mn,k}C_{mn,k}-C_{kk,n}C_{mm,n} \\
    &\quad +4 C_{mn}(C_{mn,kk}+C_{kk,mn}-C_{mk,nk}-C_{kn,mk}) + 2(2C_{kk,j}C_{jn,n} - C_{kn,m}C_{mn,k})\,.\nonumber
\end{align}
For the separate universe approach, the intrinsic spatial curvature 
from the long-wavelength mode gives the local FRW curvature in Eq.~(\ref{eq:KL})
\begin{equation}
 R_3 =6 K_\SU/a_\SU^2 \,.
\end{equation}
As usual, the first-order intrinsic curvature is given by the Laplacian of the curvature perturbation $-\Psi-\frac{1}{3}\hE$. 
In synchronous gauge, the second-order intrinsic curvature can be expressed as
\begin{align}
  a^2	R_3 &=4\nabla^2(\Psi+\tfrac{1}{3}\hat E)\nn
  &\quad+6\Psi_{,i}^2-\frac{5}{3}\hat E_{,i}^2+ E_{,ijk}^2 +4(4\Psi+\tfrac{1}{3}\hat E)(\Psi+\tfrac{1}{3}\hat E)_{,kk} -4E_{,ij}(\Psi +\tfrac{1}{3}\hat E)_{,ij}\,. \label{R3}
\end{align}
While the specific form of the intrinsic curvature on superhorizon scales is independent of the background, for concreteness let us consider the MD case. The second-order solutions for the metric and density perturbations are given by 
\begin{align}
   \hat E^{(2)} &= \convimpl{\bigg[\frac{21(u^6+v^6-u^2v^2(u^2+v^2+10))-23(u^4+v^4)-17(u^2+v^2)+19}{11200}y^4 }\nn
   &\qquad \convimpl{+  c_{\hat E}^{(2)}(u,v) y^2 +  f_{\hat E}^{(2)}(u,v)\bigg]\conv}\,,\\
    \Psi^{(2)} &= \convimpl{\bigg[\frac{5(u^4+v^4)+4u^2v^2-10(u^2+v^2)+5}{1400}y^4}\nn
    &\qquad \convimpl{+  c_\Psi^{(2)}(u,v) y^2 +  f_\Psi^{(2)}(u,v)\bigg]\conv}\,,\\
    \delta^{(2)} &= \convimpl{\bigg[\frac{u^4+v^4+12u^2v^2-2(u^2+v^2)+1}{700}y^4 +  c_\delta^{(2)}(u,v) y^2\bigg]\conv}\,,
\end{align}
where we have split the solutions into their inhomogeneous and homogeneous parts. The Einstein equations impose the following relations between the coefficient functions:
\begin{align}
    c_\Psi^{(2)}(u,v) &= c_{\hE}^{(2)}(u,v) +\frac{3(u^2-v^2+1)(u^2-v^2-1)}{40}\nn
    &\quad-9f\frac{((u-v)^2-1)((u+v)^2-1)(u^2+v^2)(u^2+v^2-1)}{40u^2v^2}\,,\\
    f_\Psi^{(2)}(u,v) &= -\frac{1}{3}f_{\hE}^{(2)}(u,v)+\frac{3(u^4+v^4)-6u^2v^2-2(u^2+v^2)-9+40c_{\hE}^{(2)}}{12}\nn
    &\quad -3f\frac{3(u^8+v^8)-6u^4v^4-5(u^6+v^6)-3u^2v^2(u^2+v^2+4)+u^4+v^4+u^2+v^2}{16u^2v^2}\nn
    &\quad + 9f^2\frac{((u-v)^2-1)((u+v)^2-1)(u^2+v^2-1)}{16u^2v^2}\,,\\
    c_{\hE}^{(2)} (u,v)&= -c_{\delta}^{(2)}(u,v) -\frac{3(u^4+v^4)-6u^2v^2+8(u^2+v^2)-3}{40} \nn
    &\quad+ 3f\frac{(3(u^4+v^4)-6u^2v^2+2(u^2+v^2)-5)(u^2+v^2)(u^2+v^2-1)}{160u^2v^2}\,.
\end{align}
In \eqref{R3}, $f_\Psi^{(2)}$ and $f_{\hE}^{(2)}$ cancel, and we can eliminate $c_\Psi^{(2)}$ and $c_{\hE}^{(2)}$ in favor of $c_\delta^{(2)}$. 
Plugging these solutions into \eqref{R3} gives
\begin{align}
 	R_3^{(2)} &= \frac{k^2}{a^2} \convimpl{\bigg[ 40c_\delta^{(2)} (u,v) + \frac{u^4+v^4+6u^2v^2-2(u^2+v^2)+1}{10}y^2\bigg]\conv}\,,
\end{align}
where we have also shown the full $y$-dependent part. For $f\ne 0$, we thus see that $R_3$ only differs due to the change in spatial coordinates assigned to the same invariant curvature, in the same way as for the density perturbation itself, hence all of the dependence on $f$ can be absorbed into $c_\delta^{(2)}$.

We can also compare the above to the second-order curvature perturbation in uniform-density gauge. Note  the quantities that are associated with the expansion of 3-curvature related  metric quantities at first order are no longer directly related to the 3-curvature at second order.  In this case, the ``curvature perturbation'' $\zeta_\delta$ is defined to be the trace of the spatial metric in uniform density gauge. 
From perturbations defined with alternate gauge conditions of $B=\hE=0$, it reads in the superhorizon limit~\cite{Malik:2003mv, Malik:2008im} 
\begin{align}
    \lim_{y\to 0}\zeta_\delta^{(2)}&= -\Psi^{(2)}-\cH\frac{\bar\rho}{\bar\rho'}\delta^{(2)}\nn
    &\  +\cH\frac{\partial_\eta(\bar\rho\delta^{(1)})^2}{\bar\rho'^2}+2\frac{\bar\rho}{\bar\rho'}\delta^{(1)}(\Psi^{(1)'}+2\cH\Psi^{(1)}) + \frac{\bar\rho^2}{\bar\rho'^2}(\delta^{(1)})^2\left(\cH'+2\cH^2-\cH\frac{\bar\rho''}{\bar\rho'}\right).\label{eq:R3y0}
\end{align}
In synchronous gauge where $B=0$, the density perturbation scales as $y^2$, and thus does not survive in the $y\to 0$ limit. In the same limit, the choice of $f=0$ for the threading sets $\hE=0$. This implies that
\begin{align}
    \lim_{y\to 0}\zeta_\delta^{(2)}&= -\Psi^{(2)}|_{f=0}\nn
    &=\convimpl{\bigg(10c_\delta^{(2)}(u,v)|_{f=0}+\frac{3+5(u^2+v^2)}{2}\bigg)\conv}\,,\label{eq:zeta_so}
\end{align}
which gives the relation \eqref{eq:curvpertrelation} for the case of local non-Gaussianity.

\section{Notation and Conventions}
\label{app:notation}

\begin{center}
\renewcommand*{\arraystretch}{1.11}
\begin{longtable}{c p{10cm} c}
\toprule
\multicolumn{1}{c}{\textbf{Symbol}} &
\multicolumn{1}{l}{\textbf{Meaning}} &
\multicolumn{1}{c}{\textbf{Reference}} \\
\midrule
\endfirsthead
\multicolumn{3}{c}
{} \\
\toprule
\multicolumn{1}{c}{\textbf{Symbol}} &
\multicolumn{1}{l}{\textbf{Meaning}} &
\multicolumn{1}{c}{\textbf{Reference}} \\
\midrule
\endhead
\bottomrule
\endfoot
\bottomrule
\endlastfoot
$\zeta$ & Comoving curvature perturbation as a function of ${\bf k},\eta$  &   (\ref{eq:zetasynch}) \\ 
$\zeta_I$ & Initial comoving curvature perturbation $\zeta({\bf k},0)$ &   (\ref{eq:original_cons_rel}) \\ 
$\zeta_\delta$ & Uniform-density curvature perturbation  &   (\ref{eq:bnl}) \\ 
$\Psi$& Trace part of the metric perturbation &   (\ref{eq:def_metric_pertb}) \\ 
$\hat E$& Traceless scalar part of the metric perturbation &   (\ref{eq:e_ij_e_hat}) \\ 
$f$& Ratio between the traceless part and the curvature perturbation in the superhorizon limit, $\lim_{q\rightarrow 0}  \frac{\hat E}{3\zeta} =	- f$&  (\ref{eq:f_def}) \\ 
$\delta$& Density perturbation&  (\ref{eq:delta_def}) \\ 
$\delta_D$& Dirac delta function&  (\ref{eq:original_cons_rel}) \\ 
\midrule
$\mathbf q$ & Soft momentum, $q\ll k_1,k_2$  &   (\ref{eq:original_cons_rel}) \\ 
$\mathbf k_S$ & Hard momentum, $\mathbf k_S = (\mathbf k_1 - \mathbf k_2)/2$  &   (\ref{eq:original_cons_rel}) \\ 
$\mu$ & Angle between soft and hard momenta, $\mu \equiv \hat{k}_S \cdot \hat q$  &   (\ref{eq:general_gauge_cons_rel}) \\ 
$\eta$ & Conformal time  &   (\ref{eq:def_metric_pertb}) \\ 
$\eta_\mm$ & Conformal time at the beginning of the MD era &   (\ref{eq:MDinitial}) \\ 
$\eta_\rr$ & Conformal time at the beginning of the RD era &   (\ref{eq:RDinitial}) \\ 
$y$ & Dimensionless time variable, $y \equiv k \eta$  &   (\ref{eq:delta_zeta_rel}) \\  
$y_{1,2}$ & Dimensionless time variable with $k_S$ and $\eta_{1,2}$, $y_{1,2} \equiv k_S \eta_{1,2}$  &   (\ref{eq:3pt_with_ignorance}) \\  
$\yy$ & Dimensionless time variable (rescaled by sound speed), \mbox{$\yy \equiv k \eta/\sqrt{3}$}  &   (\ref{eq:Tdrr}) \\  
$\yy_{1,2}$ & Dimensionless time variable (rescaled by sound speed) with $k_S$ and $\eta_{1,2}$, $\yy_{1,2} \equiv k_S \eta_{1,2}/\sqrt{3}$  &   (\ref{eq:cb_00_f}) \\  
$\widetilde \Ci(x)$ & Modified cosine integral, $\widetilde \Ci(x) \equiv \Ci(x) - \log x - \gamma_\text{E} + 1$ &   (\ref{eq:hat_e_r}) \\ 
$\convimpl{F}$ & Implicit convolution of $F$ & (\ref{eq:convimpl}) \\
\midrule
$n_s$ & Tilt of curvature power spectrum, $n_s - 1 \equiv \dd \ln k^3 P_{\zeta_I}/\dd \ln k$  &   (\ref{eq:original_cons_rel}) \\ 
$n_\delta$ & Tilt of  density power spectrum,  $n_\delta \equiv \dd \ln k^3 P_{\delta}/\dd \ln k$ &   (\ref{eq:general_gauge_cons_rel}) \\ 
$\tilde n_s$ & Tilt of curvature power spectrum with the transfer function in the MD era, $\tilde n_s - 1 \equiv \dd \ln k^3 \Tm ^2P_{\zeta_I}/\dd \ln k$ 
&   (\ref{eq:nstns}) \\ 
 $b_\NL$ & Local non-Gaussianity parameter & (\ref{eq:bnl})  \\ 
 \midrule
$D_{\mm/\rr}$ & Growth function in synchronous gauge
&   (\ref{eq:delta_zeta_rel}), (\ref{eq:trans_delta_mm_syn}), (\ref{eq:Tdrr}) \\ 
$K_{\mm/\rr}$ & Growth function in Newtonian gauge  &   (\ref{eq:sct_new}) \\ 
$T_{\mm/\rr}$ & Transfer function &   (\ref{eq:delta_zeta_rel}) \\ 
$\mathcal I_{\mm/\rr}$ & Second-order density kernel in synchronous gauge &  (\ref{delta2nd}) \\ 
$\mathcal I_{\mm/\rr}^{\rm inhom.}$ & Inhomogeneous part of $\mathcal I_{\mm/\rr}$
 &  (\ref{eq:delta_so_final_exp}), (\ref{eq:inhomogeneousMD}), (\ref{eq:i_r_inhom}) \\ 
$\mathcal J_{\mm/\rr}$ & Second-order density kernel in Newtonian gauge &  (\ref{eq:delta_so_final_exp_newton}) \\ 
 \midrule
 ${\cal H}_{\mm/\rr}$ &  Homogeneous part of $\langle\zeta\delta\delta\rangle$ in synchronous gauge & (\ref{eq:RDinitial}), (\ref{eq:MDinitial}) \\
$\mathcal B^{[s,t]}_{\mm/\rr}$ & Inhomogeneous part of $\langle\zeta\delta\delta\rangle$ in the expansion of $(q/k_S)^s \mu^t$ in synchronous gauge &  (\ref{eq:3pt_with_ignorance}) \\
$\tilde{\mathcal Q}^{[s,t]}_{\mm/\rr}$ & $\langle\zeta\delta\delta\rangle$ in the expansion of $(q/k_S)^s \mu^t$ with the local non-Gaussianity ansatz in synchronous gauge &  (\ref{eq:3ptmd_syn_from_new}) \\
$\tilde {\mathcal C}^{[s,t]}_{\mm/\rr}$ & $\langle\zeta\delta\delta\rangle$ in the expansion of $(q/k_S)^s \mu^t$ with the local non-Gaussianity ansatz in Newtonian gauge&  (\ref{eq:3pt_func_so_cal_new}) \\
\end{longtable}
\enlargethispage{.75cm}
\end{center}

\bibliographystyle{JHEP}
\bibliography{cons_rel_newtonian}

\providecommand{\href}[2]{#2}\begingroup\raggedright\begin{thebibliography}{10}

\bibitem{Mukhanov:1981xt}
V.~F. Mukhanov and G.~V. Chibisov, \emph{{Quantum Fluctuations and a
  Nonsingular Universe}}, {\emph{JETP Lett.} {\bfseries 33} (1981) 532--535}.

\bibitem{Mukhanov:1982nu}
V.~F. Mukhanov and G.~V. Chibisov, \emph{{The Vacuum energy and large scale
  structure of the universe}}, {\emph{Sov. Phys. JETP} {\bfseries 56} (1982)
  258--265}.

\bibitem{Starobinsky:1982ee}
A.~A. Starobinsky, \emph{{Dynamics of Phase Transition in the New Inflationary
  Universe Scenario and Generation of Perturbations}},
  \href{https://doi.org/10.1016/0370-2693(82)90541-X}{\emph{Phys. Lett. B}
  {\bfseries 117} (1982) 175--178}.

\bibitem{Guth:1982ec}
A.~H. Guth and S.~Y. Pi, \emph{{Fluctuations in the New Inflationary
  Universe}}, \href{https://doi.org/10.1103/PhysRevLett.49.1110}{\emph{Phys.
  Rev. Lett.} {\bfseries 49} (1982) 1110--1113}.

\bibitem{Panagiotakopoulos:1982rn}
C.~Panagiotakopoulos, A.~Salam and J.~A. Strathdee, \emph{{Supersymmetric Field
  Theory of Monopoles}},
  \href{https://doi.org/10.1016/0370-2693(82)90508-1}{\emph{Phys. Lett. B}
  {\bfseries 115} (1982) 29--32}.

\bibitem{Bardeen:1983qw}
J.~M. Bardeen, P.~J. Steinhardt and M.~S. Turner, \emph{{Spontaneous Creation
  of Almost Scale - Free Density Perturbations in an Inflationary Universe}},
  \href{https://doi.org/10.1103/PhysRevD.28.679}{\emph{Phys. Rev. D} {\bfseries
  28} (1983) 679}.

\bibitem{Planck:2018vyg}
{\scshape Planck} collaboration, N.~Aghanim et~al., \emph{{Planck 2018 results.
  VI. Cosmological parameters}},
  \href{https://doi.org/10.1051/0004-6361/201833910}{\emph{Astron. Astrophys.}
  {\bfseries 641} (2020) A6},
  [\href{https://arxiv.org/abs/1807.06209}{{\ttfamily 1807.06209}}].

\bibitem{Achucarro:2022qrl}
A.~Ach\'ucarro et~al., \emph{{Inflation: Theory and Observations}},
  \href{https://arxiv.org/abs/2203.08128}{{\ttfamily 2203.08128}}.

\bibitem{Spergel:1999xn}
D.~N. Spergel and D.~M. Goldberg, \emph{{Microwave background bispectrum. 1.
  Basic formalism}},
  \href{https://doi.org/10.1103/PhysRevD.59.103001}{\emph{Phys. Rev. D}
  {\bfseries 59} (1999) 103001},
  [\href{https://arxiv.org/abs/astro-ph/9811252}{{\ttfamily
  astro-ph/9811252}}].

\bibitem{Goldberg:1999xm}
D.~M. Goldberg and D.~N. Spergel, \emph{{Microwave background bispectrum. 2. A
  probe of the low redshift universe}},
  \href{https://doi.org/10.1103/PhysRevD.59.103002}{\emph{Phys. Rev. D}
  {\bfseries 59} (1999) 103002},
  [\href{https://arxiv.org/abs/astro-ph/9811251}{{\ttfamily
  astro-ph/9811251}}].

\bibitem{Komatsu:2001rj}
E.~Komatsu and D.~N. Spergel, \emph{{Acoustic signatures in the primary
  microwave background bispectrum}},
  \href{https://doi.org/10.1103/PhysRevD.63.063002}{\emph{Phys. Rev. D}
  {\bfseries 63} (2001) 063002},
  [\href{https://arxiv.org/abs/astro-ph/0005036}{{\ttfamily
  astro-ph/0005036}}].

\bibitem{Bartolo:2004if}
N.~Bartolo, E.~Komatsu, S.~Matarrese and A.~Riotto, \emph{{Non-Gaussianity from
  inflation: Theory and observations}},
  \href{https://doi.org/10.1016/j.physrep.2004.08.022}{\emph{Phys. Rept.}
  {\bfseries 402} (2004) 103--266},
  [\href{https://arxiv.org/abs/astro-ph/0406398}{{\ttfamily
  astro-ph/0406398}}].

\bibitem{Dalal:2007cu}
N.~Dalal, O.~Dore, D.~Huterer and A.~Shirokov, \emph{{The imprints of
  primordial non-gaussianities on large-scale structure: scale dependent bias
  and abundance of virialized objects}},
  \href{https://doi.org/10.1103/PhysRevD.77.123514}{\emph{Phys. Rev. D}
  {\bfseries 77} (2008) 123514},
  [\href{https://arxiv.org/abs/0710.4560}{{\ttfamily 0710.4560}}].

\bibitem{Slosar:2008hx}
A.~Slosar, C.~Hirata, U.~Seljak, S.~Ho and N.~Padmanabhan, \emph{{Constraints
  on local primordial non-Gaussianity from large scale structure}},
  \href{https://doi.org/10.1088/1475-7516/2008/08/031}{\emph{JCAP} {\bfseries
  08} (2008) 031}, [\href{https://arxiv.org/abs/0805.3580}{{\ttfamily
  0805.3580}}].

\bibitem{Maldacena:2002vr}
J.~M. Maldacena, \emph{{Non-Gaussian features of primordial fluctuations in
  single field inflationary models}},
  \href{https://doi.org/10.1088/1126-6708/2003/05/013}{\emph{JHEP} {\bfseries
  05} (2003) 013}, [\href{https://arxiv.org/abs/astro-ph/0210603}{{\ttfamily
  astro-ph/0210603}}].

\bibitem{Creminelli:2004yq}
P.~Creminelli and M.~Zaldarriaga, \emph{{Single field consistency relation for
  the 3-point function}},
  \href{https://doi.org/10.1088/1475-7516/2004/10/006}{\emph{JCAP} {\bfseries
  10} (2004) 006}, [\href{https://arxiv.org/abs/astro-ph/0407059}{{\ttfamily
  astro-ph/0407059}}].

\bibitem{Cheung:2007sv}
C.~Cheung, A.~L. Fitzpatrick, J.~Kaplan and L.~Senatore, \emph{{On the
  consistency relation of the 3-point function in single field inflation}},
  \href{https://doi.org/10.1088/1475-7516/2008/02/021}{\emph{JCAP} {\bfseries
  02} (2008) 021}, [\href{https://arxiv.org/abs/0709.0295}{{\ttfamily
  0709.0295}}].

\bibitem{Creminelli:2012ed}
P.~Creminelli, J.~Nore\~na and M.~Simonovi\'c, \emph{{Conformal consistency
  relations for single-field inflation}},
  \href{https://doi.org/10.1088/1475-7516/2012/07/052}{\emph{JCAP} {\bfseries
  07} (2012) 052}, [\href{https://arxiv.org/abs/1203.4595}{{\ttfamily
  1203.4595}}].

\bibitem{Hinterbichler:2012nm}
K.~Hinterbichler, L.~Hui and J.~Khoury, \emph{{Conformal Symmetries of
  Adiabatic Modes in Cosmology}},
  \href{https://doi.org/10.1088/1475-7516/2012/08/017}{\emph{JCAP} {\bfseries
  08} (2012) 017}, [\href{https://arxiv.org/abs/1203.6351}{{\ttfamily
  1203.6351}}].

\bibitem{Hinterbichler:2013dpa}
K.~Hinterbichler, L.~Hui and J.~Khoury, \emph{{An Infinite Set of Ward
  Identities for Adiabatic Modes in Cosmology}},
  \href{https://doi.org/10.1088/1475-7516/2014/01/039}{\emph{JCAP} {\bfseries
  01} (2014) 039}, [\href{https://arxiv.org/abs/1304.5527}{{\ttfamily
  1304.5527}}].

\bibitem{Creminelli:2013mca}
P.~Creminelli, J.~Nore\~na, M.~Simonovi\'c and F.~Vernizzi, \emph{{Single-Field
  Consistency Relations of Large Scale Structure}},
  \href{https://doi.org/10.1088/1475-7516/2013/12/025}{\emph{JCAP} {\bfseries
  12} (2013) 025}, [\href{https://arxiv.org/abs/1309.3557}{{\ttfamily
  1309.3557}}].

\bibitem{Horn:2014rta}
B.~Horn, L.~Hui and X.~Xiao, \emph{{Soft-Pion Theorems for Large Scale
  Structure}}, \href{https://doi.org/10.1088/1475-7516/2014/09/044}{\emph{JCAP}
  {\bfseries 09} (2014) 044},
  [\href{https://arxiv.org/abs/1406.0842}{{\ttfamily 1406.0842}}].

\bibitem{Kehagias:2013yd}
A.~Kehagias and A.~Riotto, \emph{{Symmetries and Consistency Relations in the
  Large Scale Structure of the Universe}},
  \href{https://doi.org/10.1016/j.nuclphysb.2013.05.009}{\emph{Nucl. Phys. B}
  {\bfseries 873} (2013) 514--529},
  [\href{https://arxiv.org/abs/1302.0130}{{\ttfamily 1302.0130}}].

\bibitem{Peloso:2013zw}
M.~Peloso and M.~Pietroni, \emph{{Galilean invariance and the consistency
  relation for the nonlinear squeezed bispectrum of large scale structure}},
  \href{https://doi.org/10.1088/1475-7516/2013/05/031}{\emph{JCAP} {\bfseries
  05} (2013) 031}, [\href{https://arxiv.org/abs/1302.0223}{{\ttfamily
  1302.0223}}].

\bibitem{Baldauf:2011bh}
T.~Baldauf, U.~Seljak, L.~Senatore and M.~Zaldarriaga, \emph{{Galaxy Bias and
  non-Linear Structure Formation in General Relativity}},
  \href{https://doi.org/10.1088/1475-7516/2011/10/031}{\emph{JCAP} {\bfseries
  10} (2011) 031}, [\href{https://arxiv.org/abs/1106.5507}{{\ttfamily
  1106.5507}}].

\bibitem{Jeong:2011as}
D.~Jeong, F.~Schmidt and C.~M. Hirata, \emph{{Large-scale clustering of
  galaxies in general relativity}},
  \href{https://doi.org/10.1103/PhysRevD.85.023504}{\emph{Phys. Rev. D}
  {\bfseries 85} (2012) 023504},
  [\href{https://arxiv.org/abs/1107.5427}{{\ttfamily 1107.5427}}].

\bibitem{dePutter:2015vga}
R.~de~Putter, O.~Dor\'e and D.~Green, \emph{{Is There Scale-Dependent Bias in
  Single-Field Inflation?}},
  \href{https://doi.org/10.1088/1475-7516/2015/10/024}{\emph{JCAP} {\bfseries
  10} (2015) 024}, [\href{https://arxiv.org/abs/1504.05935}{{\ttfamily
  1504.05935}}].

\bibitem{Sirko:2005uz}
E.~Sirko, \emph{{Initial conditions to cosmological N-body simulations, or how
  to run an ensemble of simulations}},
  \href{https://doi.org/10.1086/497090}{\emph{Astrophys. J.} {\bfseries 634}
  (2005) 728--743}, [\href{https://arxiv.org/abs/astro-ph/0503106}{{\ttfamily
  astro-ph/0503106}}].

\bibitem{Gnedin:2011kj}
N.~Y. Gnedin, A.~V. Kravtsov and D.~H. Rudd, \emph{{Implementing the DC Mode in
  Cosmological Simulations with Supercomoving Variables}},
  \href{https://doi.org/10.1088/0067-0049/194/2/46}{\emph{Astrophys. J. Suppl.}
  {\bfseries 194} (2011) 46},
  [\href{https://arxiv.org/abs/1104.1428}{{\ttfamily 1104.1428}}].

\bibitem{Li:2014sga}
Y.~Li, W.~Hu and M.~Takada, \emph{{Super-Sample Covariance in Simulations}},
  \href{https://doi.org/10.1103/PhysRevD.89.083519}{\emph{Phys. Rev. D}
  {\bfseries 89} (2014) 083519},
  [\href{https://arxiv.org/abs/1401.0385}{{\ttfamily 1401.0385}}].

\bibitem{Hu:2016ssz}
W.~Hu, C.-T. Chiang, Y.~Li and M.~LoVerde, \emph{{Separating the Universe into
  real and fake energy densities}},
  \href{https://doi.org/10.1103/PhysRevD.94.023002}{\emph{Phys. Rev. D}
  {\bfseries 94} (2016) 023002},
  [\href{https://arxiv.org/abs/1605.01412}{{\ttfamily 1605.01412}}].

\bibitem{Pajer:2013ana}
E.~Pajer, F.~Schmidt and M.~Zaldarriaga, \emph{{The Observed Squeezed Limit of
  Cosmological Three-Point Functions}},
  \href{https://doi.org/10.1103/PhysRevD.88.083502}{\emph{Phys. Rev. D}
  {\bfseries 88} (2013) 083502},
  [\href{https://arxiv.org/abs/1305.0824}{{\ttfamily 1305.0824}}].

\bibitem{Dai:2015rda}
L.~Dai, E.~Pajer and F.~Schmidt, \emph{{Conformal Fermi Coordinates}},
  \href{https://doi.org/10.1088/1475-7516/2015/11/043}{\emph{JCAP} {\bfseries
  11} (2015) 043}, [\href{https://arxiv.org/abs/1502.02011}{{\ttfamily
  1502.02011}}].

\bibitem{Hu:1993tc}
W.~Hu, D.~Scott and J.~Silk, \emph{{Reionization and cosmic microwave
  background distortions: A Complete treatment of second order Compton
  scattering}}, \href{https://doi.org/10.1103/PhysRevD.49.648}{\emph{Phys. Rev.
  D} {\bfseries 49} (1994) 648--670},
  [\href{https://arxiv.org/abs/astro-ph/9305038}{{\ttfamily
  astro-ph/9305038}}].

\bibitem{Dodelson:1993xz}
S.~Dodelson and J.~M. Jubas, \emph{{Reionization and its imprint on the cosmic
  microwave background}},
  \href{https://doi.org/10.1086/175191}{\emph{Astrophys. J.} {\bfseries 439}
  (1995) 503--516}, [\href{https://arxiv.org/abs/astro-ph/9308019}{{\ttfamily
  astro-ph/9308019}}].

\bibitem{Pyne:1995bs}
T.~Pyne and S.~M. Carroll, \emph{{Higher order gravitational perturbations of
  the cosmic microwave background}},
  \href{https://doi.org/10.1103/PhysRevD.53.2920}{\emph{Phys. Rev. D}
  {\bfseries 53} (1996) 2920--2929},
  [\href{https://arxiv.org/abs/astro-ph/9510041}{{\ttfamily
  astro-ph/9510041}}].

\bibitem{Mollerach:1997up}
S.~Mollerach and S.~Matarrese, \emph{{Cosmic microwave background anisotropies
  from second order gravitational perturbations}},
  \href{https://doi.org/10.1103/PhysRevD.56.4494}{\emph{Phys. Rev. D}
  {\bfseries 56} (1997) 4494--4502},
  [\href{https://arxiv.org/abs/astro-ph/9702234}{{\ttfamily
  astro-ph/9702234}}].

\bibitem{Matarrese:1997ay}
S.~Matarrese, S.~Mollerach and M.~Bruni, \emph{{Second order perturbations of
  the Einstein-de Sitter universe}},
  \href{https://doi.org/10.1103/PhysRevD.58.043504}{\emph{Phys. Rev. D}
  {\bfseries 58} (1998) 043504},
  [\href{https://arxiv.org/abs/astro-ph/9707278}{{\ttfamily
  astro-ph/9707278}}].

\bibitem{Creminelli:2004pv}
P.~Creminelli and M.~Zaldarriaga, \emph{{CMB 3-point functions generated by
  non-linearities at recombination}},
  \href{https://doi.org/10.1103/PhysRevD.70.083532}{\emph{Phys. Rev. D}
  {\bfseries 70} (2004) 083532},
  [\href{https://arxiv.org/abs/astro-ph/0405428}{{\ttfamily
  astro-ph/0405428}}].

\bibitem{Bartolo:2004ty}
N.~Bartolo, S.~Matarrese and A.~Riotto, \emph{{Gauge-invariant temperature
  anisotropies and primordial non-Gaussianity}},
  \href{https://doi.org/10.1103/PhysRevLett.93.231301}{\emph{Phys. Rev. Lett.}
  {\bfseries 93} (2004) 231301},
  [\href{https://arxiv.org/abs/astro-ph/0407505}{{\ttfamily
  astro-ph/0407505}}].

\bibitem{Tomita:2005et}
K.~Tomita, \emph{{Relativistic second-order perturbations of nonzero-lambda
  flat cosmological models and CMB anisotropies}},
  \href{https://doi.org/10.1103/PhysRevD.71.083504}{\emph{Phys. Rev. D}
  {\bfseries 71} (2005) 083504},
  [\href{https://arxiv.org/abs/astro-ph/0501663}{{\ttfamily
  astro-ph/0501663}}].

\bibitem{Bartolo:2005fp}
N.~Bartolo, S.~Matarrese and A.~Riotto, \emph{{Non-Gaussianity of Large-Scale
  Cosmic Microwave Background Anisotropies beyond Perturbation Theory}},
  \href{https://doi.org/10.1088/1475-7516/2005/08/010}{\emph{JCAP} {\bfseries
  08} (2005) 010}, [\href{https://arxiv.org/abs/astro-ph/0506410}{{\ttfamily
  astro-ph/0506410}}].

\bibitem{Bartolo:2005kv}
N.~Bartolo, S.~Matarrese and A.~Riotto, \emph{{The full second-order radiation
  transfer function for large-scale cmb anisotropies}},
  \href{https://doi.org/10.1088/1475-7516/2006/05/010}{\emph{JCAP} {\bfseries
  05} (2006) 010}, [\href{https://arxiv.org/abs/astro-ph/0512481}{{\ttfamily
  astro-ph/0512481}}].

\bibitem{Bartolo:2006cu}
N.~Bartolo, S.~Matarrese and A.~Riotto, \emph{{CMB Anisotropies at Second Order
  I}}, \href{https://doi.org/10.1088/1475-7516/2006/06/024}{\emph{JCAP}
  {\bfseries 06} (2006) 024},
  [\href{https://arxiv.org/abs/astro-ph/0604416}{{\ttfamily
  astro-ph/0604416}}].

\bibitem{Bartolo:2006fj}
N.~Bartolo, S.~Matarrese and A.~Riotto, \emph{{CMB Anisotropies at
  Second-Order. 2. Analytical Approach}},
  \href{https://doi.org/10.1088/1475-7516/2007/01/019}{\emph{JCAP} {\bfseries
  01} (2007) 019}, [\href{https://arxiv.org/abs/astro-ph/0610110}{{\ttfamily
  astro-ph/0610110}}].

\bibitem{Pitrou:2008ak}
C.~Pitrou, J.-P. Uzan and F.~Bernardeau, \emph{{Cosmic microwave background
  bispectrum on small angular scales}},
  \href{https://doi.org/10.1103/PhysRevD.78.063526}{\emph{Phys. Rev. D}
  {\bfseries 78} (2008) 063526},
  [\href{https://arxiv.org/abs/0807.0341}{{\ttfamily 0807.0341}}].

\bibitem{Khatri:2008kb}
R.~Khatri and B.~D. Wandelt, \emph{{Crinkles in the last scattering surface:
  Non-Gaussianity from inhomogeneous recombination}},
  \href{https://doi.org/10.1103/PhysRevD.79.023501}{\emph{Phys. Rev. D}
  {\bfseries 79} (2009) 023501},
  [\href{https://arxiv.org/abs/0810.4370}{{\ttfamily 0810.4370}}].

\bibitem{Tomita1967a}
K.~Tomita, \emph{{Non-Linear Theory of Gravitational Instability in the
  Expanding Universe}},
  \href{https://doi.org/10.1143/PTP.37.831}{\emph{Progress of Theoretical
  Physics} {\bfseries 37} (05, 1967) 831--846},
  [\href{https://arxiv.org/abs/https://academic.oup.com/ptp/article-pdf/37/5/831/5234391/37-5-831.pdf}{{\ttfamily
  https://academic.oup.com/ptp/article-pdf/37/5/831/5234391/37-5-831.pdf}}].

\bibitem{1981MNRAS.197..931J}
R.~{Juszkiewicz}, \emph{{On the evolution of cosmological adiabatic
  perturbations in the weakly non-linear regime}},
  \href{https://doi.org/10.1093/mnras/197.4.931}{\emph{Mon. Not. Roy. Astron.
  Soc.} {\bfseries 197} (Dec., 1981) 931--940}.

\bibitem{1983MNRAS.203..345V}
E.~T. {Vishniac}, \emph{{Why weakly non-linear effects are small in a
  zero-pressure cosmology}},
  \href{https://doi.org/10.1093/mnras/203.2.345}{\emph{Mon. Not. Roy. Astron.
  Soc.} {\bfseries 203} (Apr., 1983) 345--349}.

\bibitem{1984MNRAS.209..139J}
R.~{Juszkiewicz}, D.~H. {Sonoda} and J.~D. {Barrow}, \emph{{Non-linear
  gravitational clustering}},
  \href{https://doi.org/10.1093/mnras/209.2.139}{\emph{Mon. Not. Roy. Astron.
  Soc.} {\bfseries 209} (July, 1984) 139--144}.

\bibitem{Suto:1990wf}
Y.~Suto and M.~Sasaki, \emph{{Quasi nonlinear theory of cosmological
  selfgravitating systems}},
  \href{https://doi.org/10.1103/PhysRevLett.66.264}{\emph{Phys. Rev. Lett.}
  {\bfseries 66} (1991) 264--267}.

\bibitem{Makino:1991rp}
N.~Makino, M.~Sasaki and Y.~Suto, \emph{{Analytic approach to the perturbative
  expansion of nonlinear gravitational fluctuations in cosmological density and
  velocity fields}}, \href{https://doi.org/10.1103/PhysRevD.46.585}{\emph{Phys.
  Rev. D} {\bfseries 46} (1992) 585--602}.

\bibitem{Bartolo:2005xa}
N.~Bartolo, S.~Matarrese and A.~Riotto, \emph{{Signatures of primordial
  non-Gaussianity in the large-scale structure of the Universe}},
  \href{https://doi.org/10.1088/1475-7516/2005/10/010}{\emph{JCAP} {\bfseries
  10} (2005) 010}, [\href{https://arxiv.org/abs/astro-ph/0501614}{{\ttfamily
  astro-ph/0501614}}].

\bibitem{Koyama:2018ttg}
K.~Koyama, O.~Umeh, R.~Maartens and D.~Bertacca, \emph{{The observed galaxy
  bispectrum from single-field inflation in the squeezed limit}},
  \href{https://doi.org/10.1088/1475-7516/2018/07/050}{\emph{JCAP} {\bfseries
  07} (2018) 050}, [\href{https://arxiv.org/abs/1805.09189}{{\ttfamily
  1805.09189}}].

\bibitem{Dai:2013kra}
L.~Dai, D.~Jeong and M.~Kamionkowski, \emph{{Anisotropic imprint of
  long-wavelength tensor perturbations on cosmic structure}},
  \href{https://doi.org/10.1103/PhysRevD.88.043507}{\emph{Phys. Rev. D}
  {\bfseries 88} (2013) 043507},
  [\href{https://arxiv.org/abs/1306.3985}{{\ttfamily 1306.3985}}].

\bibitem{Malik:2008im}
K.~A. Malik and D.~Wands, \emph{{Cosmological perturbations}},
  \href{https://doi.org/10.1016/j.physrep.2009.03.001}{\emph{Phys. Rept.}
  {\bfseries 475} (2009) 1--51},
  [\href{https://arxiv.org/abs/0809.4944}{{\ttfamily 0809.4944}}].

\bibitem{Hu:2016wfa}
W.~Hu and A.~Joyce, \emph{{Separate Universes beyond General Relativity}},
  \href{https://doi.org/10.1103/PhysRevD.95.043529}{\emph{Phys. Rev. D}
  {\bfseries 95} (2017) 043529},
  [\href{https://arxiv.org/abs/1612.02454}{{\ttfamily 1612.02454}}].

\bibitem{Baumann:2009ds}
D.~Baumann, \emph{{Inflation}},  in \emph{{Theoretical Advanced Study Institute
  in Elementary Particle Physics}: {Physics of the Large and the Small}},
  pp.~523--686, 2011, \href{https://arxiv.org/abs/0907.5424}{{\ttfamily
  0907.5424}}, \href{https://doi.org/10.1142/9789814327183_0010}{DOI}.

\bibitem{Pajer:2017hmb}
E.~Pajer and S.~Jazayeri, \emph{{Systematics of Adiabatic Modes: Flat
  Universes}}, \href{https://doi.org/10.1088/1475-7516/2018/03/013}{\emph{JCAP}
  {\bfseries 03} (2018) 013},
  [\href{https://arxiv.org/abs/1710.02177}{{\ttfamily 1710.02177}}].

\bibitem{Creminelli:2011rh}
P.~Creminelli, G.~D'Amico, M.~Musso and J.~Norena, \emph{{The (not so) squeezed
  limit of the primordial 3-point function}},
  \href{https://doi.org/10.1088/1475-7516/2011/11/038}{\emph{JCAP} {\bfseries
  11} (2011) 038}, [\href{https://arxiv.org/abs/1106.1462}{{\ttfamily
  1106.1462}}].

\bibitem{Mitsou:2021mgv}
E.~Mitsou and J.~Yoo, \emph{{The spatial gauge-dependence of single-field
  inflationary bispectra}},
  \href{https://doi.org/10.1016/j.physletb.2022.137018}{\emph{Phys. Lett. B}
  {\bfseries 828} (2022) 137018},
  [\href{https://arxiv.org/abs/2109.13154}{{\ttfamily 2109.13154}}].

\bibitem{Endlich:2012pz}
S.~Endlich, A.~Nicolis and J.~Wang, \emph{{Solid Inflation}},
  \href{https://doi.org/10.1088/1475-7516/2013/10/011}{\emph{JCAP} {\bfseries
  10} (2013) 011}, [\href{https://arxiv.org/abs/1210.0569}{{\ttfamily
  1210.0569}}].

\bibitem{Endlich:2013jia}
S.~Endlich, B.~Horn, A.~Nicolis and J.~Wang, \emph{{Squeezed limit of the solid
  inflation three-point function}},
  \href{https://doi.org/10.1103/PhysRevD.90.063506}{\emph{Phys. Rev. D}
  {\bfseries 90} (2014) 063506},
  [\href{https://arxiv.org/abs/1307.8114}{{\ttfamily 1307.8114}}].

\bibitem{Baldauf:2015vio}
T.~Baldauf, U.~Seljak, L.~Senatore and M.~Zaldarriaga, \emph{{Linear response
  to long wavelength fluctuations using curvature simulations}},
  \href{https://doi.org/10.1088/1475-7516/2016/09/007}{\emph{JCAP} {\bfseries
  09} (2016) 007}, [\href{https://arxiv.org/abs/1511.01465}{{\ttfamily
  1511.01465}}].

\bibitem{Wagner:2014aka}
C.~Wagner, F.~Schmidt, C.-T. Chiang and E.~Komatsu, \emph{{Separate Universe
  Simulations}}, \href{https://doi.org/10.1093/mnrasl/slu187}{\emph{Mon. Not.
  Roy. Astron. Soc.} {\bfseries 448} (2015) L11--L15},
  [\href{https://arxiv.org/abs/1409.6294}{{\ttfamily 1409.6294}}].

\bibitem{Ip:2016jji}
H.~Y. Ip and F.~Schmidt, \emph{{Large-Scale Tides in General Relativity}},
  \href{https://doi.org/10.1088/1475-7516/2017/02/025}{\emph{JCAP} {\bfseries
  02} (2017) 025}, [\href{https://arxiv.org/abs/1610.01059}{{\ttfamily
  1610.01059}}].

\bibitem{Schmidt:2018hbj}
A.~S. Schmidt, S.~D.~M. White, F.~Schmidt and J.~St\"ucker, \emph{{Cosmological
  N-Body Simulations with a Large-Scale Tidal Field}},
  \href{https://doi.org/10.1093/mnras/sty1430}{\emph{Mon. Not. Roy. Astron.
  Soc.} {\bfseries 479} (2018) 162--170},
  [\href{https://arxiv.org/abs/1803.03274}{{\ttfamily 1803.03274}}].

\bibitem{Akitsu:2020fpg}
K.~Akitsu, Y.~Li and T.~Okumura, \emph{{Cosmological simulation in tides: power
  spectra, halo shape responses, and shape assembly bias}},
  \href{https://doi.org/10.1088/1475-7516/2021/04/041}{\emph{JCAP} {\bfseries
  04} (2021) 041}, [\href{https://arxiv.org/abs/2011.06584}{{\ttfamily
  2011.06584}}].

\bibitem{Masaki:2020drx}
S.~Masaki, T.~Nishimichi and M.~Takada, \emph{{Anisotropic separate universe
  simulations}}, \href{https://doi.org/10.1093/mnras/staa1579}{\emph{Mon. Not.
  Roy. Astron. Soc.} {\bfseries 496} (2020) 483--496},
  [\href{https://arxiv.org/abs/2003.10052}{{\ttfamily 2003.10052}}].

\bibitem{Stucker:2020fhk}
J.~St\"ucker, A.~S. Schmidt, S.~D.~M. White, F.~Schmidt and O.~Hahn,
  \emph{{Measuring the tidal response of structure formation: anisotropic
  separate universe simulations using treepm}},
  \href{https://doi.org/10.1093/mnras/stab473}{\emph{Mon. Not. Roy. Astron.
  Soc.} {\bfseries 503} (2021) 1473--1489},
  [\href{https://arxiv.org/abs/2003.06427}{{\ttfamily 2003.06427}}].

\bibitem{Eisenstein:1997jh}
D.~J. Eisenstein and W.~Hu, \emph{{Power spectra for cold dark matter and its
  variants}}, \href{https://doi.org/10.1086/306640}{\emph{Astrophys. J.}
  {\bfseries 511} (1997) 5},
  [\href{https://arxiv.org/abs/astro-ph/9710252}{{\ttfamily
  astro-ph/9710252}}].

\bibitem{Zegeye:2021yml}
D.~Zegeye, K.~Inomata and W.~Hu, \emph{{Spectral Distortion Anisotropy from
  Inflation for Primordial Black Holes}},
  \href{https://arxiv.org/abs/2112.05190}{{\ttfamily 2112.05190}}.

\bibitem{Wagner:2015gva}
C.~Wagner, F.~Schmidt, C.-T. Chiang and E.~Komatsu, \emph{{The angle-averaged
  squeezed limit of nonlinear matter N-point functions}},
  \href{https://doi.org/10.1088/1475-7516/2015/08/042}{\emph{JCAP} {\bfseries
  08} (2015) 042}, [\href{https://arxiv.org/abs/1503.03487}{{\ttfamily
  1503.03487}}].

\bibitem{Li:2014jra}
Y.~Li, W.~Hu and M.~Takada, \emph{{Super-Sample Signal}},
  \href{https://doi.org/10.1103/PhysRevD.90.103530}{\emph{Phys. Rev. D}
  {\bfseries 90} (2014) 103530},
  [\href{https://arxiv.org/abs/1408.1081}{{\ttfamily 1408.1081}}].

\bibitem{Sherwin:2012nh}
B.~D. Sherwin and M.~Zaldarriaga, \emph{{The Shift of the Baryon Acoustic
  Oscillation Scale: A Simple Physical Picture}},
  \href{https://doi.org/10.1103/PhysRevD.85.103523}{\emph{Phys. Rev. D}
  {\bfseries 85} (2012) 103523},
  [\href{https://arxiv.org/abs/1202.3998}{{\ttfamily 1202.3998}}].

\bibitem{Creminelli:2011sq}
P.~Creminelli, C.~Pitrou and F.~Vernizzi, \emph{{The CMB bispectrum in the
  squeezed limit}},
  \href{https://doi.org/10.1088/1475-7516/2011/11/025}{\emph{JCAP} {\bfseries
  11} (2011) 025}, [\href{https://arxiv.org/abs/1109.1822}{{\ttfamily
  1109.1822}}].

\bibitem{Pitrou:2013hga}
C.~Pitrou, X.~Roy and O.~Umeh, \emph{{xPand: An algorithm for perturbing
  homogeneous cosmologies}},
  \href{https://doi.org/10.1088/0264-9381/30/16/165002}{\emph{Class. Quant.
  Grav.} {\bfseries 30} (2013) 165002},
  [\href{https://arxiv.org/abs/1302.6174}{{\ttfamily 1302.6174}}].

\bibitem{Goldberger:2013rsa}
W.~D. Goldberger, L.~Hui and A.~Nicolis, \emph{{One-particle-irreducible
  consistency relations for cosmological perturbations}},
  \href{https://doi.org/10.1103/PhysRevD.87.103520}{\emph{Phys. Rev. D}
  {\bfseries 87} (2013) 103520},
  [\href{https://arxiv.org/abs/1303.1193}{{\ttfamily 1303.1193}}].

\bibitem{Inomata:2020cck}
K.~Inomata, \emph{{Analytic solutions of scalar perturbations induced by scalar
  perturbations}},
  \href{https://doi.org/10.1088/1475-7516/2021/03/013}{\emph{JCAP} {\bfseries
  03} (2021) 013}, [\href{https://arxiv.org/abs/2008.12300}{{\ttfamily
  2008.12300}}].

\bibitem{Weinberg:2008zzc}
S.~Weinberg, \emph{{Cosmology}}.
\newblock Oxford University Press, Oxford, UK, 2008.

\bibitem{Weinberg:2003sw}
S.~Weinberg, \emph{{Adiabatic modes in cosmology}},
  \href{https://doi.org/10.1103/PhysRevD.67.123504}{\emph{Phys. Rev. D}
  {\bfseries 67} (2003) 123504},
  [\href{https://arxiv.org/abs/astro-ph/0302326}{{\ttfamily
  astro-ph/0302326}}].

\bibitem{Malik:2003mv}
K.~A. Malik and D.~Wands, \emph{{Evolution of second-order cosmological
  perturbations}},
  \href{https://doi.org/10.1088/0264-9381/21/11/L01}{\emph{Class. Quant. Grav.}
  {\bfseries 21} (2004) L65--L72},
  [\href{https://arxiv.org/abs/astro-ph/0307055}{{\ttfamily
  astro-ph/0307055}}].

\end{thebibliography}\endgroup

\end{document}